# SiO and a super-stellar C/O ratio in the atmosphere of the giant exoplanet WASP-121b

**Thomas M. Evans-Soma[1,2,†], David K. Sing[3,4], Joanna K. Barstow[5], Anjali A. A. Piette[6,7], Jake Taylor[8], Joshua D. Lothringer[9], Henrique Reggiani[10], Jayesh M. Goyal[11], Eva-Maria Ahrer[2], Nathan J. Mayne[12], Zafar Rustamkulov[3], Tiffany Kataria[13], Duncan A. Christie[2], Cyril Gapp[2], Jiayin Dong[14,15], Daniel Foreman-Mackey[14], Soichiro Hattori[16,17], Mark S. Marley[18]**

[1]School of Information and Physical Sciences, University of Newcastle, Callaghan, NSW, Australia
[2]Max-Planck-Institut für Astronomie, Heidelberg, Germany
[3]Department of Earth & Planetary Sciences, Johns Hopkins University, Baltimore, MD, USA
[4]Department of Physics & Astronomy, Johns Hopkins University, Baltimore, MD, USA
[5]School of Physical Sciences, The Open University, Milton Keynes, UK
[6]School of Physics and Astronomy, University of Birmingham, Birmingham, UK
[7]Earth and Planets Laboratory, Carnegie Institution for Science, Washington, DC, USA
[8]Department of Physics, University of Oxford, Oxford, UK
[9]Space Telescope Science Institute, Baltimore, MD, USA
[10]Gemini South, Gemini Observatory, NSF's NOIRLab, La Serena, Chile
[11]School of Earth and Planetary Sciences (SEPS), National Institute of Science Education and Research (NISER), Odisha, India
[12]Department of Physics and Astronomy, Faculty of Environment Science and Economy, University of Exeter, Exeter, UK
[13]NASA Jet Propulsion Laboratory, California Institute of Technology, Pasadena, CA, USA
[14]Center for Computational Astrophysics, Flatiron Institute, New York, NY, USA
[15]Department of Astronomy, University of Illinois at Urbana-Champaign, Urbana, IL, USA
[16]Department of Astronomy, Columbia University, New York, NY, USA
[17]Department of Astrophysics, American Museum of Natural History, New York, NY, USA
[18]Department of Lunar and Planetary Sciences, University of Arizona, Tucson, Arizona, USA
[†]Corresponding author: tom.evans-soma@newcastle.edu.au

**Refractory elements such as iron, magnesium, and silicon can be detected in the atmospheres of ultrahot giant planets. This provides an opportunity to quantify the amount of refractory material accreted during formation, along with volatile gases and ices. However, simultaneous detections of refractories and volatiles have proved challenging, as the most prominent spectral features of associated atoms and molecules span a broad wavelength range. Here, using a single JWST observation of the ultrahot giant planet WASP-121b, we report detections of $H_2O$ (5.5–13.5σ), CO (10.8–12.8σ), and SiO (5.7–6.2σ) in the planet's dayside atmosphere, and $CH_4$ (3.1–5.1σ) in the nightside atmosphere. We measure super-stellar values for the atmospheric C/H, O/H, Si/H, and C/O ratios, which point to the joint importance of pebbles and planetesimals in giant planet formation. The $CH_4$-rich nightside composition is also indicative of dynamical processes, such as strong vertical mixing, having a profound influence on the chemistry of ultrahot giant planets.**

WASP-121b is an 'ultrahot' giant planet with dayside temperatures exceeding ~2,000 K, high enough to prevent refractory elements from condensing. Provided the refractories are not cold-trapped on the cooler nightside, this makes it possible to constrain the refractory and volatile contents of the atmospheric envelope by measuring the dayside gas composition[1]. When combined with key elemental ratios such as the C/O ratio, the derived refractory and volatile abundances can provide insights into the planet's formation history[2]. A practical challenge is that the wavelength coverages of available spectrographs have typically allowed detections of either refractory or volatile species, but not both, with the same observation. This is because the strongest spectral features of major refractory species—including Fe, Mg, Ca, SiO, TiO, and VO—are at near-ultraviolet (near-UV) and optical wavelengths, whereas those of major volatile species such as $H_2O$ and CO occur at infrared wavelengths[3,4].

We have used JWST to observe a 'phase curve' for WASP-121b, in which thermal emission from the system was continuously recorded as the planet completed an orbit around the host star[5]. As expected, the infrared wavelength range spanned by the measurement provides sensitivity to important volatiles such as $H_2O$ and CO in the planetary atmosphere. The precision of JWST also allows us to detect a relatively weak SiO band, delivering abundance constraints for a key refractory species. In addition, the complete orbital phase coverage enables separate characterization of the planet's dayside and nightside temperature profiles and chemical compositions. This, in turn, allows us to explore how dynamical processes shape the global atmospheric chemistry, which can be driven out of equilibrium with local thermobaric conditions if gas transport occurs on sufficiently short timescales[6-9].

## Results

### Observations

We observed WASP-121b using the Near Infrared Spectrograph (NIRSpec) instrument on JWST with the G395H grism, covering the wavelength (λ) range 2.73–5.17 µm. The observation lasted 37.8 h and encompassed two consecutive secondary eclipses and a primary transit. We generated 349 wavelength-dependent 'spectroscopic' phase curves by binning each measured spectrum into 10-pixel-wide bins along the dispersion axis, translating to spectroscopic bin widths of Δλ=6.7 nm and a mean spectral resolving power of λ/Δλ≈600 across the G395H passband. Further details of the observation and data reduction are provided in 'Observations and data reduction' in Methods.

### Phase-curve fitting

To fit the spectroscopic phase curves, we adopted a spherical harmonic dipole to describe the planetary brightness map and a quadratic limb-darkening law for the host star. A linear polynomial in time and the pixel coordinates of the target spectrum on the detector was simultaneously fitted to account for drift in the instrument baseline. We extracted the following three observables to characterize the planetary atmosphere from each spectroscopic phase-curve fit: the ratio of the disk-integrated planetary brightness to the disk-integrated stellar brightness ($F_p/F_\star$) as a function of orbital phase ($\phi$); the phase offset of the planetary brightness map ($\Delta\phi$); and the effective planetary radius ($R_p$). Further details of the phase-curve fitting are given in 'White phase-curve fitting' and 'Spectroscopic phase-curve fitting' in Methods.

We binned the measured $F_p/F_\star$ values into 36 phase bins that avoided the two eclipses and transit, with each phase bin covering 46 min of the planetary orbit (Extended Data Fig. 1). By doing this for each spectroscopic phase curve, we were able to obtain the planetary emission spectrum as a function of orbital phase (see 'Generating the phase-resolved emission spectra' in Methods). Figure 1 shows the resulting emission spectra for the phase bins immediately preceding the secondary eclipse and primary transit, when the dayside and nightside hemispheres of WASP-121b were maximally visible, respectively. To verify the robustness of these spectra, we confirmed that they were in good agreement with those produced by a second independent analysis (see 'Independent Data Reduction' in Methods). Emission spectra for the full set of orbital phase bins are shown in Extended Data Fig. 2.

### Atmospheric retrieval analyses

To infer the atmospheric properties of the dayside and nightside hemispheres, we performed retrieval analyses for the two spectra shown in Fig. 1. We adopted widely used parametric forms for the dayside and nightside pressure-temperature ($PT$) profiles and assumed thermochemical equilibrium. We allowed the elemental abundances of carbon, oxygen, and silicon to vary individually as free parameters, following tests that indicated molecules comprising these atoms accounted for the spectral features in the data. We set the relative abundances of all other heavy elements equal to the solar values and allowed them to vary collectively in the retrievals via an additional metallicity parameter. Further details are given in 'Atmospheric retrieval analyses' in Methods.

The outputs of these retrieval analyses are shown in Fig. 2. The $PT$ profile for the dayside hemisphere exhibits a thermal inversion, in line with previous measurements for WASP-121b[10]. We find that the temperature increases from approximately 2,600 to 3,200 K over the $10^{-1}$ to $10^{-3}$ bar pressure range. Across the same pressure range, the inferred nightside $PT$ profile cools from approximately 1,400 to 1,100 K. To interpret the retrieved abundances for carbon, oxygen, and silicon, we compared them to updated elemental abundances of the host star obtained from high-resolution spectroscopy (see 'Stellar abundances and relative planetary abundances' in Methods). For the dayside spectrum, we measure planetary ratios relative to the equivalent host-star ratios of $(C/H)/(C/H)_\star=23.96^{+6.13}_{-4.75}$, $(O/H)/(O/H)_\star=12.19^{+2.78}_{-2.24}$, and $(Si/H)/(Si/H)_\star=9.89^{+5.99}_{-2.89}$, where quoted values are the medians with $1\sigma$ uncertainties defined by the 16th and 84th percentiles (68% credible interval). The abundances for all three elements are super-stellar, with $(C/H)/(C/H)_\star>11.92$, $(O/H)/(O/H)_\star>7.27$, and $(Si/H)/(Si/H)_\star>2.49$ at 99.99% probability. We derive a C/O ratio of $0.92^{+0.02}_{-0.03}$, which is >1.63 times the stellar ratio of $(C/O)_\star=0.47\pm0.06$ at 99.99% probability. We also determine upper limits for the volatile/refractory ratios of $(C/Si)/(C/Si)_\star<8.07$, $(O/Si)/(O/Si)_\star<4.02$, and $((C+O)/Si)/((C+O)/Si)_\star<5.24$ at 99.99% probability. Posterior distributions for these elemental abundances and the derived ratios are shown in Fig. 3.

The abundances for different chemical species implied by the retrievals are shown in Fig. 2 and the corresponding contributions to the dayside and nightside emission spectra are illustrated in Extended Data Fig. 3. On the dayside, the data trace out spectral features of $H_2O$, SiO, and CO in emission, with the $H_2O$ and SiO features muted considerably owing to thermal dissociation (Fig. 2). The measured amplitudes of the $H_2O$ and CO dayside emission features are comparable to those predicted by three-dimensional (3D) general circulation model (GCM) simulations for WASP-121b published in ref. [11] (Fig. 1), although the GCM spectra do not show an SiO feature. On the nightside, the data exhibit spectral features of $CH_4$ in absorption, rather than $H_2O$ in absorption as had previously been reported for WASP-121b on the basis of HST observations at shorter wavelengths[12]. Using 'free chemistry' retrieval analyses that did not enforce the requirement of thermochemical equilibrium (see 'Atmospheric retrieval analyses' in Methods), we obtain statistically significant detections of dayside $H_2O$ ($5.5$–$13.5\sigma$), dayside CO ($10.8$–$12.8\sigma$), dayside SiO ($5.7$–$6.2\sigma$), and nightside $CH_4$ ($3.1$–$5.1\sigma$).

## Discussion

We have obtained a conclusive detection of SiO in the atmosphere of WASP-121b. Evidence of SiO had previously been reported in the atmosphere of WASP-121b and another ultrahot gas giant, WASP-178b, on the basis of absorption signals measured at near-UV wavelengths[13]. However, these near-UV data could also be explained by MgI and FeII absorption, preventing an unambiguous identification of SiO. With NIRSpec, we are now able to verify the presence of SiO in the atmosphere of WASP-121b and, due to its properties as a strong absorber of shortwave radiation[13], the important role that it is likely playing in driving the dayside thermal inversion. Furthermore, silicon is expected to be one of the primary cloud constituents in hot (~1,000–2,000 K) atmospheres[14] and can provide valuable information about the overall formation history of a planet, as it allows the volatile/refractory ratio to be constrained[1,2]. By delivering a robust measurement of a planetary atmosphere's silicon abundance, our SiO detection stands out from previous detections of atomic silicon[15,16] and silicate clouds[17,18] that lacked abundance constraints.

A GCM simulation for WASP-121b that assumed solar metallicity and included SiO opacity was presented in ref. [19]. The predicted dayside spectrum did not show the 4–4.3 μm SiO band, as it was obscured by overlapping $H_2O$ and CO bands. Our NIRSpec observation demonstrates that in reality this will not always be the case, particularly if an atmosphere is enriched in silicon and oxygen, as we find for WASP-121b. It raises the possibility of further SiO detections with JWST in the future, which would complement efforts at UV wavelengths while being far less hampered by interstellar dust extinction and low UV fluxes of host stars.

Our finding that the C/H, O/H, and C/O of WASP-121b are all super-stellar provides evidence that the planet accreted most of its volatiles as gas that was enriched by inward-drifting pebbles[20-22]. The super-stellar C/O and C/H ratios in particular may indicate that WASP-121b accreted most of its gaseous envelope beyond the $H_2O$ ice line but within the $CH_4$ ice line, where the gas-phase carbon abundance could have been locally elevated by the evaporation of inward-drifting $CH_4$-rich pebbles[22]. If WASP-121b instead accreted most of its volatiles as icy planetesimals between the $H_2O$ and CO ice lines, the oxygen-rich composition of these planetesimals should have produced a sub-stellar C/O ratio[23]. Alternatively, a sub-stellar volatile inventory would be expected if the volatiles were accreted from gas that had not been enriched by evaporating pebbles[21].

While gas accretion seems to have delivered the bulk of volatiles, our finding of a super-stellar Si/H ratio (Fig. 3) suggests that there was also substantial enrichment by refractory solids following accretion of the gaseous envelope[24,25]. We calculate that our measured silicon abundance is consistent with the addition of approximately $21.2^{+11.8}_{-6.6}$ $M_\oplus$ (where $M_\oplus$ is the mass of the Earth) of rocky material to the atmosphere (see 'Accretion of rocky material' in Methods), which is comparable to predictions of planetesimal accretion models[26,27]. By contrast, it may not be possible for pebbles to deliver this quantity of refractory material, as the accretion of solid pebbles should largely cease once the pebble isolation mass is reached[28]. After this point, the planet would continue to accrete volatile-rich gas and obtain an atmosphere that has a sub-stellar silicon abundance or a higher volatile/refractory ratio than we have observed (Fig. 3). For example, in the pebble accretion simulations of ref. [24] that produced super-stellar C/H, O/H, and C/O values, the (C+O)/Si ratio rises to at least ten times the stellar ratio.

Super-stellar C/O constraints have recently been obtained for WASP-121b using ground-based spectroscopy[3,4], underscoring the robustness of this particular result. Combined with their own measurement of a super-stellar volatile/refractory ratio, the authors of ref. [4] proposed a broadly similar formation scenario to the one that we have outlined, with WASP-121b accreting most of its atmospheric envelope in a volatile-rich environment beyond the $H_2O$ ice line. In contrast, the authors of ref. [3] recovered a sub-stellar volatile/refractory ratio, which they interpreted as evidence for formation within the $H_2O$ ice line and accretion dominated by refractory-rich planetesimals rather than volatile ices. Ultimately, these tensions between available datasets will need to be resolved to uncover a clearer picture of how WASP-121b formed.

We also stress the broader difficulties in linking a planet's observed composition to its formation history. Various complications are reviewed in refs. [1,2,22,29] and include: accretion from compositionally distinct regions within the protoplanetary disk due to orbital migration; unknown dust opacity and self-shadowing affecting the thermal structure and associated chemistry of the disk; time-dependent chemical and physical evolution of the disk; and erosion of the planetary core raising the heavy element content of the atmospheric envelope. Further investigation of such effects will be required to rigorously assess the basic qualitative picture we have presented here for the formation of WASP-121b.

Regarding the chemistry of the atmosphere, the evidence we have uncovered for $CH_4$ on the nightside of WASP-121b was not anticipated. One reason for this is that models of hot giant planet atmospheres have tended to assume atmospheric C/O ratios close to the solar value of approximately 0.55 (ref. [30]). This leads to dayside and nightside spectra sculpted by $H_2O$ and CO, as seen for the GCM of ref. [11] (Fig. 1). When the C/O ratio is increased from 0.55 to our measured value of 0.92, the nightside CO abundance changes little, but the $CH_4$ abundance increases by 2–4 orders of magnitude and the $H_2O$ abundance decreases by a similarly large amount (Extended Data Fig. 4). Sequestration of oxygen into the various oxygen-bearing condensates shown in Fig. 2 can further reduce the $H_2O$ abundance[31]. As a result, the atmospheric opacity becomes dominated by $CH_4$, explaining why $H_2O$, CO, and SiO are not detected on the nightside (Extended Data Fig. 3). Although the CO opacity is stronger than the $CH_4$ opacity within narrow line cores across the 4–4.5 μm wavelength range, the effect this has on the emission spectrum is too subtle to be detected at the resolution of the data (Fig. 1).

When the dynamics of the atmosphere are taken into account, a high C/O ratio by itself cannot easily explain our measured abundance of $CH_4$ on the nightside of WASP-121b. Simulations of hot giant planets find that horizontal winds with speeds of ~1–10 km $s^{-1}$ advect gas between the dayside and nightside hemispheres at pressures below ~1 bar (refs. [8,19,32,33]). This would mean that gas traverses the nightside hemisphere of WASP-121b on typical timescales of $\sim 10^4$–$10^6$ s, which is substantially shorter than the timescales required to reach chemical equilibrium. Specifically, we find that the nightside infrared photosphere coincides with pressures of $\sim 10^{-2}$–$10^{-4}$ bar and temperatures of ~1,200 K (Fig. 2), where the timescales for $CH_4$ and CO to reach their local equilibrium abundances would be $\sim 10^{10}$ s and $\sim 10^{15}$ s, respectively[9,34]. Conversely, on the dayside, high temperatures translate to chemical reaction timescales that are shorter than dynamical timescales throughout the observable atmosphere, allowing chemical equilibrium to be established. If horizontal winds dominate the transport of gas throughout the observable atmosphere[8], we would expect gas to reach equilibrium with the dayside conditions and then be quenched as it travels across the nightside, preserving its $CH_4$-poor composition.

We propose instead that vertical mixing on the nightside hemisphere is efficiently transporting $CH_4$-rich gas up to the infrared photosphere from deeper layers of the atmosphere that are in chemical equilibrium. To test this hypothesis, we performed 1D model calculations for the nightside chemistry with vertical mixing parameterized by an eddy diffusion coefficient ($K_{zz}$). We adopted the retrieved nightside $PT$ profile shown in Fig. 2 but started with the dayside chemical composition. The latter can be thought of as a conservative assumption, equivalent to instantaneous advection of gas from the dayside to nightside. We then evolved the composition of the gas over timescales ranging from $10^4$–$10^6$ s for a range of $K_{zz}$ values between $10^9$ $cm^2$ $s^{-1}$ and $10^{13}$ $cm^2$ $s^{-1}$ using a C–H–N–O–S non-equilibrium chemical network[35]. Figure 4 shows the resulting $CH_4$ abundances predicted at the infrared photosphere. We find that the measured $CH_4$ abundance can be reproduced with $K_{zz}$ values of $\sim 10^9$–$10^{12}$ $cm^2$ $s^{-1}$, which are consistent with available constraints[36,37]. The predicted $CH_4$ abundance is also higher than the $H_2O$ abundance, in line with our detection of $CH_4$ and non-detection of $H_2O$ on the nightside. Similar results were obtained when we repeated the calculations using the $PT$ profiles from the GCM simulation of ref. [11] (Extended Data Fig. 5). From these analyses, we deduce that gas mixed vertically from deeper layers of the atmosphere is setting the composition at the infrared photosphere, rather than horizontally quenched $CH_4$-poor gas transported from the dayside, as is often seen in dynamical simulations[7,8].

Vertical mixing also offers a natural explanation for why silicon is not cold-trapped on the nightside hemisphere of WASP-121b, as demonstrated by our detection of SiO on the dayside. Recent microphysical modelling has found that cloud formation should occur rapidly in hot giant planet atmospheres, on timescales of ~1 s (ref. [38]). We thus expect clouds of silicates to form efficiently on the nightside of WASP-121b at pressures of $\sim 10^{-1}$–1 bar, where the measured temperature profile crosses the condensation curves of $MgSiO_3$, $Mg_2SiO_4$, and $SiO_2$ (Fig. 2). As mentioned above, iron and magnesium have also been detected in the atmosphere of WASP-121b, along with numerous other refractory metals such as vanadium, chromium, and nickel[39–41]. The same mixing that we hypothesize is transporting $CH_4$ upwards to the observable atmosphere is likely preventing these refractory species from condensing on the nightside and becoming permanently cold-trapped in the deep atmosphere. A notable exception is titanium, for which non-detections provide strong evidence of being efficiently cold-trapped[40,41]. However, despite conditions being favourable for the formation of various refractory clouds on the nightside, we find that models including clouds do not appreciably improve the fit to the available data (see 'Discussion of the retrieval analyses' in Methods). At the same time, we are unable to rule out clouds at pressures above $\sim 10^{-1}$ bar, as such clouds would not necessarily produce an observable signature (Extended Data Fig. 6).

GCM simulations will be valuable for further investigating the potential role played by nightside clouds. For example, existing cloud-free GCMs predict nightside temperatures that are warmer than we measure, particularly at pressures $>10^{-2}$ bar (Extended Data Fig. 5), resulting in strong emission peaks between absorption bands that are not observed in the data (Fig. 1). Nightside clouds may help to reduce this discrepancy by blocking emission from the deepest, warmest layers of the atmosphere[42–45]. Other effects, such as magnetic influences through Lorentz forces, could also play an important role in lowering nightside temperatures by reducing the efficiency of day-night heat recirculation[46–48].

## FIGURES

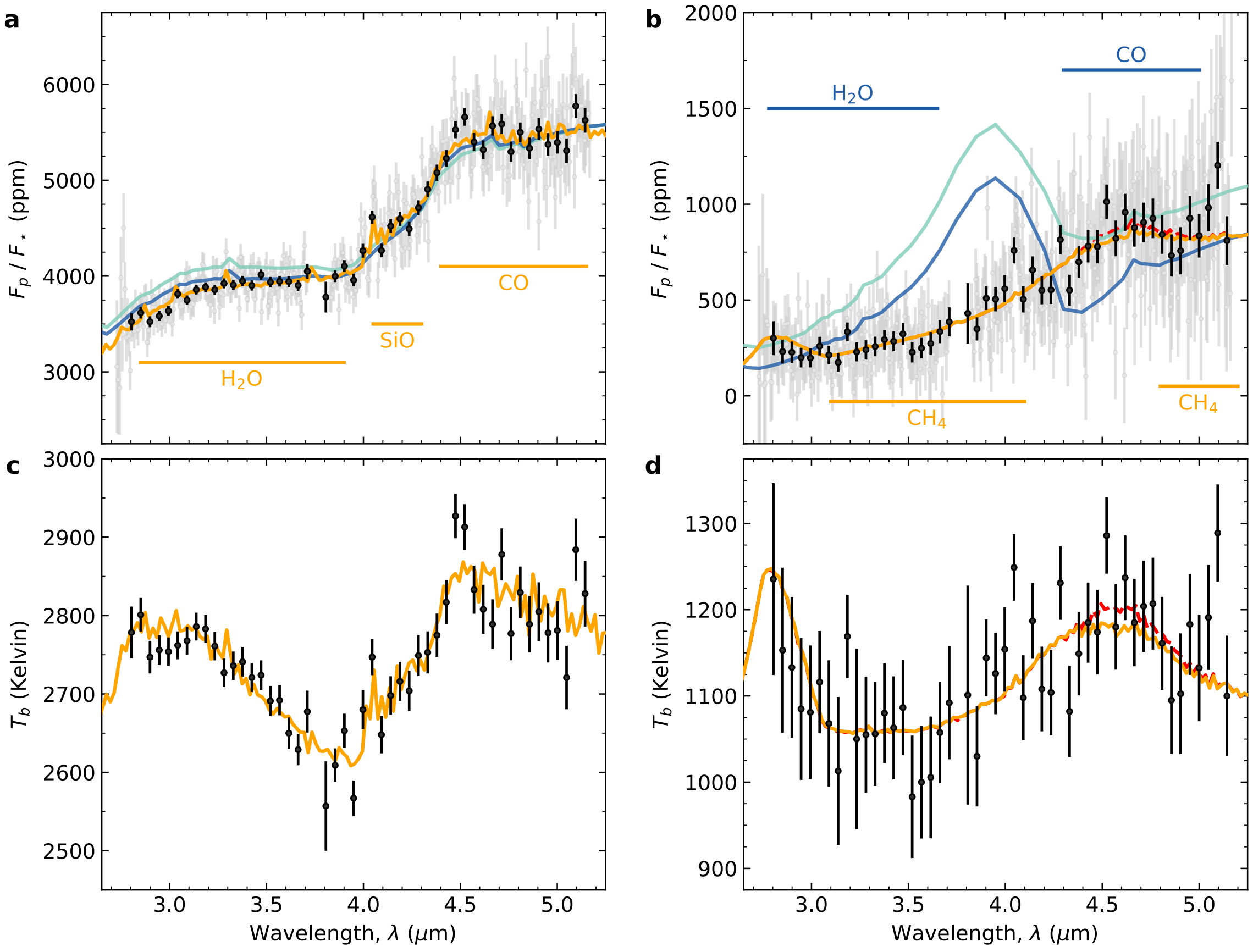

**Fig. 1 | Dayside and nightside emission spectra of WASP-121b. a,** Dayside planet-to-star emission ($F_p/F_\star$) based on $n = 5{,}000$ posterior samples from each of the 349 spectroscopic phase-curve fits. The grey circles show median values and the grey error bars indicate $1\sigma$ uncertainties. The black circles show the median values in 50 equally spaced bins, with black error bars showing $\tilde{\sigma}_i/\sqrt{N_i}$, where $\tilde{\sigma}_i$ is the median $1\sigma$ uncertainty and $N_i$ is the number of data points in the $i$th bin. The orange line shows the maximum a posteriori model that assumes thermochemical equilibrium, with detected molecules labelled below. Predicted dayside spectra from the GCM simulations published in ref. [11] for atmospheric metallicities of $1\times$ solar (light blue line) and $5\times$ solar (dark blue line) are also shown. **b,** The same as **a** for the nightside spectrum. The $H_2O$ and CO absorption features in the GCM spectra are labelled but not seen in the data. The dashed red line shows the spectrum predicted by the maximum a posteriori model (orange line) when CO opacity is switched off. **c,** Wavelength-dependent brightness temperatures ($T_b$) derived from the binned dayside emission data and model shown in **a**. **d,** Wavelength-dependent $T_b$ derived from the binned nightside emission data and models shown in **b**.

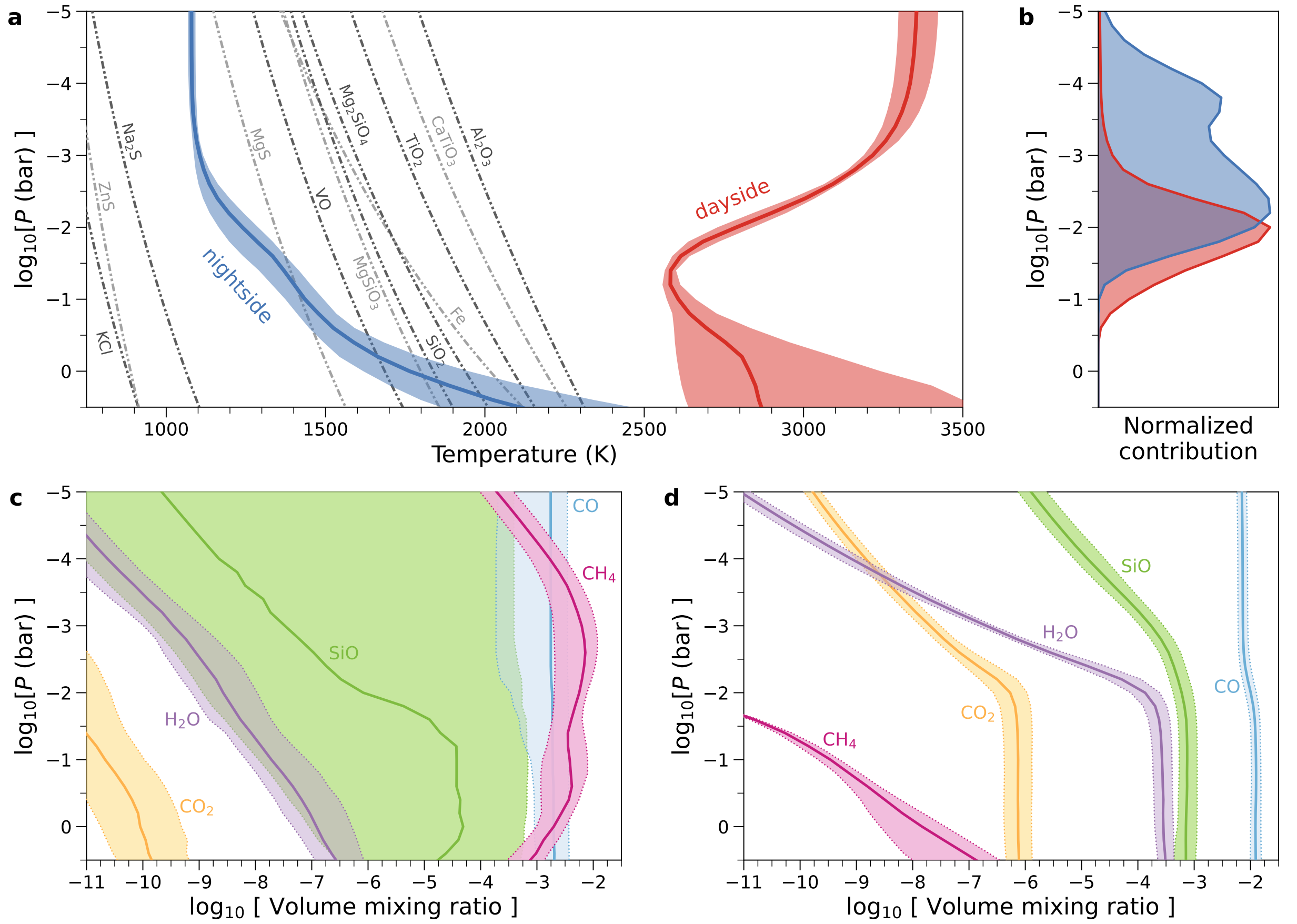

**Fig. 2 | Properties of the atmosphere of WASP-121b. a,** $PT$ profiles inferred from the retrieval analysis for the dayside (red) and nightside (blue) atmospheres assuming thermochemical equilibrium. Condensation curves for refractory species are shown as labelled dash-dot lines. **b,** Normalized contribution functions for the dayside (red) and nightside (blue) atmospheres integrated across the G395H passband, illustrating the relative amount of radiation originating from each pressure level. **c,** Volume mixing ratios (VMRs) inferred for key molecules in the nightside atmosphere. **d,** The same as **c** for the dayside atmosphere. Solid lines in **a**, **c**, and **d** are the median values obtained at each pressure level and the shaded regions indicate the associated $1\sigma$ uncertainties defined by the 16th and 84th percentiles (68% credible interval) based on $n = 1{,}000$ posterior samples.

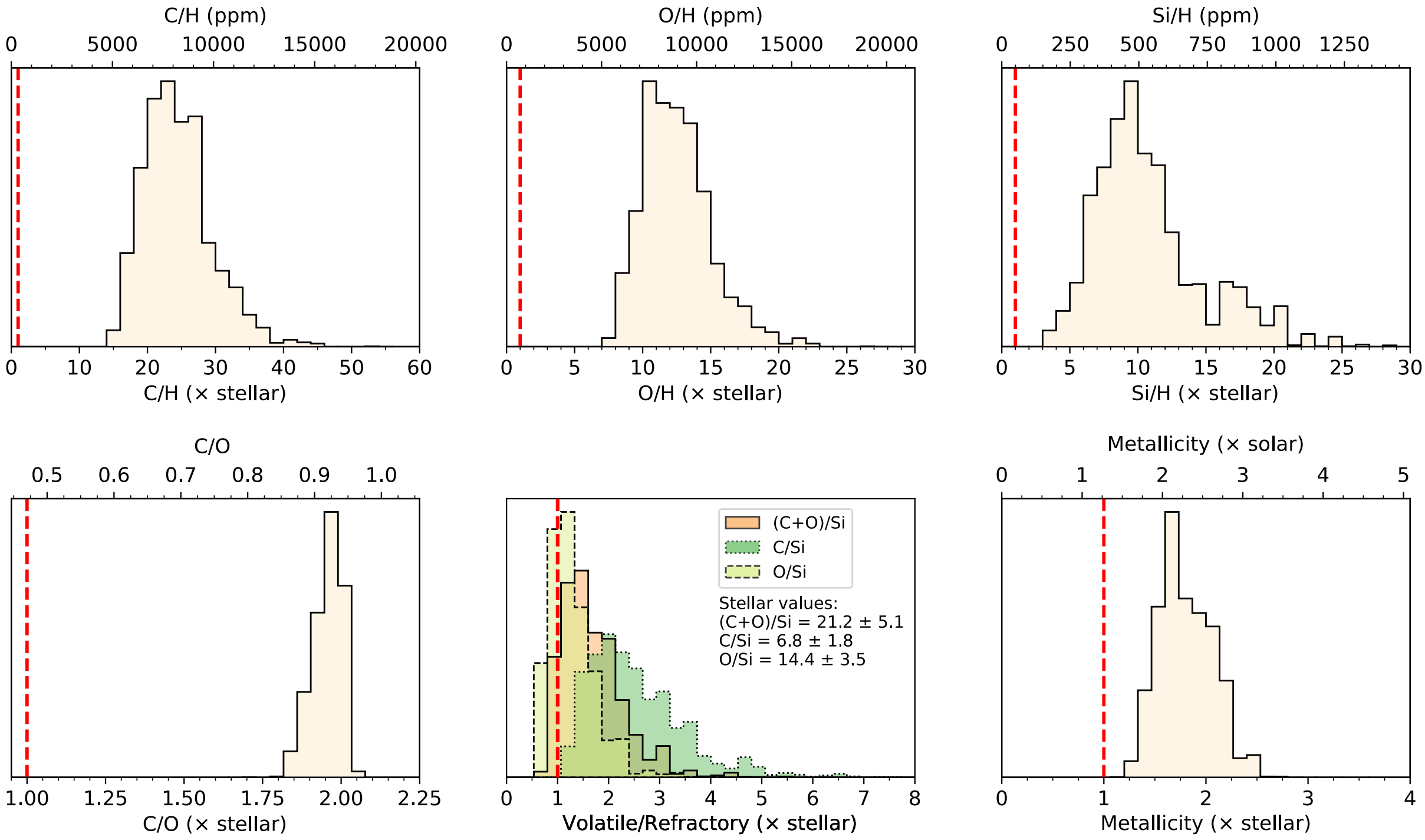

**Fig. 3 l Elemental abundances and derived ratios.** Top: posterior distributions obtained for C/H (left), O/H (middle), and Si/H (right) ratios from the retrieval analysis for the dayside atmosphere assuming thermochemical equilibrium. Bottom: posterior histograms derived for the C/O ratio (left), volatile/refractory ratios (middle), and metallicity (right). The metallicity regulated the abundances of heavy elements besides C, O, and Si, as described in the text. Vertical dashed lines indicate the corresponding values measured for the host star using high-resolution spectroscopy.

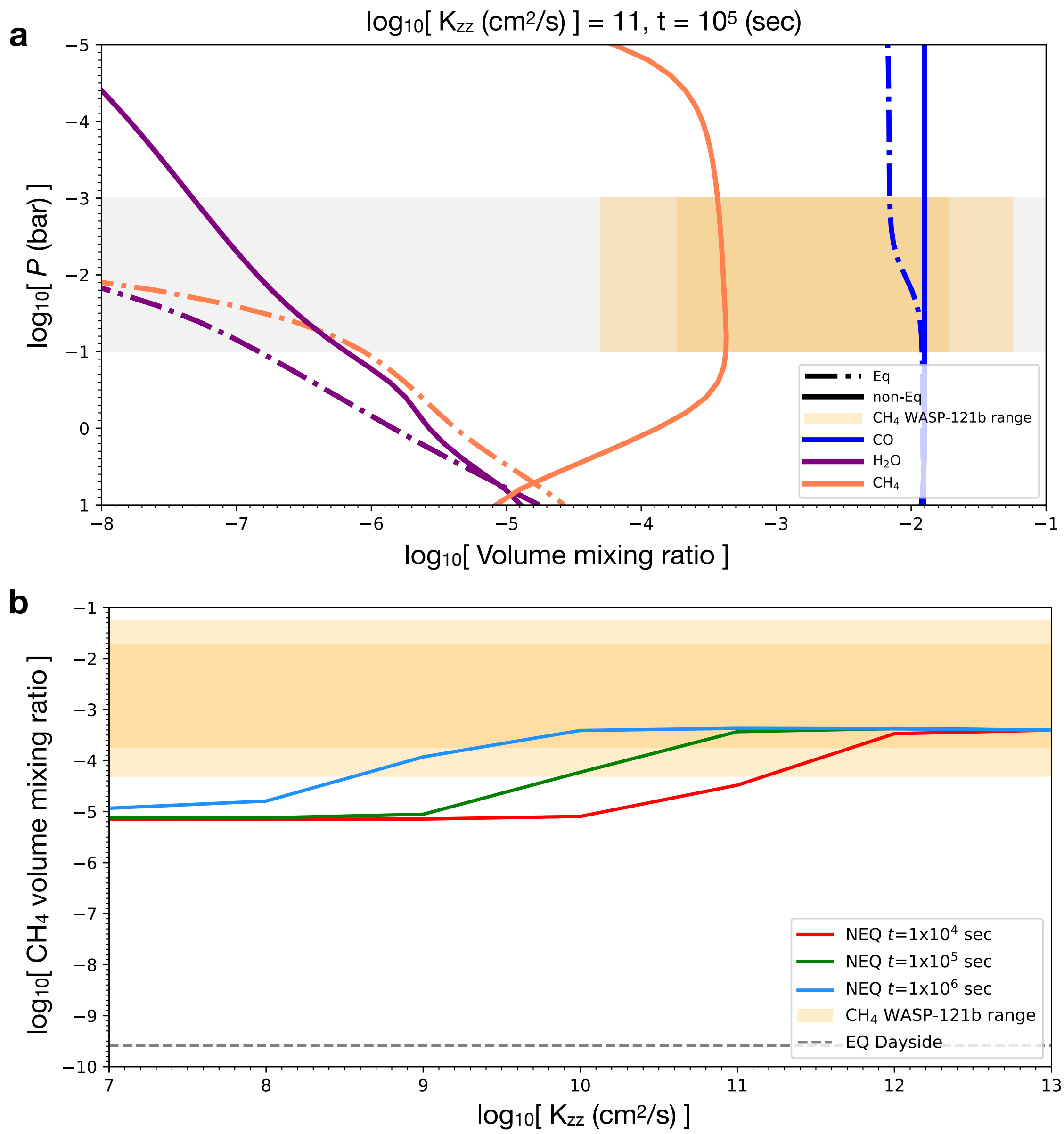

**Fig. 4 | Effect of vertical mixing on nightside abundances. a,** Pressure-dependent VMR profiles for $CH_4$, $H_2O$, and CO assuming equilibrium chemistry (dashed lines) and non-equilibrium chemistry (solid lines). The non-equilibrium chemistry calculations adopted $K_{zz} = 10^{11}$ cm$^2$ s$^{-1}$ and were evolved over a timescale of $10^5$ s. The light grey shading indicates the pressure levels probed by the data. Dark orange shading shows the range of median $CH_4$ VMRs obtained from the retrieval analyses and light orange shading shows the combined extent of the associated $1\sigma$ uncertainties defined by the 16th and 84th percentiles. **b,** Mean $CH_4$ VMR across the $10^{-1}$–$10^{-3}$ bar pressure range as a function of $K_{zz}$ over the timescales indicated. The orange shading is the same as in **a**. The dashed grey line shows the retrieved $CH_4$ VMR for the dayside, which was adopted as the starting abundance.

## METHODS

### Observations and data reduction

The WASP-121 system was continuously monitored for 37.8 h on 2022 October 14–15 using the Bright Object Time Series mode of the NIRSpec instrument on JWST. The observation began 145 min before secondary eclipse ingress and finished 105 min after egress of the following secondary eclipse. As the planet is obscured during secondary eclipses, this allowed us to calibrate the brightness level of the host star at both the start and end of the observation. Light received from the WASP-121 system was dispersed across the two 2,048 × 2,048 pixel HgCdTe detectors at the NIRSpec focal plane by the G395H grism. The two detectors, denoted NRS1 and NRS2, provide continuous wavelength coverage between 2.70–3.72 µm and 3.82–5.15 µm, respectively. For each integration, 2,048 × 32 pixel subarrays were read from both detectors using the NRSRAPID readout pattern with 42 groups per integration, translating to individual integration times of 38.8 s. A total of 3,504 integrations was acquired over the full observation. There were five high-gain antenna movements during the observation, which caused small but noticeable deviations in the measured fluxes for five of the integrations. We retained these impacted integrations in our analysis, as we expect them to have a negligible effect on the overall results.

To analyze the data, we started by using the `FIREFly` code[49] to extract the source spectrum from each integration. Further details of our `FIREFly` reduction can be found in ref. [5]. From the resulting time series of spectra, separate 'white' phase curves were generated for the NRS1 and NRS2 detectors by summing the spectra from each detector across the full wavelength range. We also binned the spectra into narrower 10-pixel-wide channels along the dispersion axis, producing 349 spectroscopic phase curves.

### White phase-curve fitting

Before fitting the spectroscopic phase curves, we refined our estimates for various system parameters by performing a joint fit to the higher-signal-to-noise white phase curves. For each white phase curve, our model $\Pi$ took the form $\Pi = \alpha \times \beta$, where $\alpha$ is the astrophysical signal and $\beta$ is the instrument baseline.

We used the `starry` package[50] to model the time-dependent astrophysical signal $\alpha$. `starry` accounts for emission from both the star and the planet, as well as primary transits and secondary eclipses. We adopted a degree $\ell = 1$ spherical harmonic expansion for the planetary brightness map and a quadratic limb-darkening profile for the star.

For the instrument baseline $\beta$, we investigated three forms. First, a linear trend with respect to time $t$:

$$\beta_1(t) = c_0 + c_1 t \,.$$

Second, a linear trend with respect to $t$ and the $xy$ pointing jitter coordinates described in ref. [5]:

$$\beta_2(t, x, y) = c_0 + c_1 t + c_x x + c_y y \,.$$

Third, a linear trend with respect to $t$, plus an exponential function:

$$\beta_3(t) = c_0 + c_1 t + r_0 e^{-t/r_1} \,.$$

We note that $\beta_3$ allows for an initial settling of the telescope following a new pointing. Baselines of this form have mostly been used for analyses of JWST Mid-Infrared Instrument (MIRI) light curves, which tend to exhibit prominent ramp-like systematics lasting ~30–60 min at the beginning of an exposure[51].

In the fitting, the following parameters were shared across both the NRS1 and NRS2 white phase curves: stellar mass ($M_\star$); planet mass ($M_p$); orbital inclination ($i$); logarithm of the stellar radius ($\log_{10} R_\star$); and deviation in the orbital period ($\Delta P_{\mathrm{orb}}$) from the value of $P_{\mathrm{orb}} = 1.27492504 \pm 1.5 \times 10^{-7}$ days reported in ref. [52]. We fitted the following parameters separately for each detector: the logarithm of the planetary radius ($\log_{10} R_p$); the deviation of the transit mid-time ($\Delta T_{\mathrm{mid}}$) from the value of $T_{\mathrm{mid}} = 2{,}459{,}867.64297 \pm 0.00027$ $\mathrm{BJD_{TDB}}$ predicted by the ephemerides of ref. [52]; the relative planet-to-star luminosity ($A$); the coefficient of the $Y_{1,0}$ spherical harmonic ($y_{1,0}$); the phase offset of the planetary brightness map ($\Delta\phi$); the coefficients of the stellar limb-darkening profile ($u_1$ and $u_2$); the instrument baseline parameters ($c_0$, $c_1$, $c_x$, $c_y$, $r_0$, and $r_1$); and a high-frequency systematics noise term ($\sigma_{\mathrm{syst}}$), explained below. We also tested more elaborate planetary brightness maps comprising combinations of the $Y_{1,0}$, $Y_{1,1}$, and $Y_{2,0}$ spherical harmonics, with the associated coefficients treated as free parameters in the model fitting. However, we found that a planetary brightness map described by $Y_{1,0}$ alone was favoured by the Bayesian information criterion (BIC).

For each white phase-curve fit, a multivariate normal distribution was adopted for the likelihood function and marginalized using affine-invariant Markov chain Monte Carlo (MCMC), as implemented by the `emcee` package[53]. We used 150 walkers and took 5,000 steps to obtain samples from the posterior distribution, following an initial burn-in phase of 500 steps. Note that when evaluating the likelihood function, the error bar on the $i$th data point was taken to be the quadrature sum $\sigma_i = \sqrt{\sigma^2_{\mathrm{Poiss},i} + \sigma^2_{\mathrm{syst}}}$, where $\sigma_{\mathrm{Poiss},i}$ was the estimated Poisson noise for the $i$th data point returned by the instrument pipeline and $\sigma_{\mathrm{syst}}$ was a free parameter allowing for additional high-frequency systematics.

This overall methodology for fitting the white phase curves is similar to that used in ref. [5], but with a few variations. First, we adopted different priors for a subset of parameters. In particular, normal priors were adopted for $M_\star$ and $M_p$ based on a radial velocity analysis of the G395H data[54]. Normal priors obtained using the `exoTiC-LD` package[55] were also adopted for the wavelength-dependent limb-darkening coefficients ($u_1$ and $u_2$). Details of all adopted priors are provided in Supplementary Fig. 1. Second, we allowed $\Delta T_{\mathrm{mid}}$ to vary separately for the NRS1 and NRS2 detectors, whereas ref. [5] adopted a shared transit mid-time for both detectors. Third, although the beginning of the exposure was not visibly affected by strong systematics, we discarded the first 45 min of data before fitting. This precautionary step was taken to provide ample time for the telescope to settle into its new pointing, while still retaining 100 min of pre-eclipse baseline.

Of the three instrumental baselines tested ($\beta_1$, $\beta_2$, and $\beta_3$), we found that $\beta_2$ gave the lowest BIC. The results of the corresponding fit are summarized in Supplementary Table 1 and Supplementary Fig. 1, and are generally consistent with those reported in ref. [5] to within $1\sigma$. However, $>1\sigma$ differences are observed for a subset of parameters, namely: $\Delta\phi$ for NRS1 is $3.7\sigma$ lower; $\Delta\phi$ for NRS2 is $2.3\sigma$ lower; $y_{1,0}$ for NRS1 is $1.7\sigma$ lower; $\sigma_{\mathrm{syst}}$ for NRS1 is $1.5\sigma$ lower; and $\sigma_{\mathrm{syst}}$ for NRS2 is $3.3\sigma$ lower. These differences could arise from one or more of the analysis variations noted above. For example, the adoption of `exoTiC-LD` priors for the limb-darkening coefficients seems to have driven $u_1$ and $u_2$ away from the values found in ref. [5], which adopted substantially broader priors for those parameters. Nonetheless, the primary purpose of the white phase-curve fit was to refine the values of $M_\star$, $\log_{10} R_\star$, $M_p$, $\Delta P_{\rm orb}$, and $i$ in particular, as these parameters were subsequently held fixed for the spectroscopic phase-curve fits (see next section). The updated values for the latter all agree to better than $0.8\sigma$ with those reported in ref. [5] and ref. [56].

We also note that for the maximum a posteriori model—which in general we refer to as the 'best-fit' model—the root-mean-square (rms) of the residuals is found to be 126 ppm for the NRS1 phase curve and 153 ppm for the NRS2 phase curve, compared to Poisson uncertainties derived from the data reduction pipeline of 92 and 137 ppm, respectively. The scatter in the residuals bins down more slowly than expected for uncorrelated noise (Supplementary Fig. 2), which suggests that our instrument baseline model has not captured all of the systematics in the data. This should be inconsequential for the final atmospheric spectra that we obtain for WASP-121b, as the white phase-curve fit is only used to refine our estimates for $M_\star$, $\log_{10} R_\star$, $M_p$, $\Delta P_{\rm orb}$, and $i$. Our posterior distributions for these parameters are fully consistent with the priors (Supplementary Fig. 1), which were adapted from the independent measurements cited above and in ref. [5]. The sole exception is $i$, for which our refined value of $87.97^{+0.17}_{-0.15}$ ° is significantly lower than the prior value of $88.49 \pm 0.16$ ° taken from ref. [52]. However, even if this $\sim 0.5$ ° difference between the prior and posterior for $i$ is caused by uncorrected systematics in our white phase-curve fit, this would have a negligible effect on the final atmospheric spectra that we derive from the spectroscopic phase-curve fits described in the next section.

## Spectroscopic phase-curve fitting

The spectroscopic phase curves were fitted in a similar fashion to the white phase curves. The main difference was that for each spectroscopic phase-curve fit, a subset of model parameters that are not expected to vary with wavelength were held fixed to the best-fit solution from the white phase-curve fit reported in Supplementary Table 1, namely: $M_\star$, $\log_{10} R_\star$, $M_p$, $\Delta P_{\rm orb}$, and $i$. All other parameters were allowed to vary separately in each spectroscopic phase-curve fit, including $\Delta T_{\rm mid}$ and $\Delta\phi$. As for the white phase-curve fit, `emcee` was used to marginalize the posterior distribution for each spectroscopic phase curve, with 150 walkers, 500 burn-in steps, and 5,000 sampling steps.

Supplementary Fig. 3 shows a selection of spectroscopic phase-curve fits spanning a range of fit qualities as measured by the $\chi^2$ statistic. We find that the fit residuals are well-behaved and consistent with white noise. This is conveyed in the bottom panels of Supplementary Fig. 3, which provide a summary of the noise properties for the spectroscopic phase-curve fit residuals. The bottom left panel shows that the histograms of residuals taken across all wavelength channels are very close to normally distributed, with means close to zero and standard deviations within 2% of the pipeline Poisson noise. The bottom right panel shows the Allan deviations for the phase-curve residuals, with the rms of the residuals reducing at a rate of $\sqrt{N}$ where $N$ is the number of points per bin.

## Generating the phase-resolved emission spectra

To extract the phase-resolved planetary emission spectra, we started by randomly selecting 5,000 posterior samples from each spectroscopic phase-curve fit. For each of these posterior samples, we generated the corresponding systematics baseline ($\beta$) and divided the associated data by this baseline to give a 'corrected' phase curve. We then binned the corrected data into 36 phase bins, each of 46 min duration. In defining the locations of the phase bins, we were careful to avoid the secondary eclipses and primary transit. Our chosen bins are illustrated in Extended Data Fig. 1, along with simplified representations of the relative fractions of the dayside and nightside hemispheres of the planet that were visible at each phase. The bin numbers are denoted $n_\phi$, which increases in chronological order from $n_\phi = 1$ for the first phase bin to $n_\phi = 36$ for the last phase bin. In each phase bin, we took the median of the binned values generated from the 5,000 posterior samples as our measurement of $F_p/F_\star$. To calculate the uncertainty, we took the quadrature sum $\sqrt{\sigma_{\rm r}^2 + \sigma_{\rm s}^2}$, where $\sigma_{\rm r}$ denotes a random component and $\sigma_{\rm s}$ denotes a systematic component. For $\sigma_{\rm r}$, we calculated the standard deviation of the model residuals in a given phase bin, then divided by $\sqrt{N_j}$ where $N_j$ is the number of data points in the phase bin. We repeated this for all of the 5,000 posterior samples and took the median of these values as $\sigma_{\rm r}$. To calculate $\sigma_{\rm s}$, we first binned the phase-curve models ($\alpha$) computed from each of the 5,000 posterior samples into the 36 phase bins. We then took the standard deviation of these binned $\alpha$ values in each phase bin as the $\sigma_{\rm s}$ value for that phase bin. This process was undertaken separately for each of the 349 wavelength channels to generate the full set of phase-resolved, wavelength-dependent $F_p/F_\star$ measurements with uncertainties.

We repeated this process for independent sets of spectroscopic phase-curve fits performed using each of the instrument baselines ($\beta_1$, $\beta_2$, and $\beta_3$). We found that the resulting emission spectra obtained with $\beta_1$ and $\beta_2$ were consistent to within $1\sigma$ in all wavelength channels. The emission spectra obtained with $\beta_1$ and $\beta_3$ were consistent to within $1\sigma$ in all but one of the wavelength channels, with the values obtained for the latter channel still being consistent to within $1.2\sigma$. $\beta_1$ was favoured by the BIC over $\beta_2$ and $\beta_3$ in 322 and 345 of the 349 wavelength channels, respectively. For the subset of channels in which either $\beta_2$ or $\beta_3$ was favoured by the BIC, the emission spectrum value obtained for each channel was always consistent to within $1\sigma$ of the corresponding value obtained with $\beta_1$. Given the overall BIC preference for $\beta_1$, along with the consistency of results obtained for each of the baselines tested, we adopt the emission spectra obtained using $\beta_1$ from this point onwards in our analysis.

**Independent data reduction**

As a check, we performed a second data reduction using the `Eureka!` pipeline[57]. The steps we took were similar to those of ref. [58], with a few modifications. For Stage 1, we set the up-the-ramp jump-detection threshold to 10$\sigma$, rather than the default $4\sigma$, to avoid pixels being erroneously flagged as outliers. We corrected $1/f$ noise at the group-level within `Eureka!`'s Stage 1 by masking the trace and performing a column-by-column subtraction. For each column, we determined the offset to be subtracted by first excluding outliers >3 times the median of the unmasked pixels and fitting a zeroth-order polynomial to the remaining pixels in the column. We also applied a scale factor for the bias correction of each integration and both detectors, using a smoothing filter calculated from the first group. Before extracting the spectrum from each integration, we corrected for the curvature of the G395H trace using the method described in ref. [58] and performed a column-by-column background subtraction. For the background subtraction, we fitted a flat line to the pixels more than seven pixels away from the central row of the trace in that column. Before fitting to the background pixels, we performed two iterations of outlier masking, adopting outlier thresholds of $5\sigma$ along the time axis and >5 times the median along the spatial axis. The time-series spectra were then extracted from the background-subtracted integrations using optimal extraction[59] with an aperture of 7 pixels.

We generated phase curves from the `Eureka!` time-series spectra and fitted them using the same methodology described above for the `FIREFly` data. Supplementary Fig. 4 compares the resulting nightside ($n_\phi = 18$) and dayside ($n_\phi = 34$) spectra to those obtained from the `FIREFly` analysis. For the dayside and nightside, the pairs of spectra agree to within $1\sigma$ for 344 and 341 of the 349 wavelength channels, respectively. Of the remaining channels, the agreement is better than $2\sigma$ for seven of the nightside channels and four of the dayside channels. This leaves a single channel of each spectrum for which the difference between the two analyses is > $2\sigma$. Namely, we obtain a $3.5\sigma$ difference for the 2.755 µm channel of the nightside spectrum, which is close to the short-wavelength edge of the NRS1 passband, and a similar $3.4\sigma$ difference for the 3.860 µm channel of the dayside spectrum. As these > $2\sigma$ discrepancies are restricted to a single channel at each phase, they should not affect the interpretation of the spectra, which overall exhibit excellent agreement (Supplementary Fig. 4).

**Atmospheric retrieval analyses**

To characterize the dayside and nightside hemispheres of WASP-121b, we performed retrieval analyses using five independent codes: ATMO[60–64], NEMESIS[65,66], CHIMERA[67], HyDRA[68–70], and PETRA[71]. The varying setups used for each of these codes are described below. Opacity sources considered by the retrieval codes[72–108] are summarized in Supplementary Table 2. Further details of each retrieval code can be found in previous studies that have used them for exoplanet emission retrieval analyses, of which we cite a number of relevant examples here: ATMO was used in ref. [109] and ref. [110] to analyze the dayside emission of WASP-121b measured with HST and Spitzer; NEMESIS was used in ref. [51] to analyze the dayside and nightside spectra of WASP-43b measured with JWST; CHIMERA was used in ref. [111] to analyze the dayside and nightside spectra of the warm sub-Neptune GJ1214b measured with JWST; HyDRA was used in ref. [112] to analyze the dayside emission spectrum of the ultrahot giant planet WASP-18b measured with JWST; and PETRA was used in ref. [71] to analyze the HST and Spitzer emission spectra of the hot giant planets HD 209458b and WASP-43b, as well as a synthetic JWST emission spectrum for the ultrahot giant planet KELT-9b.

For this study of WASP-121b, we have restricted our retrieval analyses to the $n_\phi = 34$ and $n_\phi = 18$ spectra, which are shown in Fig. 1. As can be seen in Extended Data Fig. 1, the $n_\phi = 34$ spectrum corresponds to the phase bin immediately preceding the second secondary eclipse, while the $n_\phi = 18$ spectrum corresponds to the phase bin immediately preceding the primary transit. In practice, we chose the $n_\phi = 34$ and $n_\phi = 18$ spectra to characterize the dayside and nightside hemispheres, respectively (Extended Data Fig. 1). However, it is important to note that retrieval results can potentially be biased if the measured emission at a given phase is modelled without allowing for inhomogeneous properties across the visible hemisphere[113,114]. This is particularly relevant for the $n_\phi = 18$ phase bin, where the nightside hemisphere dominates the fractional area coverage of the planetary disk, but the thin crescent of visible dayside hemisphere may still contribute meaningfully to the measured emission. We took two approaches to investigate how contamination by the dayside crescent affected the retrieval results for the $n_\phi = 18$ spectrum. The first approach, taken for the NEMESIS retrieval, was to explicitly model each spectrum as a weighted sum of dayside and nightside contributions. The second approach, taken for the CHIMERA and HyDRA retrievals, was to include an additional free parameter in the form of the 'dilution factor' described in ref. [114].

The retrieval analyses presented in the following sections were performed on the 349-channel spectra (grey points in Fig. 1) and provided excellent fits to the data. Supplementary Fig. 5 shows histograms of the residuals for the best-fit models from each retrieval. In all cases, the residuals closely follow normal distributions. The reduced $\chi^2$ values ($\chi^2_v$) are also printed in each panel. For the ATMO equilibrium chemistry retrievals, we obtain $\chi^2_v = 1.12$ for the dayside spectrum ($n_\phi = 34$) and $\chi^2_v = 1.09$ for the nightside spectrum ($n_\phi = 18$). The $\chi^2_v$ values for all of the free chemistry retrievals (NEMESIS, PETRA, HyDRA, and CHIMERA) fall between 1.04 and 1.10. Unless stated otherwise, reported results are median values with $1\sigma$ uncertainties defined by the 16th and 84th percentiles (68% credible interval).

**ATMO retrieval.** We used ATMO to perform independent retrievals for the $n_\phi = 18$ and $n_\phi = 34$ spectra. For each of the two retrievals, a single $PT$ profile was assumed. Specifically, we adopted three-parameter[115] and five-parameter[116–118] analytical forms to model the $PT$ profiles for the $n_\phi = 18$ and $n_\phi = 34$ spectra, respectively. The five-parameter $PT$ profile was adopted for the $n_\phi = 34$ spectrum to allow more flexibility in capturing the shape of the dayside thermal inversion. We fixed the internal temperature to 500 K based on the results of ref. [119]. By adopting a single $PT$ profile for each phase, we did not explicitly allow for the possibility of temperature inhomogeneities across the visible hemisphere. However, as described below, additional tests performed with NEMESIS, CHIMERA, and HyDRA demonstrate that this should not significantly bias the final results.

For the chemistry, we allowed the carbon, oxygen, and silicon abundances to vary as separate free parameters, while the relative abundances of all other heavy elements were held fixed to the solar values and varied jointly via an additional metallicity parameter. ATMO adopts the solar abundances of ref. [120], except for C, N, O, P, S, K, and Fe, which are taken from ref. [30]. Given the $PT$ profile and elemental abundances, the abundances of chemical species were then calculated assuming thermochemical equilibrium, which was done by minimizing the Gibbs free energy. The adopted priors for each model parameter are listed in Supplementary Table 3. The posterior distribution was sampled using Nested Sampling[121].

The results of the ATMO retrieval are shown in Figs. 1–3. As already noted in the main text, spectral features of $H_2O$, CO, SiO, and $CH_4$ allow the carbon, oxygen, and silicon abundances to be constrained, as well as the atmospheric $PT$ profile (Fig. 1). We confirm the dayside thermal inversion and infer a nightside $PT$ profile that cools with decreasing pressure (Fig. 2), in agreement with previous studies of WASP-121b (refs. [10,12]). For the dayside ($n_\phi = 34$), we obtain VMRs of C/H=$8{,}036^{+1{,}771}_{-1{,}442}$ ppm and O/H=$8{,}741^{+1{,}897}_{-1{,}516}$ ppm, and derive a tight constraint for the atmospheric C/O ratio of $0.92^{+0.02}_{-0.03}$ (Fig. 3). We also obtain Si/H=$491^{+244}_{-138}$ ppm for the dayside, which, combined with the C/H and O/H measurements, allows us to derive C/Si=$15.78^{+7.22}_{-4.69}$, O/Si=$17.07^{+7.49}_{-4.74}$, and (C+O)/Si=$32.84^{+14.73}_{-9.42}$. For the nightside ($n_\phi = 18$), we obtain C/H=$6{,}794^{+10{,}399}_{-5083}$ ppm, which is consistent with the dayside constraint, but relatively weak owing to the lower signal-to-noise of the nightside emission. As described below for the free chemistry NEMESIS, PETRA, HyDRA, and CHIMERA retrievals, we do not detect any O-bearing or Si-bearing species on the nightside, preventing us from obtaining useful constraints on the absolute O/H and Si/H abundances. However, we are able to constrain the relative nightside abundances to be C/O>0.72 and C/Si>0.74 at 95% probability, which are consistent with the dayside results.

The pressure-dependent VMRs for $H_2O$, CO, $CO_2$, SiO, and $CH_4$ derived under the assumption of thermochemical equilibrium are shown for the nightside and dayside atmospheres in Fig. 2. At a pressure of $10^{-2}$ bar on the nightside, we obtain base-10 logarithmic VMRs of $-2.25^{+0.14}_{-0.44}$ dex for $CH_4$ and $-2.73^{+0.29}_{-0.73}$ dex for CO. The equivalent logarithmic VMRs at a pressure of $10^{-2}$ bar on the dayside are $-12.38^{+0.10}_{-0.12}$ dex for $CH_4$ and $-2.00^{+0.10}_{-0.10}$ dex for CO. For $H_2O$, $CO_2$, and SiO, thermal dissociation reduces the dayside abundances at pressures below approximately $10^{-2}$ bar for $H_2O$ and $CO_2$, and below approximately $10^{-3}$ bar for SiO. Note that photodissociation—which was not modelled—may further reduce molecular abundances at pressures below $10^{-5}$ bar (ref. [122]). Deeper in the dayside atmosphere at a pressure of $10^{-1}$ bar, where neither thermal dissociation nor photodissociation reduce the abundances, the logarithmic VMRs are $-3.57^{+0.18}_{-0.16}$ dex for $H_2O$, $-6.13^{+0.24}_{-0.25}$ dex for $CO_2$, and $-3.12^{+0.18}_{-0.14}$ dex for SiO. The nightside SiO abundance is weakly constrained, but nonetheless consistent with the inferred dayside abundance, while the nightside abundances of $H_2O$ and $CO_2$ are found to be less than 1 ppm and 1 ppb, respectively.

Although abundances for many other species are predicted by the ATMO retrievals under the assumption of thermochemical equilibrium, we focus on $H_2O$, CO, $CO_2$, SiO, and $CH_4$ because (with the exception of $CO_2$) they are the only ones that the free chemistry retrievals find direct evidence for in the data (see below). The detection of these molecules is also supported by the wavelength-dependent contribution functions for the ATMO retrievals, which are shown in Extended Data Fig. 3. The dayside contribution function traces out spectral features of $H_2O$, SiO, and CO, while the nightside contribution function exhibits the distinctive spectral signature of $CH_4$. Supplementary Fig. 6 highlights the dayside SiO emission signal in particular, as it is relatively subtle compared with the dayside $H_2O$ and CO signals. We discuss the nightside $CH_4$ signal in more detail below. From the contribution functions, we also note that the data appear to be probing a greater range of pressures on the nightside than on the dayside. The absolute gradient inferred for the nightside temperature profile is much lower than that inferred for the dayside (Fig. 3), meaning that a relatively large range of pressures must be sampled to reproduce the spectral features observed in the nightside spectrum.

Extended Data Fig. 5 compares our retrieved $PT$ profiles to those of various models that self-consistently treat the radiative transfer and chemistry of the atmosphere. Specifically, we show $PT$ profiles for a range of dayside and nightside longitudes from published MITgcm[11] and Exo-FMS[19] 3D GCM simulations, as well as 1D radiative-convective thermochemical equilibrium models computed with ATMO for internal temperatures of 100 K and 500 K, assuming the elemental abundances derived from the retrieval. Although the 3D models adopted solar elemental abundances, it is still interesting to make a preliminary comparison between the thermal profiles that they predict and those that we infer from the data. On the dayside, the retrieved $PT$ profile is broadly consistent with the temperatures spanned by the self-consistent 1D and 3D models across the $\sim 10^{-1}$ to $\sim 10^{-3}$ bar pressure range probed by the data (Fig. 2). On the nightside, the retrieved temperatures are similar to those predicted by the MITgcm simulation for pressures below $\sim 10^{-2}$ bar, but cooler at higher pressures. The Exo-FMS simulation predicts considerably warmer nightside temperatures at all pressures compared with our retrieved temperatures.

**NEMESIS retrieval.** We used NEMESIS to perform a joint retrieval for the $n_\phi = 18$ and $n_\phi = 34$ spectra. We considered the measured $F_p$ at each phase to be the weighted sum of dayside and nightside components:

$$F_p = wF_{\mathrm{day}} + (1 - w)F_{\mathrm{night}} \,, \quad (1)$$

where $F_{\mathrm{day}}$ and $F_{\mathrm{night}}$ denote the emission from the dayside and nightside hemispheres, respectively, and $w$ is a geometric term giving the fraction of the visible hemisphere at orbital phase $\phi$ taken up by the dayside hemisphere[123]. For the latter, we used:

$$w(\phi) = [\,1 - \cos(2\pi\phi - \Delta\phi)]/2 \quad (2)$$

and fixed $\Delta\phi = -3°$, namely, the measured phase offset for the planetary brightness map rounded to the nearest degree (Supplementary Table 1 and Supplementary Fig. 1). We also allowed for the difference in the average emission angle from the dayside segment between the two phases. This could be particularly important for the dayside segment of the $n_\phi = 18$ spectrum, which will be limb-darkened or limb-brightened in different spectral regions depending on the $PT$ profile. We calculate the weighted-average emission angle of the dayside segment to be 76° at phase $n_\phi = 18$ and 45° at phase $n_\phi = 34$, and took these differences into account when computing the measured emission at each phase according to equation (1). We did not consider a similar correction for the nightside segment of phase $n_\phi = 34$ to be necessary, because it makes a very small contribution to the overall spectrum measured at that phase.

We adopted a free chemistry approach for the NEMESIS retrieval, in which abundances of select chemical species were allowed to vary as free parameters without requiring thermochemical equilibrium to be satisfied. The chemical species considered were $H_2O$, $H^-$, CO, $CO_2$, $CH_4$, and SiO. Abundances for these species were varied separately for the $F_{\mathrm{day}}$ and $F_{\mathrm{night}}$ model components. We allowed for the possibility of $H_2O$ being thermally dissociated on the dayside by varying its abundance with pressure according to $P^\tau$ above some base pressure level $P_0$, which has been shown to provide a reasonable approximation for the effects of thermal dissociation[11]. In addition to the $H_2O$ abundance for $P > P_0$, we varied $\tau$ and $P_0$ as free parameters in the retrieval. All other abundances—including the nightside $H_2O$ abundance—were assumed to be uniform in pressure. We adopted separate six-parameter $PT$ profiles[124] for the dayside and nightside. Priors for each model parameter are listed in Supplementary Table 4. Posterior sampling was performed using the `pymultinest`[125] implementation of `MultiNest`[121,126] Nested Sampling.

As with the ATMO retrieval, the NEMESIS retrieval recovers a thermal inversion on the dayside and a temperature profile that cools with decreasing pressure on the nightside (Supplementary Fig. 7). Tight abundance constraints indicate the presence of $H_2O$, CO, and SiO on the dayside, and $CH_4$ on the nightside (Supplementary Fig. 7). We infer logarithmic VMRs of $-3.31^{+0.23}_{-0.26}$ dex for the deep-atmosphere $H_2O$, $-1.29^{+0.13}_{-0.19}$ dex for CO, and $-2.66^{+0.16}_{-0.21}$ dex for SiO on the dayside, and $-1.75^{+0.44}_{-0.66}$ dex for $CH_4$ on the nightside. We also find that the inferred dayside and nightside spectra extrapolate well to the other orbital phases (Extended Data Fig. 2).

To evaluate the significance of the $CH_4$ detection, we performed two nightside-only NEMESIS retrievals, with $CH_4$ included and with $CH_4$ removed. In these additional retrievals, we also included $C_2H_2$, HCN, and $NH_3$, to ensure that the signal we observe really is due to $CH_4$. By comparing the Bayesian evidence for this 'null hypothesis' retrieval without $CH_4$ ($\mathcal{Z}_0$) to the Bayesian evidence of the model that included all of the molecules ($\mathcal{Z}_1$), we obtain a log Bayes factor ($\ln \mathcal{B}_{10} = \ln \mathcal{Z}_1/\mathcal{Z}_0$) of $\ln \mathcal{B}_{10} = 10.4$. Under numerous empirically calibrated evidence scales, this corresponds to the highest band of evidence in favor of a $CH_4$ detection[127–129]. Following refs. [130–132], we can also translate our $\ln \mathcal{B}_{10}$ value into an equivalent frequentist detection significance, as is common in the exoplanet literature. Using this approach, we obtain a detection significance of $4.9\sigma$ for $CH_4$ on the nightside of WASP-121b. Similar dayside-only retrievals were performed to obtain decisive detections of $H_2O$ ($\ln \mathcal{B}_{10} = 88.3$, $13.5\sigma$), CO ($\ln \mathcal{B}_{10} = 56.0$, $10.8\sigma$), and SiO ($\ln \mathcal{B}_{10} = 14.5$, $5.7\sigma$). Plots of the best-fit NEMESIS retrievals with and without each molecule are shown in Supplementary Figs. 8–11.

We additionally used NEMESIS to fit for a nightside cloud layer using the cloud parameterization of ref. [133], as was done previously in ref. [134]. We fitted for the top pressure of an opaque, grey cloud ($P_{\mathrm{top}}$), and a scattering power-law index ($\gamma$) and opacity scaling ($\sigma_0$) for a 'haze' above the cloud. From this retrieval, we are only able to provide a weak constraint for the cloud-top pressure being $>10^{-3}$ bar, while the remaining cloud parameters are effectively unconstrained.

**CHIMERA retrieval.** We used CHIMERA to perform free chemistry retrievals for the $n_\phi = 18$ spectrum, focusing on characterization of the nightside atmosphere only. The chemical species considered were $H_2O$, CO, $CO_2$, $CH_4$, $C_2H_2$, HCN, and $NH_3$. We assumed that all VMRs were uniform with pressure, as the temperatures on the nightside are too low for thermal dissociation to be important. We performed separate retrievals adopting widely used three-parameter[115], five-parameter[116–118], and six-parameter[124] functions for the $PT$ profile. The three-parameter profile was the same as that used for the ATMO nightside analysis, the five-parameter profile was the same as that used for the ATMO dayside analysis, and the six-parameter profile was the same as that used for the NEMESIS analyses. We also performed retrievals with and without the dilution factor described in ref. [114] to account for contribution to the observed spectrum by the visible dayside crescent. The priors used for each model parameter are listed in Supplementary Table 5. Posterior samples were drawn using `pymultinest`[125].

The retrievals performed for each $PT$ profile support the presence of $CH_4$, with $\ln \mathcal{B}_{10} = 5.60$ ($3.8\sigma$) for the three-parameter profile, $\ln \mathcal{B}_{10} = 9.32$ ($4.7\sigma$) for the five-parameter profile, and $\ln \mathcal{B}_{10} = 11.0$ ($5.1\sigma$) for the six-parameter profile. The inferred $PT$ profiles are plotted in Supplementary Fig. 7 and all cool with decreasing pressure across the $1$–$10^{-3}$ bar pressure range. The six-parameter profile is essentially identical to the NEMESIS six-parameter profile and is also in good agreement with the HyDRA six-parameter profile (see below). The three-parameter profile is fully consistent with the ATMO three-parameter profile. Interestingly, the five-parameter profile exhibits an inversion for pressures below approximately $10^{-3}$ bar, producing a narrow $CH_4$ emission spike at 3.35 μm while the broader $CH_4$ features remain in absorption, which marginally improves the fit to the data (Supplementary Fig. 12). However, on the basis of the evaluated Bayesian evidence values, we do not find a clear preference for any one of the profiles in particular. Specifically, we obtain $\ln \mathcal{Z}_1 = -192.72 \pm 0.07$ for the three-parameter profile, $\ln \mathcal{Z}_1 = -190.39 \pm 0.08$ for the five-parameter profile, and $\ln \mathcal{Z}_1 = -191.56 \pm 0.06$ for the six-parameter profile. A comparison of the CHIMERA best-fit models with and without $CH_4$ for the five-parameter profile is shown in Supplementary Fig. 12.

As shown in Supplementary Fig. 7, the corresponding logarithmic VMRs for $CH_4$ are found to be $-2.43^{+0.91}_{-1.67}$ dex (three-parameter), $-3.73^{+0.49}_{-0.57}$ dex (five-parameter), and $-1.73^{+0.48}_{-0.78}$ dex (six-parameter). Besides $CH_4$, no evidence is obtained for any of the other chemical species that were considered. We also find that the inclusion of a dilution factor does not significantly affect the results and is not justified by the Bayesian evidence.

**HyDRA retrieval.** We used HyDRA to perform separate free chemistry retrievals for the $n_\phi = 18$ and $n_\phi = 34$ spectra. The chemical species considered for the dayside ($n_\phi = 34$) were $H_2O$, CO, $CO_2$, SiO, $SiO_2$, HCN, $H^-$, OH, FeH, TiO, and VO. For the nightside ($n_\phi = 18$), we considered $H_2O$, CO, $CO_2$, SiO, $SiO_2$, HCN, $H^-$, and $CH_4$. Thermal dissociation was modelled for the dayside abundances of $H_2O$, TiO, VO, and $H^-$ using the methods of ref. [11]. The deep abundances of each of these species were free parameters in the retrieval, while their pressure- and temperature-dependent abundance profiles were calculated using the coefficients in table 1 of ref. [11] as inputs to equations (1) and (2) of the same work. All other VMRs were assumed to be uniform in pressure. The six-parameter function[124] was used to model the $PT$ profiles at both phases. We tested the inclusion of clouds in the retrievals, which were parameterized by a modal particle size ($a_c$), the cloud base pressure ($P_c$), a pressure exponent ($\varepsilon$), the cloud particle abundance ($f_c$), and the cloud covering fraction ($\psi_c$). The cloud particle abundance was assumed to be zero at pressures greater than $P_c$, and to decrease at pressures less than $P_c$, such that at pressure $P$ the abundance was $f_c(P/P_c)^\varepsilon$. We assumed an $MgSiO_3$ cloud composition, and used the cloud absorption cross sections from ref. [135]. The priors adopted for each model parameter are listed in Supplementary Table 6. Posterior sampling was performed using `pymultinest`[125].

The HyDRA retrieval again recovers a thermal inversion for the dayside and a cooling $PT$ profile for the nightside. On the dayside, decisive detections are made for CO with $\ln \mathcal{B}_{10} = 78.7$ ($12.8\sigma$), SiO with $\ln \mathcal{B}_{10} = 17.5$ ($6.2\sigma$), and $H_2O$ with $\ln \mathcal{B}_{10} = 13.4$ ($5.5\sigma$). The inferred logarithmic VMRs for CO and SiO on the dayside are $-2.17^{+0.35}_{-0.37}$ dex and $-3.17^{+0.33}_{-0.34}$ dex, respectively, while the logarithmic VMR for the deep atmosphere $H_2O$ is found to be $-4.24^{+0.35}_{-0.32}$ dex. We caution that the inferred SiO abundance may be biased, as it was assumed to be uniform in pressure but in reality should be thermally dissociated (Fig. 2). On the nightside, the detection significance for $CH_4$ is found to be $\ln \mathcal{B}_{10} = 3.3$ $(3.1\sigma)$, with an inferred logarithmic VMR of $-2.05^{+0.55}_{-0.81}$ dex. However, we also ran a separate retrieval analysis for the nightside spectrum with $CH_4$ as the only opacity source and compared the Bayesian evidence to that obtained for a simple blackbody fit. In this restricted scenario, HyDRA decisively favors a model with $CH_4$ absorption over a blackbody with $\ln B_{10} = 31.5$ $(8.1\sigma)$. Plots of the best-fit HyDRA retrievals with and without each detected molecule are shown in Supplementary Figs. 13–16. We additionally find that a dilution parameter is not statistically justified in either the dayside or nightside retrievals, nor are clouds.

**PETRA retrieval.** We used PETRA to perform free chemistry retrievals for the $n_\phi = 34$ spectrum, focusing on characterization of the dayside atmosphere only. The chemical species we considered were $H_2O$, CO, and SiO. Only the $H_2O$ abundance was allowed to be affected by thermal dissociation, following a similar approach to that used in the NEMESIS and HyDRA free chemistry retrievals. The $PT$ profile was modelled with the same five-parameter function used in the ATMO, CHIMERA, and HyDRA retrievals. The priors used for each model parameter are listed in Supplementary Table 7. Posterior sampling was performed using differential-evolution MCMC[136,137].

As for all other dayside retrievals, the inferred $PT$ profile exhibits a strong thermal inversion (Supplementary Fig. 7). However, the thermal inversion inferred by PETRA starts at a pressure of ~$10^{-3}$ bar, which is higher in the atmosphere than the thermal inversions inferred by the other retrievals. For the chemical abundances, we obtain logarithmic VMRs of $-2.03^{+0.32}_{-0.36}$ dex for CO, $-3.15^{+0.29}_{-0.28}$ dex for SiO, and $-3.92^{+0.28}_{-0.27}$ dex for $H_2O$. We note that the lower edge of the 68% credible interval (16th percentile) for the $H_2O$ abundance decreases with increasing pressure for pressures above approximately $10^{-2}$ bar. This corresponds to solutions where the thermal dissociation exponent is negative $(\tau < 0)$ and indicates that the data are not constraining the $H_2O$ abundance for pressures >$10^{-2}$ bar in the PETRA retrieval.

### Discussion of the retrieval analyses

The four retrievals that considered the $n_\phi = 34$ spectrum (ATMO, NEMESIS, HyDRA, and PETRA) recover dayside $PT$ profiles with thermal inversions and fit the data with emission features of $H_2O$, CO, and SiO. The exact shapes of the inferred $PT$ profiles differ somewhat and appear to be correlated with differences in the inferred pressure-dependent abundances, particularly that of $H_2O$. As shown in Supplementary Fig. 7, NEMESIS and PETRA favour the dissociation of $H_2O$ at pressures <$10^{-3}$ bar, resulting in thermal inversions at lower pressures than those obtained by ATMO and HyDRA, for which $H_2O$ dissociation is found to occur at pressures of ~$10^{-2}$ bar. These results emphasize how challenging it can be to model pressure-variable abundances with free chemistry retrievals in a purely parametric manner, even when constrained by JWST-quality data. Despite the variation in absolute abundances and thermal inversion pressures, the inferred C/O, C/Si, and O/Si ratios are found to be highly consistent across all the retrieval analyses (Supplementary Fig. 17).

Each of the retrieval analyses that were performed for the $n_\phi = 18$ spectrum (ATMO, NEMESIS, HyDRA, and CHIMERA) return nightside $PT$ profiles that cool with decreasing pressure and find evidence for $CH_4$ absorption features. The $PT$ profiles and absolute $CH_4$ abundances are in good overall agreement across the retrievals (Supplementary Fig. 7). The one mild exception is the CHIMERA retrieval that adopted the five-parameter $PT$ profile, which finds a low-pressure inversion that is not seen in the other retrieval analyses. As noted above, because this inversion occurs at low pressures it only has a minor effect on the appearance of the emission spectrum. Otherwise, the CHIMERA five-parameter $PT$ profile broadly resembles those of the other analyses but shifted to higher pressures. This reflects the relatively low $CH_4$ abundance inferred by CHIMERA for the five-parameter $PT$ profile (Supplementary Fig. 7), which has the effect of moving the photosphere to a higher pressure than for the other retrieval models.

In assessing the detection significances for each molecule, we report the corresponding range of values obtained from the various free chemistry retrievals: $5.5–13.5\sigma$ for dayside $H_2O$; $10.8–12.8\sigma$ for dayside CO; $5.7–6.2\sigma$ for dayside SiO; and $3.1–5.1\sigma$ for nightside $CH_4$. Within these ranges, we favour those detection significances obtained by the retrieval codes that used the latest line list for the corresponding molecule (Supplementary Table 2). For $H_2O$, HyDRA uses the HITEMP line list of ref. [74], which is based on the earlier ab initio line list of ref. [72], while both NEMESIS and CHIMERA use the more recent ExoMOL line list of ref. [73]. For CO, NEMESIS uses the line list of ref. [79], while HyDRA and CHIMERA use the earlier HITEMP line list of ref. [74]. For SiO, HyDRA uses the latest ExoMOL line list of ref. [82], which superseded the earlier line list of ref. [81] used by NEMESIS. For $CH_4$, CHIMERA uses the recent HITEMP line list of ref. [78], whereas NEMESIS and HyDRA use the earlier ExoMOL line lists of refs. [75–77]. On the basis of these line lists, our favoured detection significances are therefore $13.5\sigma$ for dayside $H_2O$ (NEMESIS), $10.8\sigma$ for dayside CO (NEMESIS), and $6.2\sigma$ for dayside SiO (HyDRA). For nightside $CH_4$, we favour the detection significance of $4.7\sigma$ from the CHIMERA retrieval performed with the five-parameter $PT$ profile, which gave a higher Bayesian evidence ($\ln \mathcal{Z}_1 = -190.39 \pm 0.08$) than the three-parameter ($\ln \mathcal{Z}_1 = -192.72 \pm 0.07$) and six-parameter ($\ln \mathcal{Z}_1 = -191.56 \pm 0.06$) $PT$ profiles.

We note that the lowest detection significance of $3.1\sigma$ obtained for nightside $CH_4$ by the HyDRA retrieval may be overly conservative. Higher detection significances of $5.1\sigma$ and $4.9\sigma$, respectively, were obtained by the CHIMERA and NEMESIS retrievals for the same six-parameter $PT$ profile adopted by HyDRA. We suspect that the difference in detection significances may be due to HyDRA using the early ExoMOL line lists of refs. [75,77] for $CH_4$, whereas NEMESIS uses the expanded ExoMOL line list of ref. [76], and CHIMERA uses the more recent HITEMP line list of ref. [78], which (unlike the ab initio ExoMOL line lists) is largely based on experimental data. Differences in the adopted $H_2O$ line list could also potentially affect the inferred $CH_4$ detection significance, given that $H_2O$ and $CH_4$ have broad features at similar wavelengths within the G395H passband. As noted above, NEMESIS and CHIMERA both use the $H_2O$ line list of ref. [73], which is more recent than the line list of ref. [74] used by HyDRA. Indeed, the dayside $H_2O$ detection significance obtained by NEMESIS ($13.5\sigma$) is considerably higher than that obtained by HyDRA ($5.5\sigma$), as would be expected if the $H_2O$ line list used by NEMESIS is better suited for the conditions of the WASP-121b atmosphere than the older line list used by HyDRA. Furthermore, the fact that NEMESIS obtains lower detection significances than HyDRA for the other molecules (that is, CO and SiO on the dayside) demonstrates that NEMESIS does not systematically overestimate detection significances relative to HyDRA.

In addition to the formal detection significances, there are several other reasons that we favour nightside models including $CH_4$. First, when $CH_4$ is removed from the free chemistry retrievals (NEMESIS, HyDRA, and CHIMERA), a thermal inversion at pressures $>10^{-3}$ bar is required to reproduce the observed nightside spectrum with emission features of $H_2O$ and CO. A thermal inversion this deep in the non-irradiated nightside atmosphere is not expected, on the basis of state-of-the-art GCM simulations (Extended Data Fig. 5). It would also be in tension with HST measurements that have previously revealed a decreasing $PT$ profile on the nightside of WASP-121b (ref. [12]). Second, the inference of an atmospheric C/O=$0.92^{+0.02}_{-0.03}$ from the dayside spectrum (ATMO) is a robust result (Fig. 3). With an atmospheric C/O ratio this high, equilibrium chemistry predicts a logarithmic VMR above $-4$ dex for $CH_4$ at the temperatures of the nightside (Extended Data Fig. 4), which is broadly consistent with the abundances inferred by the free chemistry retrievals (Supplementary Fig. 7). Although horizontal quenching of $CH_4$-poor gas from the dayside could in principle dominate the nightside composition, this nonetheless demonstrates that the $CH_4$ abundances we have inferred are physically plausible. We also reiterate that we did allow for dayside contamination of the $n_\phi = 18$ spectrum using two approaches: including a dilution factor as a free parameter (HyDRA and CHIMERA) and by explicitly modelling the dayside contribution (NEMESIS). The evidence for $CH_4$ and a decreasing $PT$ profile on the nightside persists with both of these approaches.

We do not make a statistically significant detection of CO on the nightside, despite equilibrium chemistry predicting a CO abundance that is comparable to the $CH_4$ abundance (Fig. 2 and Extended Data Fig. 4). This is because the abundance-weighted CO opacity is only higher than that of $CH_4$ within narrow line cores across the ~4–4.5 µm wavelength range. At the spectral resolution of the nightside data, the corresponding opacity enhancement provided by CO over $CH_4$ is quite modest (Extended Data Fig. 3) and we find that the inclusion or exclusion of CO has only a minor effect on the spectrum relative to the measurement uncertainties (Fig. 1). In particular, when we simply switch off the CO opacity in the best-fit ATMO model shown in Fig. 1, the $\chi^2$ increases from 370.6 to 372.4, corresponding to a negligible $\chi_v^2$ increase from 1.087 to 1.092. Consequently, the nightside CO abundance is only weakly constrained by the free chemistry retrievals, while the CO abundance displayed in Fig. 2 primarily reflects the chemical equilibrium constraint imposed for the ATMO retrieval.

We also do not find evidence for clouds on the nightside of WASP-121b, based on the Bayes factor computed from the HyDRA retrievals performed with and without clouds, as well as the lack of cloud constraints obtained by NEMESIS. This is despite the inferred nightside $PT$ profile crossing numerous condensation curves of refractory species (Fig. 2), including silicates which are expected to be the dominant source of cloud opacity on the hot nightsides of planets like WASP-121b (ref. [14]). However, as noted in the main text, our non-detection of clouds does not necessarily imply the absence of nightside clouds. In particular, the inferred $PT$ profile for the nightside hemisphere crosses the $MgSiO_3$, $Mg_2SiO_4$, and $SiO_2$ condensation curves at pressures of ~1 bar, whereas the data are only sensitive to pressures below ~$10^{-1}$ bar (Fig. 2 and Extended Data Fig. 6). It therefore remains possible that a thick cloud layer composed of silicates and/or other refractory species, such as Fe, is present on the nightside of WASP-121b at pressures >$10^{-1}$ bar.

## Stellar abundances and relative planetary abundances

We derived photospheric and fundamental stellar parameters for the WASP-121 host star using the same algorithm described in ref. [138]. For our isochrone fitting, we include multiwavelength photometry from the UV to the near-infrared: Tycho-2 BT and VT[139], Gaia Data Release 2 (refs. [140–143]), the G, J, H, and Ks bands from the Two Micron All Sky Survey (2MASS) All-Sky Point Source Catalog[144], and the W1 and W2 bands from the Wide-field Infrared Survey Explorer (WISE) AllWISE mid-infrared data[145,146]. We also included the Gaia Data Release 3 (ref. [143]) parallax-based distance to the WASP-121 system[147]. We included the extinction $A_V$ inference based on 3D maps of extinction in the solar neighbourhood from the Structuring by Inversion the Local Interstellar Medium (Stilism) programme[148–150].

For the spectroscopic-based inferences, we used the equivalent-widths of FeI and FeII atomic absorption lines. The equivalent widths were derived from a spectrum of the WASP-121 host star measured with the ESPRESSO instrument on the Very Large Telescope. The spectrum is available on the European Southern Observatory archive under program ID 106.21QM.001. It was taken on 2021 September 6 and has a signal-to-noise ratio of ~150 at 5,000 Å. To measure the equivalent widths we used the `splot` task of IRAF (IRAF is distributed by the National Optical Astronomy Observatory, which is operated by the Association of the Universities for Research in Astronomy, Inc. under cooperative agreement with the National Science Foundation). The atomic data and measured equivalent widths for each line are reported in Supplementary Table 8. We adapted the atomic data from ref. [151] and assumed photospheric solar abundances from ref. [152].

We used the `isochrones` package[153] to fit the MESA Isochrones and Stellar Tracks[154–156] library to our photospheric stellar parameters, as well as our input multiwavelength photometry, parallax, and extinction data using `pymultinest`[125]. Further details of our approach are provided in ref. [138]. The adopted stellar parameters are given in Supplementary Table 9.

We inferred elemental abundances of CI, OI, NaI, MgI, AlI, SiI, SI, KI, CaI, ScII, TiII, VI, CrI, CrII, MnI, FeI, FeII, CoI, NiI, CuI, ZnII, YII, ZrII, BaII, CeII, NdII, and SmII using the equivalent-width method, including isotopic/hyperfine splitting details where needed. We measured the equivalent widths by fitting Gaussian profiles with the `splot` task in `IRAF`. We also used the `deblend` task to disentangle absorption lines from adjacent spectral features whenever necessary. We assumed the solar abundances of ref. [152] and local thermodynamic equilibrium (LTE) and used the 1D plane-parallel solar-composition MARCS model atmospheres[157] and the 2019 version of the LTE radiative transfer code `MOOG`[158] to infer elemental abundances based on our equivalent widths. Our abundance inferences and uncertainties are reported in Supplementary Table 10. Where possible, we updated our elemental abundances for departures from LTE by linearly interpolating published grids of non-LTE corrections. Specifically, non-LTE corrections taken from the tables of refs. [159–162] were applied to the carbon, oxygen, aluminum, calcium, silicon, potassium, and iron abundances.

We highlight the stellar abundances derived for CI, OI, and SiI in particular, with reference to the abundances in the standard $A(X)$ format. Here $A(X) = \log_{10}(N_X/N_\mathrm{H}) + 12$, where $N_X$ is the number density of species $X$ and $N_\mathrm{H}$ is the number density of hydrogen. The non-LTE (LTE) abundances we obtain are: 8.527±0.050 (8.545±0.050) for CI; 8.855±0.024 (9.238±0.037) for OI; and 7.696±0.097 (7.711±0.097) for SiI. The non-LTE and LTE abundances are consistent to well within the $1\sigma$ uncertainties for both CI and SiI. However, the non-LTE abundance for OI is significantly lower than the LTE abundance. This results in an upward revision of the stellar C/O ratio from 0.203±0.050 (LTE) to 0.470±0.061 (non-LTE). As an aside, we note that our LTE C/O ratio is consistent with the value of 0.23±0.05 reported in ref. [163].

To compute elemental ratios for the planetary atmosphere relative to those of the host star, we used the ATMO posterior samples shown in Fig. 3 for the planet abundances and randomly drew stellar abundances from normal distributions that had means and standard deviations set to the non-LTE values listed in Supplementary Table 10. We find that the C/H, O/H, Si/H, and C/O ratios of the planet normalized to the equivalent ratios of the host star are greater than the following values at >99.99% probability: $(\mathrm{C/H})/(\mathrm{C/H})_\star>11.92$, $(\mathrm{O/H})/(\mathrm{O/H})_\star>7.27$, $(\mathrm{Si/H})/(\mathrm{Si/H})_\star>2.49$, and $(\mathrm{C/O})/(\mathrm{C/O})_\star>1.63$. The corresponding medians with uncertainties reflecting the 68% credible intervals (16th–84th percentiles) are: $(\mathrm{C/H})/(\mathrm{C/H})_\star=23.96^{+6.13}_{-4.75}$, $(\mathrm{O/H})/(\mathrm{O/H})_\star=12.19^{+2.78}_{-2.24}$, $(\mathrm{Si/H})/(\mathrm{Si/H})_\star=9.89^{+5.99}_{-2.89}$, and $(\mathrm{C/O})/(\mathrm{C/O})_\star=1.96^{+0.11}_{-0.11}$. We also find that the volatile/refractory ratios of the planet normalized to the host star are less than the following values at >99.99% probability: $(\mathrm{C/Si})/(\mathrm{C/Si})_\star<8.07$, $(\mathrm{O/Si})/(\mathrm{O/Si})_\star<4.02$, and $((\mathrm{C+O})/\mathrm{Si})/((\mathrm{C+O})/\mathrm{Si})_\star<5.24$. The corresponding 68% credible intervals are $(\mathrm{C/Si})/(\mathrm{C/Si})_\star=2.30^{+1.06}_{-0.61}$, $(\mathrm{O/Si})/(\mathrm{O/Si})_\star=1.19^{+0.51}_{-0.32}$, and $((\mathrm{C+O})/\mathrm{Si})/((\mathrm{C+O})/\mathrm{Si})_\star=1.55^{+0.71}_{-0.41}$.

## Accretion of rocky material

We can use the Si/H enrichment of the planetary atmosphere relative to the host star to estimate the quantity of rocky material that WASP-121b accreted following its formation. If we assume that the planet's primordial atmospheric envelope has Si/H equal to that of the host star, and is then enriched by silicon through subsequent accretion of rocky material, then:

$$\frac{n_{\mathrm{Si,env}}}{n_{\mathrm{H,env}}} = \frac{n_{\mathrm{Si},\star}}{n_{\mathrm{H},\star}} + \frac{n_{\mathrm{Si},\epsilon}}{n_{\mathrm{H,env}}} ,$$

where $n_{\mathrm{Si,env}}$ is the total number of silicon atoms in the atmospheric envelope, $n_{\mathrm{H,env}}$ is the total number of hydrogen atoms in the atmospheric envelope, $n_{\mathrm{Si},\star}/n_{\mathrm{H},\star}$ is the ratio of silicon atoms to hydrogen atoms in the host star, and $n_{\mathrm{Si},\epsilon}$ is the number of accreted silicon atoms. Note that the above formula implicitly assumes that the accreted silicon is mixed throughout the entire atmospheric envelope.

The number of hydrogen atoms in the atmospheric envelope is given by:

$$n_{\mathrm{H,env}} = \frac{N_{\mathrm{A}} f_{\mathrm{H}} (M_{\mathrm{p}} - M_{\mathrm{core}})}{\mu_{\mathrm{H}}} ,$$

where $N_{\mathrm{A}}$ is Avogadro's constant, $f_{\mathrm{H}}$ is the hydrogen mass fraction for the envelope, and $\mu_{\mathrm{H}}$ is the molar mass of hydrogen. We have also assumed that $M_{\mathrm{p}} = M_{\mathrm{core}} + M_{\mathrm{env}}$, where $M_{\mathrm{p}}$ is the total planet mass, $M_{\mathrm{core}}$ is the mass of the core, and $M_{\mathrm{env}}$ is the mass of the atmospheric envelope.

Combining the above, we derive the following expression for the number of accreted silicon atoms:

$$n_{\mathrm{Si},\epsilon} = \frac{N_{\mathrm{A}} f_{\mathrm{H}} \left(M_{\mathrm{p}} - M_{\mathrm{core}}\right)}{\mu_{\mathrm{H}}} \left[ \left(\frac{n_{\mathrm{Si}}}{n_{\mathrm{H}}}\right)_{\mathrm{env}} - \left(\frac{n_{\mathrm{Si}}}{n_{\mathrm{H}}}\right)_{\star} \right] .$$

To calculate $n_{\mathrm{Si},\epsilon}$, we take $(n_{\mathrm{Si}}/n_{\mathrm{H}})_{\mathrm{env}} = 491^{+244}_{-138}$ ppm, $(n_{\mathrm{Si}}/n_{\mathrm{H}})_{\star} = 50$ ppm, and $M_{\mathrm{p}} = 1.2 M_{\mathrm{J}}$ (where $M_{\mathrm{J}}$ is the mass of Jupiter) from our measurements reported above, as well as the standard values for $N_{\mathrm{A}}$ ($6.022 \times 10^{23}$ atoms mol$^{-1}$) and $\mu_{\mathrm{H}}$ ($10^{-3}$ kg mol$^{-1}$). We also assume $M_{\mathrm{core}} = 15 M_{\oplus}$, based on the estimate of Jupiter's core mass[164], and $f_{\mathrm{H}} = 75\%$, based on the approximate hydrogen mass fraction of Jupiter's atmosphere[165]. With these input values, we obtain $n_{\mathrm{Si},\epsilon} = 1.63 \times 10^{50}$ atoms of silicon added to the envelope of WASP-121b by the accretion of rocky material.

If we further assume that the rocky material has an Earth-like composition, we can translate this number of accreted silicon atoms to an equivalent number of Earth masses of rocky material. The estimated silicon mass fraction for the bulk Earth is 16% (ref. [166]). Given that Earth has a mass of $5.9723 \times 10^{24}$ kg, this equates to $9.5557 \times 10^{23}$ kg of silicon in the Earth. Adopting $\mu_{\mathrm{Si}} = 0.028$ kg mol$^{-1}$ for the molar mass of silicon, the total number of silicon atoms in the Earth then works out to be approximately $n_{\mathrm{Si},\oplus} = 2.06 \times 10^{49}$. Therefore, we estimate that $n_{\mathrm{Si},\epsilon}/n_{\mathrm{Si},\oplus} = 21.2^{+11.8}_{-6.6}$, implying that the accretion of approximately 15–33 $M_{\oplus}$ of rocky material with an Earth-like composition can explain the silicon enrichment we have measured for the atmosphere of WASP-121b.

As noted above, this calculation assumes that the planetary envelope started out with a silicon enrichment equal to that of the host star. However, the gas accreted by the planet at formation may have been depleted in silicon, as silicon would have condensed out of the gas phase throughout the protoplanetary disk[2,24]. If we repeat the above calculation assuming that all of the silicon observed in the atmosphere was acquired through the accretion of rocky material by setting $n_{\mathrm{Si,env}}/n_{\mathrm{H,env}} = n_{\mathrm{Si},\epsilon}/n_{\mathrm{H,env}}$, then we estimate the quantity of accreted rocky material to be $n_{\mathrm{Si},\epsilon}/n_{\mathrm{Si},\oplus} = 23.7^{+11.8}_{-6.6}$, or approximately $17$–$36\ M_{\oplus}$.

### Other wavelength-dependent quantities of interest

Supplementary Fig. 18 shows the atmospheric transmission spectrum of the day–night terminator region given by the inferred values for $R_p$ as a function of wavelength. The data exhibit a downward slope between 2.75–4 µm, then increase sharply and peak at around 4.5–4.8 µm, before dropping off again towards longer wavelengths. These wavelength-dependent variations in the transmission spectrum indicate that the data are precise enough to identify key absorbing species at the day–night terminator. Predictions from models published in ref. [1] and ref. [167] are also shown in Supplementary Fig. 18, both of which were fitted to optical and near-infrared data obtained with HST. Neither model provides a good fit to the longer-wavelength JWST data. Further analysis of the transmission spectrum will be presented in an accompanying paper[168].

Wavelength-dependent values for $u_1$ and $u_2$, $\Delta T_{\mathrm{mid}}$, and $\Delta\phi$ are also shown in Supplementary Figs. 19–21. We find that the posterior values for $u_1$ and $u_2$ closely match the `exoTiC-LD` priors. For $\Delta T_{\mathrm{mid}}$ and $\Delta\phi$, we do not see any clear evidence for variations as a function of wavelength that can easily be explained. A flat line fitted to the $\Delta T_{\mathrm{mid}}$ values gives $\chi_v^2 = 1.00$, as would be expected if there are no wavelength-dependent variations. However, a higher $\chi_v^2 = 1.74$ is obtained when a flat line is fitted to the $\Delta\phi$ values, indicating possible evidence for wavelength-dependent variations that should be investigated further in future studies.

### Data availability

The data used in this paper are associated with JWST programme G.O. 1729 (P.I., Evans-Soma; co-P.I., Kataria) and are publicly available via the Mikulski Archive for Space Telescopes at https://mast.stsci.edu. The specific observations analyzed can be accessed via https://doi.org/10.17909/6qnn-6j23 (ref. [169]). Data products derived in this work are available via Zenodo at https://doi.org/10.5281/zenodo.15377434 (ref. [170]).

### Code availability

The codes used in this Article to extract, reduce, and analyze the data are as follows: FIREFly[49], ExoTIC-LD[55], starry[50], emcee[53], MultiNest[126], PyMultiNest[125], astropy[171,172], matplotlib[173], numpy[174], pandas[175], scipy[176], ATMO[60–64], NEMESIS[65,66], HyDRA[68–70], PETRA[71], CHIMERA[67], Eureka![57].

## EXTENDED DATA

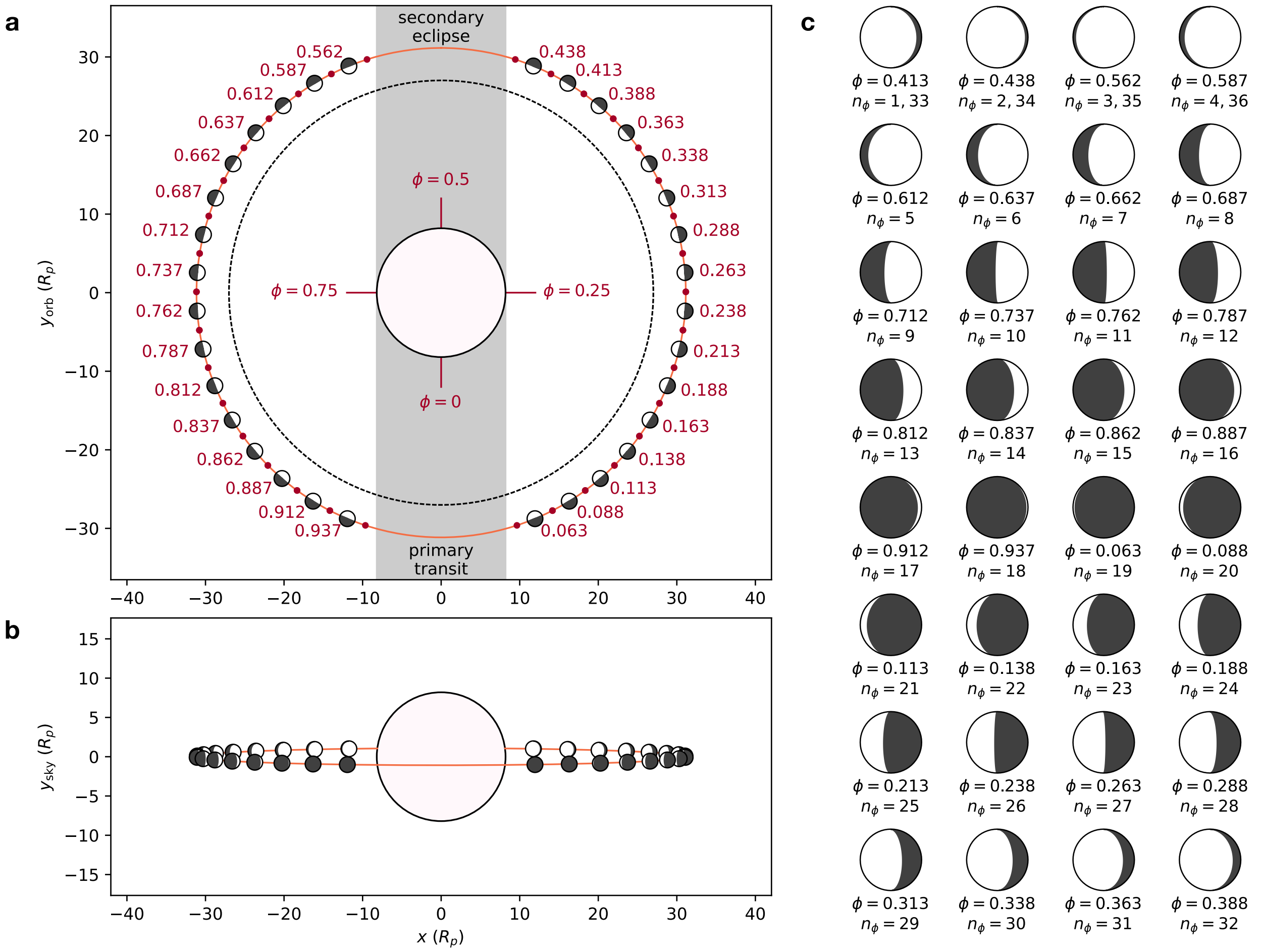

**Extended Data Fig. 1 | Adopted phase bins. a,** WASP-121 system to scale as seen from directly above the orbital plane. Central circle shows the host star, red line shows the circular orbit of WASP-121b, and dashed black line shows the Roche limit. Edges of the adopted phase ($\phi$) bins are indicated by small red circles along the orbital curve, with central phases labelled in red. Illustrations of WASP-121b are also shown at the central phase of each bin. Grey shaded region indicates the phases coinciding with the primary transit and secondary eclipse. **b,** View of the WASP-121 system projected onto the sky plane as it would appear from the solar system. **c,** Gallery of planetary phases with corresponding phase bin numbers ($n_\phi$). Two phases immediately before and two phases immediately after secondary eclipse were covered at both the beginning and end of the observation.

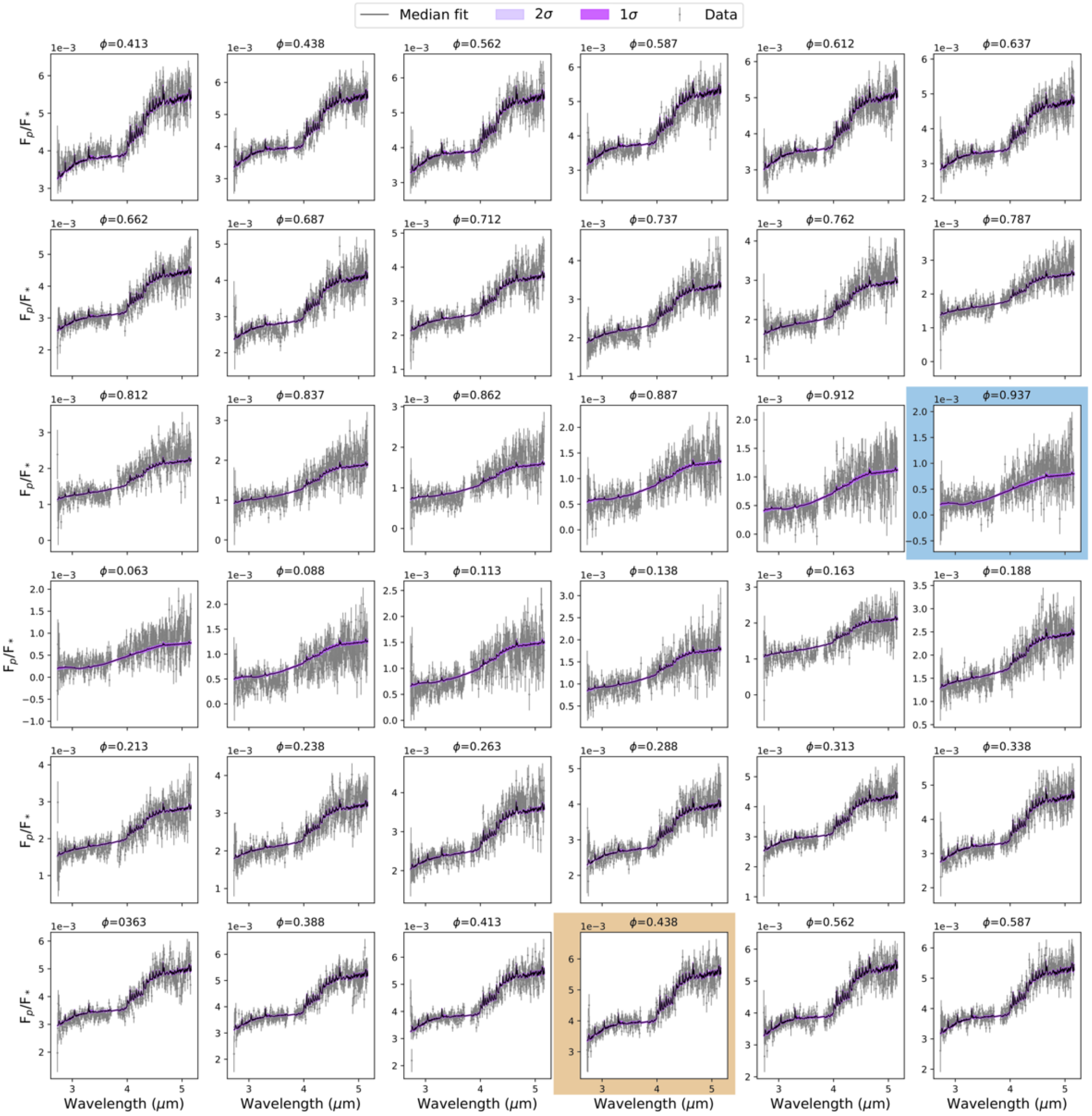

**Extended Data Fig. 2 | Emission spectra measured for all phase bins.** Phases ($\phi$) are listed at the top of each panel. Grey circles show median values and error bars indicate $1\sigma$ uncertainties based on $n = 5{,}000$ posterior samples. Purple lines show the emission spectra predicted at each phase from a weighted sum of the dayside and nightside components inferred by the NEMESIS retrieval analysis of the $n_\phi = 18$ ($\phi = 0.937$, panel with blue border) and $n_\phi = 34$ ($\phi = 0.438$, panel with brown border) spectra.

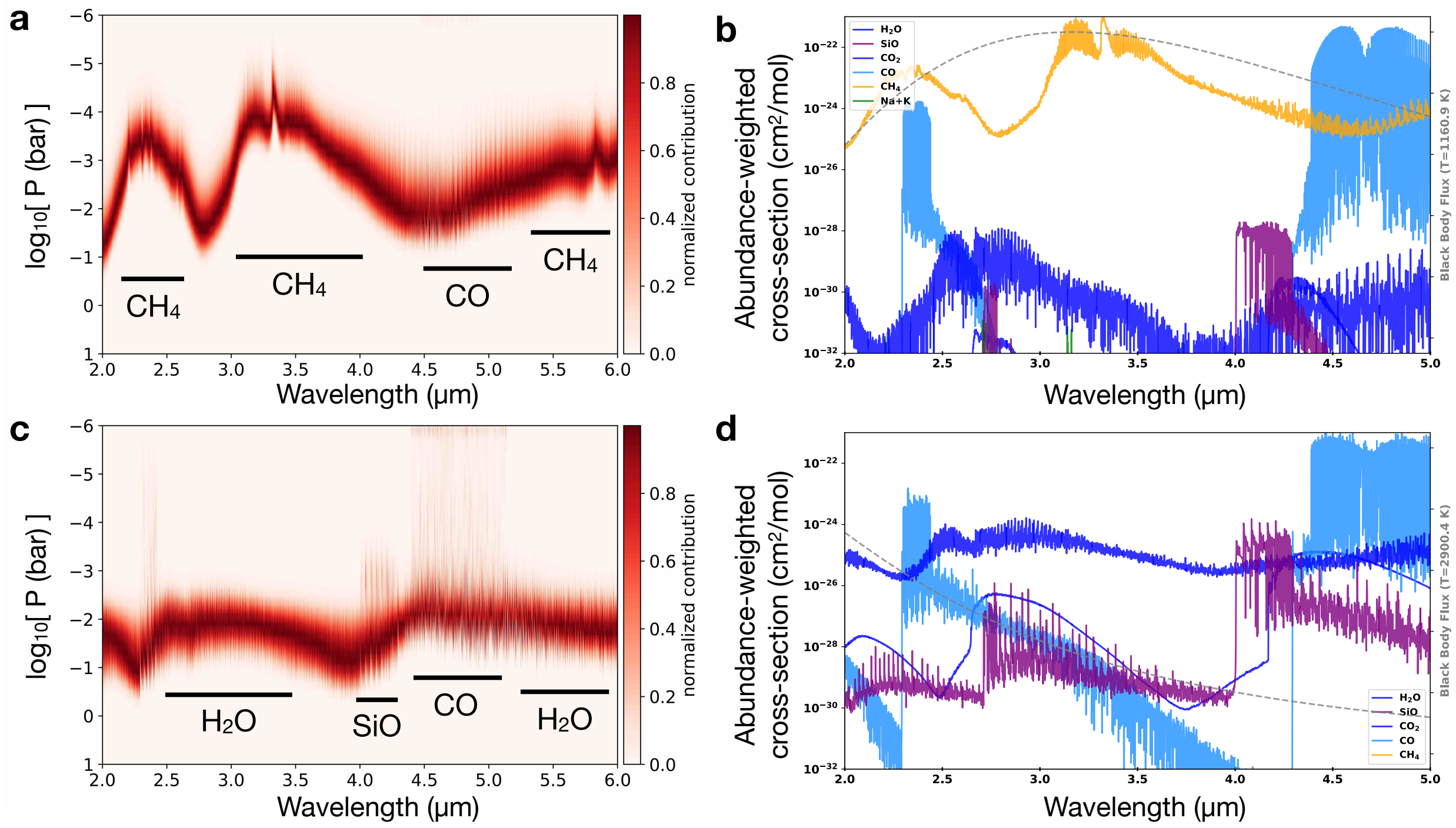

**Extended Data Fig. 3 | Atmospheric opacities. a,** Contribution function for the nightside atmosphere derived from the ATMO retrieval. The wavelength-dependent peak of the contribution function exhibits the spectral signature of $CH_4$, which dominates the nightside opacity in our model. **b,** Solid lines show the abundance-weighted absorption cross-sections for key chemical species at a pressure of 30 mbar from the best-fit ATMO model for the nightside. Dashed grey line shows a blackbody emission curve for the retrieved temperature at the same pressure level. **c**, Contribution function for the dayside atmosphere derived from the ATMO retrieval, exhibiting features of $H_2O$, SiO, and CO. **d,** The same as **b** for the dayside.

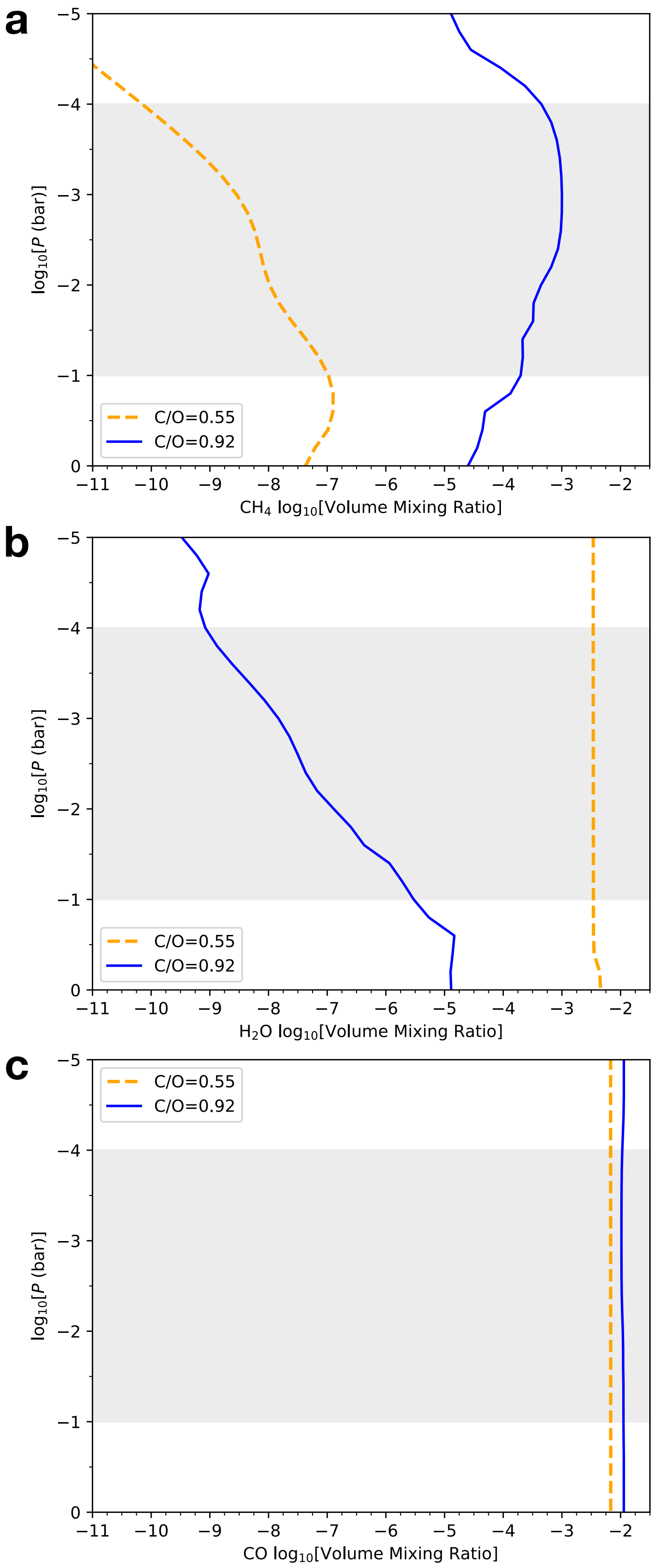

**Extended Data Fig. 4 | Nightside equilibrium chemistry abundances.** Solid blue lines show the pressure-dependent abundances predicted by equilibrium chemistry for **a,** $CH_4$, **b,** $H_2O$, and **c,** CO, assuming the nightside $PT$ profile derived from the ATMO retrieval and C/O=0.92. Dashed orange lines show the same, but for C/O=0.55. Grey shading indicates the approximate pressures probed by the data.

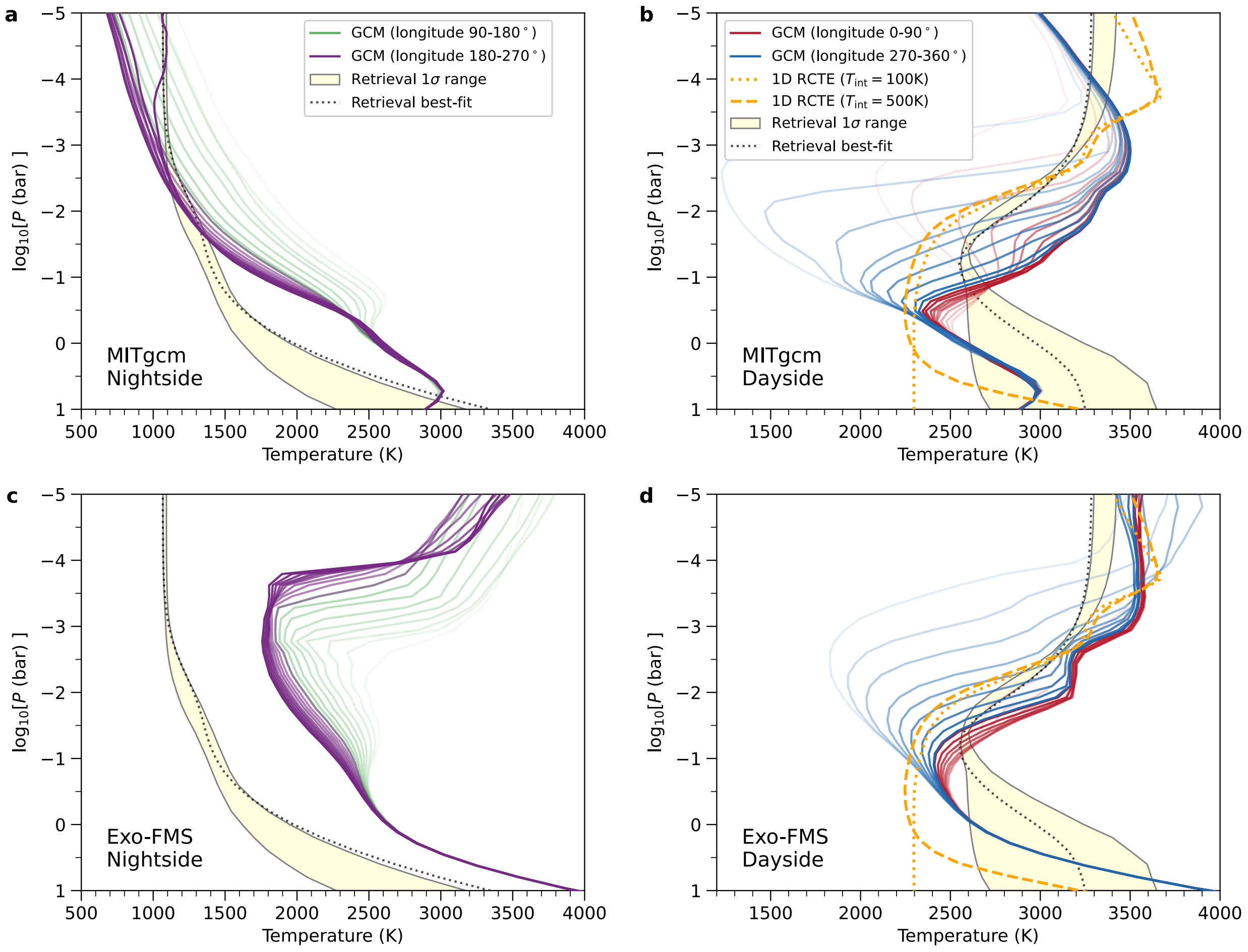

**Extended Data Fig. 5 | Retrieved *PT* profiles compared to self-consistent *PT* profiles.** **a,** Yellow shading indicates the 16th to 84th percentile (68% credible interval) temperatures inferred from the ATMO retrieval analysis at each pressure level for the nightside hemisphere. Black dotted line shows the best-fit (maximum a posteriori) profile. Green and purple lines show equatorial temperature profiles for longitudes ranging between 90–180° and 180–270°, respectively, taken from the MITgcm 3D GCM of ref. [11]. Plotted line transparency increases with increasing distance from the anti-stellar point at longitude 180°. **b,** Similar to **a**, but showing *PT* profiles for the dayside hemisphere. Red and blue lines show equatorial temperature profiles for longitudes ranging between 0–90° and 270–360°, respectively, from the GCM of ref. [11]. Line opacities decrease with increasing distance from the sub-stellar point at longitude 0°. Also shown are 1D radiative-convective thermo-chemical equilibrium models obtained with ATMO using the retrieved element abundances as inputs and assuming internal temperatures of $T_{\mathrm{int}} = 100$ K (orange dotted line) and $T_{\mathrm{int}} = 500$ K (orange dashed line). **c, d,** The same as **a, b,** but showing the Exo-FMS 3D GCM of ref. [19].

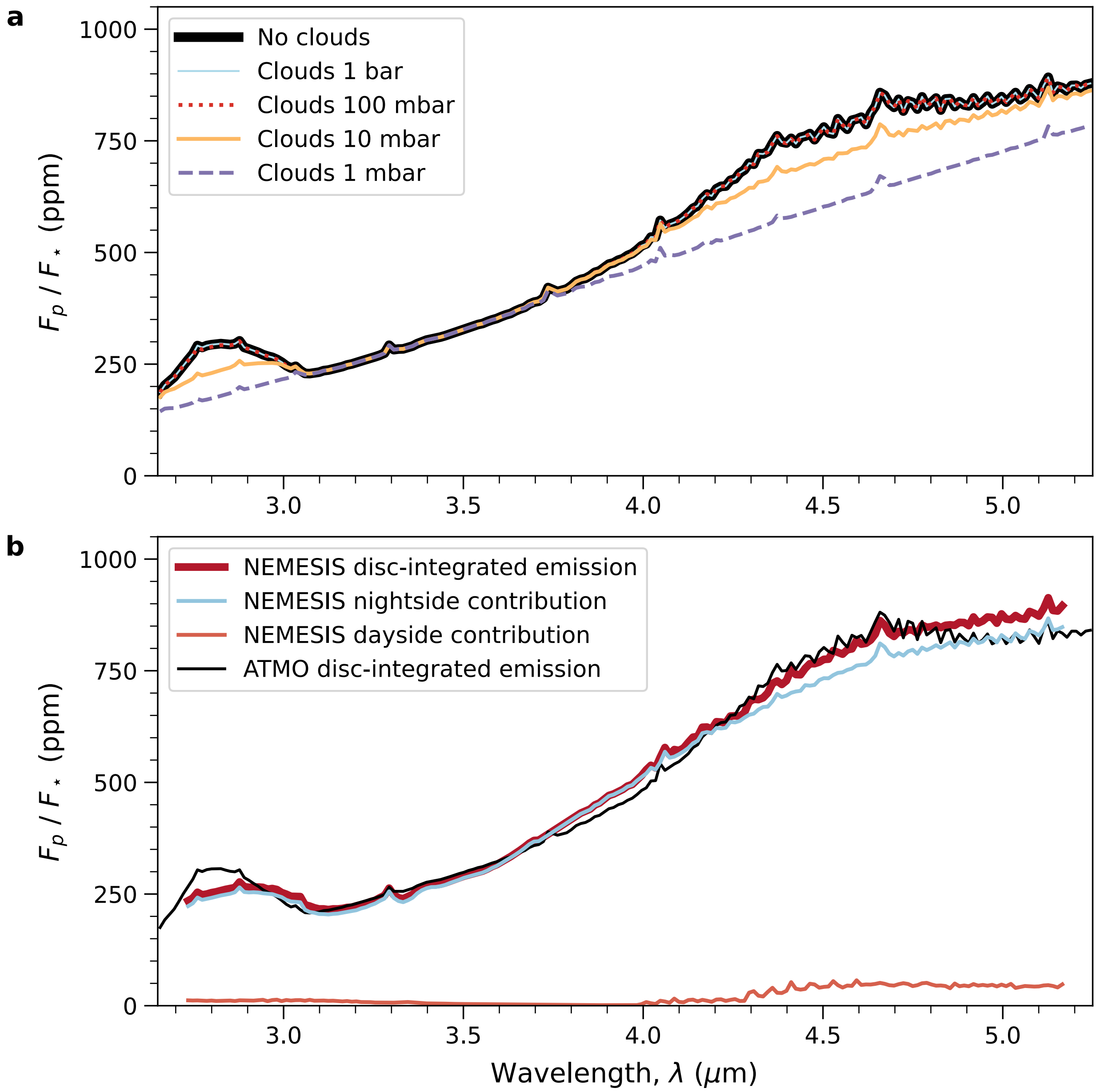

**Extended Data Fig. 6 | Effect of clouds and dayside crescent on the $n_\phi$=18 spectrum. a,** Best-fit ATMO model (thick black line) with optically thick cloud layers added at 1 bar (blue solid line), 100 mbar (red dotted line), 10 mbar (solid orange line), and 1 mbar (dashed purple line). **b,** Best-fit NEMESIS model (thick red line) with the relative contributions from the nightside (blue line) and dayside (light red line) hemispheres. Best-fit ATMO model from **a** is also shown for comparison (black line).

**Acknowledgements.** This work is based on observations made with the NASA/ESA/CSA JWST. The data were obtained from the Mikulski Archive for Space Telescopes at the Space Telescope Science Institute, which is operated by the Association of Universities for Research in Astronomy, Inc., under NASA contract NAS 5-03127 for JWST. J.K.B. is supported by an STFC Ernest Rutherford Fellowship (grant number ST/T004479/1). N.J.M. is funded by a UKRI Future Leaders Fellowship (grant number MR/T040866/1).

**Author contributions.** T.M.E.-S and T.K. co-led the observing programme. T.M.E.-S and E.-M.A. analyzed the JWST data. T.E.-S wrote the manuscript with assistance from D.K.S, J.K.B, A.A.A.P., and N.J.M. D.K.S, J.K.B, A.A.A.P., J.T., and J.D.L. performed the retrieval analyses with assistance from Z.R. and J.M.G. H.R. analyzed the high-resolution spectroscopy and extracted the stellar abundances. C.G., J.D., D.F.-M., and S.H. provided support for the JWST data analysis. D.A.C. and M.S.M. supported the interpretation of the results.

**Funding.** Open access funding provided by Max Planck Society.

**Competing interest declaration.** The authors declare no competing interests.

**SUPPLEMENTARY INFORMATION**

**Supplementary Table 1 | White phase-curve fit results.** Posterior distributions (medians with $1\sigma$ uncertainties defined by the 16th and 84th percentiles of $n = 5{,}000$ posterior samples) and best-fit values for the fitted and derived phase-curve model parameters.

| Fitted parameters | Posterior distribution | Best-fit |
|---|---|---|
| Shared: $M_\star$ ($M_\odot$) | $1.332^{+0.018}_{-0.017}$ | $1.332$ |
| Shared: $M_\mathrm{p}$ ($M_\mathrm{J}$) | $1.168^{+0.040}_{-0.041}$ | $1.170$ |
| Shared: $i$ (°) | $87.97^{+0.17}_{-0.15}$ | $87.96$ |
| Shared: log10[ $R_\star$ ($R_\odot$) ] | $0.1567^{+0.0020}_{-0.0020}$ | $0.1567$ |
| Shared: $\Delta P_\mathrm{orb}$ (s) | $-0.0009^{+0.0129}_{-0.0121}$ | $-0.0009$ |
| NRS1: log10[ $R_\mathrm{p}$ ($R_\mathrm{J}$) ] | $0.233^{+0.002}_{-0.002}$ | $0.233$ |
| NRS2: log10[ $R_\mathrm{p}$ ($R_\mathrm{J}$) ] | $0.237^{+0.002}_{-0.002}$ | $0.237$ |
| NRS1: $\Delta T_\mathrm{mid}$ (s) | $-26^{+17}_{-18}$ | $-26$ |
| NRS2: $\Delta T_\mathrm{mid}$ (s) | $-28^{+17}_{-18}$ | $-28$ |
| NRS1: $A$ (ppm) | $2031^{+6}_{-7}$ | $2031$ |
| NRS2: $A$ (ppm) | $2784^{+8}_{-9}$ | $2784$ |
| NRS1: $y_{1,0}$ | $0.804^{+0.003}_{-0.003}$ | $0.805$ |
| NRS2: $y_{1,0}$ | $0.669^{+0.002}_{-0.002}$ | $0.669$ |
| NRS1: $\Delta\phi$ (°) | $2.95^{+0.11}_{-0.12}$ | $2.96$ |
| NRS2: $\Delta\phi$ (°) | $2.39^{+0.12}_{-0.13}$ | $2.39$ |
| NRS1: $u_1$ | $0.048^{+0.006}_{-0.006}$ | $0.047$ |
| NRS2: $u_1$ | $0.031^{+0.005}_{-0.006}$ | $0.030$ |
| NRS1: $u_2$ | $0.118^{+0.010}_{-0.011}$ | $0.120$ |
| NRS2: $u_2$ | $0.090^{+0.010}_{-0.009}$ | $0.090$ |

**Supplementary Table 1** ***(continued)***

| Derived parameters | Posterior distribution | Best-fit |
|---|---|---|
| Shared: $R_\star$ ($R_\odot$) | $1.434^{+0.007}_{-0.007}$ | 1.434 |
| NRS1: $R_\mathrm{p}$ ($R_\mathrm{J}$) | $1.712^{+0.008}_{-0.008}$ | 1.721 |
| NRS2: $R_\mathrm{p}$ ($R_\mathrm{J}$) | $1.727^{+0.008}_{-0.008}$ | 1.727 |
| NRS1: $R_\mathrm{p}/R_\star$ | $0.122659^{+0.000062}_{-0.000056}$ | 0.122657 |
| NRS2: $R_\mathrm{p}/R_\star$ | $0.123732^{+0.000064}_{-0.000066}$ | 0.123736 |
| Shared: $\Delta P_\mathrm{orb}$ (day) | $1.27492503^{+0.00000015}_{-0.00000014}$ | 1.27492503 |
| NRS1: $\Delta T_\mathrm{mid}$ ($BJD_{TDB}$) | $2459867.64267^{+0.00020}_{-0.00021}$ | 2459867.64265 |
| NRS2: $\Delta T_\mathrm{mid}$ ($BJD_{TDB}$) | $2459867.64265^{+0.00020}_{-0.00021}$ | 2459867.64263 |

**Supplementary Table 2 | Atmospheric opacities.** Line and continuum opacity sources used by each retrieval code. Note that the abundances for TiO, VO, FeH, HCN, $NH_3$, $C_2H_2$ and OH were held fixed to values appropriate for a background solar composition atmosphere in the PETRA retrieval, including the opacity sources listed for the PHOENIX model in Extended Data Table 2 of ref. [58]. For further details on opacity sources adopted by the ATMO retrieval, which assumed thermochemical equilibrium including additional species not shown below, see refs [60–64].

| Opacity | ATMO | NEMESIS | HyDRA | CHIMERA | PETRA |
|---|---|---|---|---|---|
| $H_2O$ | 72 | 73 | 74 | 73 | 72 |
| $CH_4$ | 75 | 76 | 75, 77 | 78 | — |
| CO | 74 | 79 | 74 | 74 | 80 |
| SiO | 81 | 81 | 82 | — | 83 |
| $CO_2$ | 84 | 85 | 74 | 85 | 86 |
| $SiO_2$ | — | — | 87 | — | — |
| HCN | 88,89 | 89 | 88 | 89 | 86 |
| NH3 | 90 | — | — | 91 | 86 |
| $C_2H_2$ | 92 | 93 | — | 93 | 86 |
| $H^-$ | 94, 95 | 94 | 94, 95 | — | 94 |
| $H_2$-$H_2$ | 96 | 97, 98 | 96 | 96 | 99 |
| $H_2$-He | 96 | 100, 101 | 96 | 96 | 99 |
| OH | — | — | 74 | — | 86 |
| TiO | 102 | — | 103 | — | 104 |
| VO | 105 | — | 105 | — | 106 |
| FeH | 107 | — | 108 | — | 108 |

**Supplementary Table 3 | Priors adopted for the ATMO retrievals.** Listed values give the lower and upper bounds of the uniform distributions used as priors in the retrieval analyses.

| **ATMO dayside** | | | **ATMO nightside** | | |
|---|---|---|---|---|---|
| **Parameter** | **Description** | **Prior** | **Parameter** | **Description** | **Prior** |
| [C/H] | Carbon Metallicity | $[-1, 2]$ dex | [C/H] | Carbon metallicity | $[-1, 2]$ dex |
| [O/H] | Oxygen metallicity | $[-1, 2]$ dex | [O/H] | Oxygen metallicity | $[-1, 2]$ dex |
| [Si/H] | Silicon metallicity | $[-1, 2]$ dex | [Si/H] | Silicon metallicity | $[-1, 2]$ dex |
| [M/H] | Other heavy elements metallicity | $[-1, 2]$ dex | [M/H] | Other heavy elements metallicity | $[-1, 2]$ dex |
| $\log_{10} \kappa_{\mathrm{IR}}$ | Infrared opacity | $[-5, -0.5]$ dex | $\log_{10} \kappa_{\mathrm{IR}}$ | Infrared opacity | $[-5, -0.5]$ dex |
| $\log_{10} \gamma_{\mathrm{V1/IR}}$ | Visible channel 1 relative opacity | $[-4, 1.5]$ dex | $\log_{10} \gamma_{\mathrm{V/IR}}$ | Visible Relative opacity | $[-4, 1.5]$ dex |
| $\log_{10} \gamma_{\mathrm{V2/IR}}$ | Visible channel 2 relative opacity | $[-4, 1.5]$ dex | $\beta_{\mathrm{r}}$ | Redistribution factor | $[0, 2]$ |
| $\beta_{\mathrm{r}}$ | Redistribution factor | $[0, 2]$ | | | |
| $\alpha_{\mathrm{V}}$ | Visible channels partition | $[0, 1]$ | | | |

**Supplementary Table 4 | Priors adopted for the NEMESIS retrievals.** The same as Supplementary Table 3, for the NEMESIS retrievals.

| NEMESIS dayside | | | NEMESIS nightside | | |
|---|---|---|---|---|---|
| **Parameter** | **Description** | **Prior** | **Parameter** | **Description** | **Prior** |
| $\log_{10}$ $H_2O$ | Log $H_2O$ deep mixing ratio | $[-12, -1]$ dex | $\log_{10}$ $H_2O$ | Log $H_2O$ mixing ratio | $[-12, -1]$ dex |
| $\log_{10} P_0$ | Log $H_2O$ knee pressure | $[-12, -1]$ dex | $\log_{10}$ $H^-$ | Log $H^-$ mixing ratio | $[-13, -5]$ dex |
| $\log_{10} \tau$ | Log $H_2O$ power index | $[-12, -1]$ dex | $\log_{10}$ CO | Log CO mixing ratio | $[-12, -1]$ dex |
| $\log_{10}$ $H^-$ | Log $H^-$ mixing ratio | $[-13, -5]$ dex | $\log_{10}$ $CO_2$ | Log $CO_2$ mixing ratio | $[-12, -1]$ dex |
| $\log_{10}$ CO | Log CO mixing ratio | $[-12, -1]$ dex | $\log_{10}$ $CH_4$ | Log $CH_4$ mixing ratio | $[-12, -1]$ dex |
| $\log_{10}$ $CO_2$ | Log $CO_2$ mixing ratio | $[-12, -1]$ dex | $\log_{10}$ SiO | Log SiO mixing ratio | $[-12, -1]$ dex |
| $\log_{10}$ $CH_4$ | Log $CH_4$ mixing ratio | $[-12, -1]$ dex | $\log_{10}$ $C_2H_2$ | Log $C_2H_2$ mixing ratio | $[-12, -1]$ dex |
| $\log_{10}$ SiO | Log SiO mixing ratio | $[-12, -1]$ dex | $\log_{10}$ HCN | Log HCN mixing ratio | $[-12, -1]$ dex |
| $\log_{10} P_1$ | Log $PT$ pressure 1 | $[-8, 2]$ dex bar | $\log_{10} P_1$ | Log $PT$ pressure 1 | $[-8, 2]$ dex bar |
| $\log_{10} P_2$ | Log $PT$ pressure 2 | $[-8, 2]$ dex bar | $\log_{10} P_2$ | Log $PT$ pressure 2 | $[-8, 2]$ dex bar |
| $\log_{10} P_3$ | Log $PT$ pressure 3 | $[-8, 2]$ dex bar | $\log_{10} P_3$ | Log $PT$ pressure 3 | $[-8, 2]$ dex bar |
| $T_1$ | Temperature at $P_1$ | $[1000, 3500]$ K | $T_1$ | Temperature at $P_1$ | $[800, 3300]$ K |
| $\alpha_1$ | $PT$ power index 1 | $[0.02, 1]$ | $\alpha_1$ | $PT$ power index 1 | $[0.02, 1]$ |
| $\alpha_2$ | $PT$ power index 2 | $[0.02, 1]$ | $\alpha_2$ | $PT$ power index 2 | $[0.02, 1]$ |
| | | | $\log_{10} P_{\mathrm{top}}$ | Log cloud pressure | $[-6, 1]$ dex bar |
| | | | $\gamma$ | Scattering index | $[0, 14]$ |
| | | | $\sigma_0$ | Log cloud opacity | $[-10, 10]$ dex |

**Supplementary Table 5 | Priors adopted for the CHIMERA retrievals.** The same as Supplementary Tables 3–4, for the CHIMERA retrievals.

| CHIMERA nightside | | |
|---|---|---|
| **Parameter** | **Description** | **Prior** |
| $\log_{10}$ $H_2O$ | Log $H_2O$ mixing ratio | $[-12, -1]$ dex |
| $\log_{10}$ CO | Log CO mixing ratio | $[-12, -1]$ dex |
| $\log_{10}$ $CO_2$ | Log $CO_2$ mixing ratio | $[-12, -1]$ dex |
| $\log_{10}$ $CH_4$ | Log $CH_4$ mixing ratio | $[-12, -1]$ dex |
| $\log_{10}$ $C_2H_2$ | Log $C_2H_2$ mixing ratio | $[-12, -1]$ dex |
| $\log_{10}$ HCN | Log HCN mixing ratio | $[-12, -1]$ dex |
| $\log_{10}$ $NH_3$ | Log $NH_3$ mixing ratio | $[-12, -1]$ dex |
| $\log_{10} \kappa_{\mathrm{IR}}$ | Infrared opacity | $[-4, 2.5]$ dex |
| $\log_{10} \gamma_{\mathrm{V1/IR}}$ | Visible channel 1 relative opacity | $[-2.5, 4]$ dex |
| $\log_{10} \gamma_{\mathrm{V2/IR}}$ | Visible channel 2 relative opacity | $[-2.5, 4]$ dex |
| $T_{\mathrm{eff}}$ | Effective temperature | $[400, 2800]$ K |
| $\alpha_{\mathrm{V}}$ | Visible channels partition | $[0, 1]$ |
| $\delta$ | Dilution factor | $[0, 1]$ |

**Supplementary Table 6 | Priors adopted for the HyDRA retrievals.** The same as Supplementary Tables 3–5, for the HyDRA retrievals.

| HyDRA dayside | | | HyDRA nightside | | |
|---|---|---|---|---|---|
| **Parameter** | **Description** | **Prior** | **Parameter** | **Description** | **Prior** |
| $\log_{10}$ $H_2O$ | Log $H_2O$ deep mixing ratio | $[-15, -1]$ dex | $\log_{10}$ $H_2O$ | Log $H_2O$ mixing ratio | $[-15, -1]$ dex |
| $\log_{10}$ $H^-$ | Log $H^-$ mixing ratio | $[-15, -1]$ dex | $\log_{10}$ $H^-$ | Log $H^-$ mixing ratio | $[-15, -1]$ dex |
| $\log_{10}$ CO | Log CO mixing ratio | $[-15, -1]$ dex | $\log_{10}$ CO | Log CO mixing ratio | $[-15, -1]$ dex |
| $\log_{10}$ $CO_2$ | Log $CO_2$ mixing ratio | $[-15, -1]$ dex | $\log_{10}$ $CO_2$ | Log $CO_2$ mixing ratio | $[-15, -1]$ dex |
| $\log_{10}$ SiO | Log SiO mixing ratio | $[-15, -1]$ dex | $\log_{10}$ $CH_4$ | Log $CH_4$ mixing ratio | $[-15, -1]$ dex |
| $\log_{10}$ $SiO_2$ | Log $SiO_2$ mixing ratio | $[-15, -1]$ dex | $\log_{10}$ SiO | Log SiO mixing ratio | $[-15, -1]$ dex |
| $\log_{10}$ TiO | Log TiO mixing ratio | $[-15, -1]$ dex | $\log_{10}$ $SiO_2$ | Log $SiO_2$ mixing ratio | $[-15, -1]$ dex |
| $\log_{10}$ VO | Log VO mixing ratio | $[-15, -1]$ dex | $\log_{10}$ HCN | Log HCN mixing ratio | $[-15, -1]$ dex |
| $\log_{10}$ OH | Log OH mixing ratio | $[-15, -1]$ dex | $\log_{10} P_1$ | Log $PT$ pressure 1 | $[-5, 3]$ dex bar |
| $\log_{10}$ FeH | Log FeH mixing ratio | $[-15, -1]$ dex | $\log_{10} P_2$ | Log $PT$ pressure 2 | $[-5, 3]$ dex bar |
| $\log_{10}$ HCN | Log HCN mixing ratio | $[-15, -1]$ dex | $\log_{10} P_3$ | Log $PT$ pressure 3 | $[-5, 3]$ dex bar |
| $\log_{10} P_1$ | Log $PT$ pressure 1 | $[-5, 3]$ dex bar | $T_1$ | 100 mbar temperature | $[500, 4000]$ K |
| $\log_{10} P_2$ | Log $PT$ pressure 2 | $[-5, 3]$ dex bar | $\alpha_1$ | $PT$ power index 1 | $[0.02, 1]$ |
| $\log_{10} P_3$ | Log $PT$ pressure 3 | $[-5, 3]$ dex bar | $\alpha_2$ | $PT$ power index 2 | $[0.02, 1]$ |
| $T_1$ | 100 mbar temperature | $[500, 4000]$ K | $\delta$ | Dilution factor | $[0, 1]$ |
| $\alpha_1$ | $PT$ power index 1 | $[0.02, 1]$ | $\log_{10} a_c$ | Modal cloud particle size | $[-2, 1]$ dex |
| $\alpha_2$ | $PT$ power index 2 | $[0.02, 1]$ | $\log_{10} P_c$ | Cloud base pressure | $[-5, 2]$ dex bar |
| $\delta$ | Dilution factor | $[0, 1]$ | $\log_{10} f_c$ | Cloud particle abundance | $[-30, -1]$ dex |
| $\log_{10} a_c$ | Modal cloud particle size | $[-2, 1]$ dex | $\varepsilon$ | Cloud pressure exponent | $[0.3, 10]$ |
| $\log_{10} P_c$ | Cloud base pressure | $[-5, 2]$ dex bar | $\psi_c$ | Cloud covering fraction | $[0, 1]$ |
| $\log_{10} f_c$ | Cloud particle abundance | $[-30, -1]$ dex | | | |

**Supplementary Table 6** ***(continued)***

| | | | | | |
|---|---|---|---|---|---|
| $\varepsilon$ | Cloud pressure exponent | $[\,0.3, 10\,]$ | | | |
| $\psi_c$ | Cloud covering fraction | $[\,0, 1\,]$ | | | |

**Supplementary Table 7 I Priors adopted for the PETRA retrievals.** The same as Supplementary Tables 3–6, for the PETRA retrievals.

| PETRA dayside | | |
|---|---|---|
| **Parameter** | **Description** | **Prior** |
| $\log_{10}$ $H_2O$ | Log $H_2O$ deep mixing ratio | $[-20, -1\,]$ dex |
| $\log_{10}P_0$ | Log $H_2O$ knee pressure | $[-5, 5\,]$ dex bar |
| $\log_{10}\tau$ | Log $H_2O$ power index | $[-15, 15\,]$ dex |
| $\log_{10}$ CO | Log CO mixing ratio | $[-20, -1\,]$ dex |
| $\log_{10}$ SiO | Log SiO mixing ratio | $[-20, -1\,]$ dex |
| $\log_{10}\kappa_{\mathrm{IR}}$ | Infrared opacity | $[\,-4, 2.5\,]$ dex |
| $\log_{10}\gamma_{\mathrm{V1/IR}}$ | Visible channel 1 relative opacity | $[-2.5, 4\,]$ dex |
| $\log_{10}\gamma_{\mathrm{V2/IR}}$ | Visible channel 2 relative opacity | $[-2.5, 4\,]$ dex |
| $\beta_r$ | Redistribution factor | $[\,0.3, 1.75\,]$ |
| $\alpha_V$ | Visible channels partition | $[\,0, 1\,]$ |

**Supplementary Table 8 I Stellar atomic data, equivalent-width measurements, and individual line abundance inferences.** In columns one through five we give the wavelength of the atomic transition, the species, excitation potential of the transition (ep), the logarithm of the transition probability (log(gf)), the measured equivalent width (EW) and the calculated line abundance (A(X)), respectively. The atomic data are an updated version of the line list initially presented in ref. [151], with the addition of hyperfine structure and isotopic splitting for several elements. *Provided as a Supplementary File in .xlsx format.*

**Supplementary Table 9 I Properties of the WASP-121 host star.** All of our stellar parameters were inferred with the methodology described in ref. [138], which is a hybrid of classical spectroscopic analysis and an isochrone fitting technique, with the spectroscopic data being used as a prior to the isochrone fitting. Quoted values are the medians with $1\sigma$ uncertainties defined by the 16th and 84th percentiles. With our methodology the inferred stellar parameters are consistent with all available stellar information, namely: spectroscopic, photometric, astrometric and extinction data. Effective temperature is derived mostly from photometric information, surface gravity comes mostly from astrometric data, and metallicity from spectroscopic data.

| Parameter | Value | Unit |
|---|---|---|
| Tycho *B* | $11.093 \pm 0.049$ | Vega mag |
| Tycho *V* | $10.571 \pm 0.045$ | Vega mag |
| Gaia DR2 *G* | $10.3752 \pm 0.002$ | Vega mag |
| 2MASS *J* | $9.625 \pm 0.021$ | Vega mag |
| 2MASS *H* | $9.439 \pm 0.025$ | Vega mag |
| 2MASS $K_s$ | $9.374 \pm 0.022$ | Vega mag |
| WISE $W_1$ | $9.356 \pm 0.023$ | Vega mag |
| WISE $W_2$ | $9.387 \pm 0.019$ | Vega mag |
| Gaia DR3 parallax | $3.799 \pm 0.010$ | mas |
| Distance, *d* | $262.6^{+0.3}_{-0.5}$ | pc |
| Interstellar extinction, $A_V$ | $0.05^{+0.02}_{-0.03}$ | mag |
| Effective temperature, $T_{eff}$ | $6481^{+46}_{-48}$ | K |
| Surface gravity, $\log_{10} g_\star$ | $4.23^{+0.01}_{-0.02}$ | dex m s$^{-2}$ |
| Stellar mass, $M_\star$ | $1.36^{+0.03}_{-0.03}$ | $M_\odot$ |
| Stellar radius, $R_\star$ | $1.49^{+0.01}_{-0.01}$ | $R_\odot$ |
| Luminosity, $L_\star$ | $3.52 \pm 0.08$ | $L_\odot$ |
| Isochrone-based age, $t_{iso}$ | $1.65^{+0.35}_{-0.33}$ | Gyr |
| Metallicity, [Fe/H] | $0.114 \pm 0.039$ | dex |
| Microturbulence, $\xi$ | $1.57 \pm 0.11$ | km s$^{-1}$ |

**Supplementary Table 10 I WASP-121 stellar abundances.** Mean estimated abundances for each of our measured elements in standard [X/H] and *A*(X) formats, where $A(X) = \log_{10} N_X/N_H + 12$. Quoted uncertainties include both the line-by-line dispersion when multiple lines were available, and uncertainties propagated from the stellar parameter uncertainties. Non-LTE corrections obtained using a multi-variable linear interpolation technique are provided where available, with uncertainties reflecting the line-by-line dispersion.

| | LTE abundances | | Non-LTE abundances | |
|---|---|---|---|---|
| **Species** | **[X/H]** | ***A*(X)** | **[X/H]** | ***A*(X)** |
| C I | $0.085 \pm 0.063$ | $8.545 \pm 0.063$ | $0.067 \pm 0.050$ | $8.527 \pm 0.050$ |
| O I | $0.548 \pm 0.052$ | $9.238 \pm 0.052$ | $0.165 \pm 0.024$ | $8.855 \pm 0.024$ |
| Na I | $0.172 \pm 0.029$ | $6.392 \pm 0.029$ | — | — |
| Mg I | $0.046 \pm 0.147$ | $7.596 \pm 0.147$ | — | — |
| Al I | $0.110 \pm 0.030$ | $6.540 \pm 0.030$ | $0.096$ | $6.526$ |
| Si I | $0.201 \pm 0.037$ | $7.711 \pm 0.037$ | $0.195 \pm 0.097$ | $7.696 \pm 0.097$ |
| S I | $0.144 \pm 0.117$ | $7.264 \pm 0.117$ | — | — |
| K I | $0.616 \pm 0.084$ | $5.686 \pm 0.084$ | $0.099$ | $5.169$ |
| Ca I | $0.212 \pm 0.056$ | $6.512 \pm 0.056$ | $-0.304 \pm 0.111$ | $5.996 \pm 0.111$ |
| Ti II | $0.107 \pm 0.079$ | $5.077 \pm 0.079$ | — | — |
| V I | $-0.064 \pm 0.063$ | $3.836 \pm 0.063$ | — | — |
| Cr I | $0.166 \pm 0.071$ | $5.786 \pm 0.071$ | — | — |
| Cr II | $0.230 \pm 0.088$ | $5.850 \pm 0.088$ | — | — |
| Mn I | $0.016 \pm 0.060$ | $5.436 \pm 0.060$ | — | — |
| Fe I | $0.126 \pm 0.044$ | $7.586 \pm 0.044$ | $0.164 \pm 0.079$ | $7.624 \pm 0.079$ |
| Fe II | $0.169 \pm 0.070$ | $7.629 \pm 0.070$ | $0.221 \pm 0.075$ | $7.681 \pm 0.075$ |
| Co I | $0.132 \pm 0.035$ | $5.072 \pm 0.035$ | — | — |
| Ni I | $0.150 \pm 0.051$ | $6.350 \pm 0.051$ | — | — |
| Cu I | $0.021 \pm 0.091$ | $4.201 \pm 0.091$ | — | — |
| Zn I | $-0.021 \pm 0.057$ | $4.539 \pm 0.057$ | — | — |
| Sc II | $0.269 \pm 0.099$ | $3.409 \pm 0.099$ | — | — |

**Supplementary Table 10** ***(continued)***

| | | | | |
|---|---|---|---|---|
| Y II | $0.233 \pm 0.146$ | $2.443 \pm 0.146$ | — | — |
| Zr II | $0.051 \pm 0.113$ | $2.641 \pm 0.113$ | — | — |
| Ba II | $0.166 \pm 0.127$ | $2.436 \pm 0.127$ | — | — |
| Ce II | $0.212 \pm 0.117$ | $1.792 \pm 0.117$ | — | — |
| Nd II | $0.308 \pm 0.067$ | $1.728 \pm 0.067$ | — | — |
| Sm II | $0.344 \pm 0.066$ | $1.294 \pm 0.066$ | — | — |

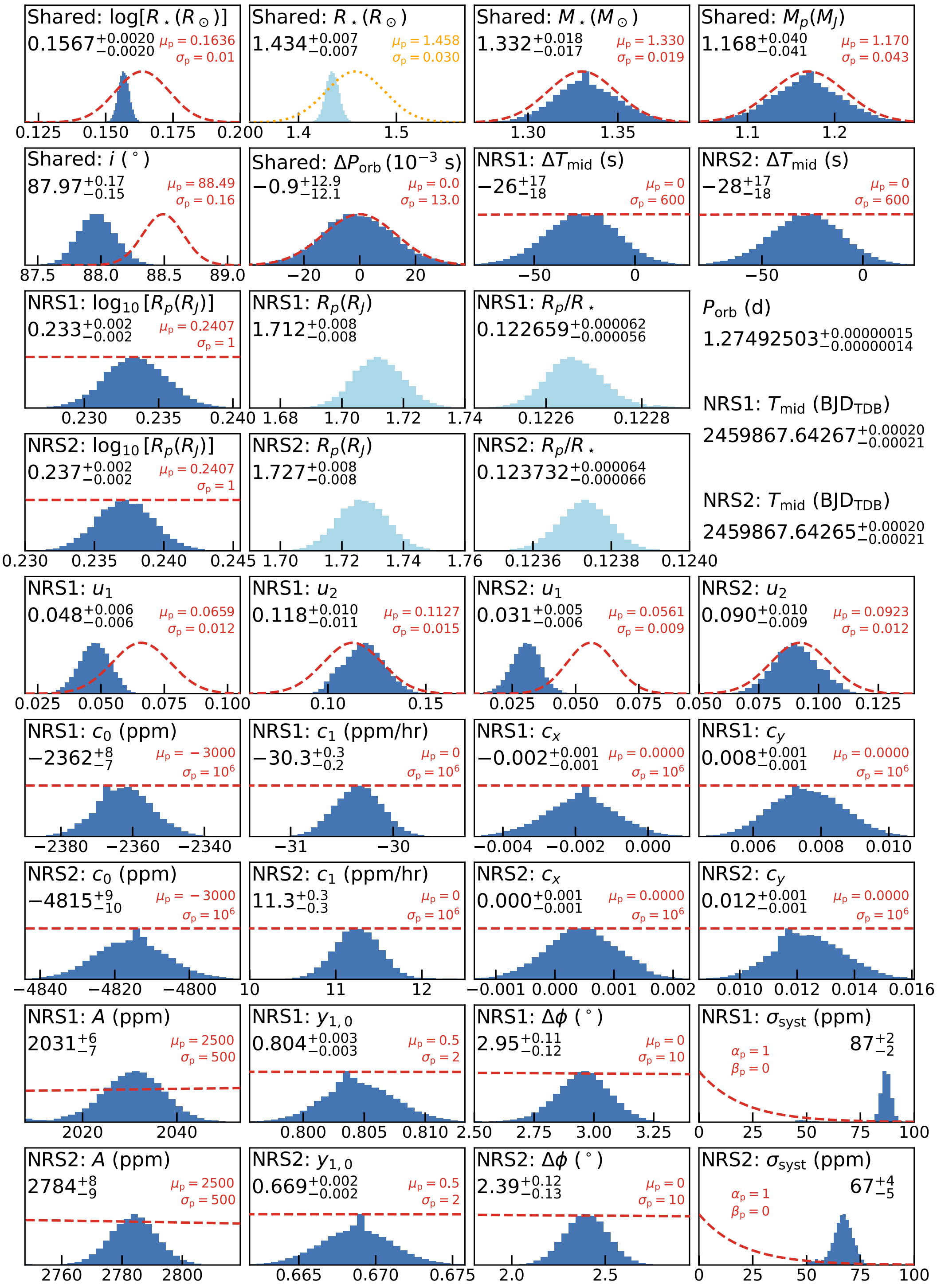

**Supplementary Fig. 1 | Results of the white phase-curve fit.** An updated version of a similar figure presented in ref. [5]. Dark (light) blue histograms show posterior distributions for fitted (derived) parameters. Posterior median values with $1\sigma$ uncertainties defined by the 16th and 84th percentiles of $n = 5{,}000$ posterior samples are displayed in black font. Prior distributions are plotted as dashed red lines, with corresponding parameters displayed in red font, namely: the mean ($\mu_{\mathrm{p}}$) and standard deviation ($\sigma_{\mathrm{p}}$) for normal priors; and the shape ($\alpha_{\mathrm{p}}$) and inverse-scale ($\beta_{\mathrm{p}}$) parameters for gamma priors. Our derived distribution for $R_\star$ is compared to the posterior distribution for the stellar radius reported by ref. [56], which is shown by the dotted orange line.

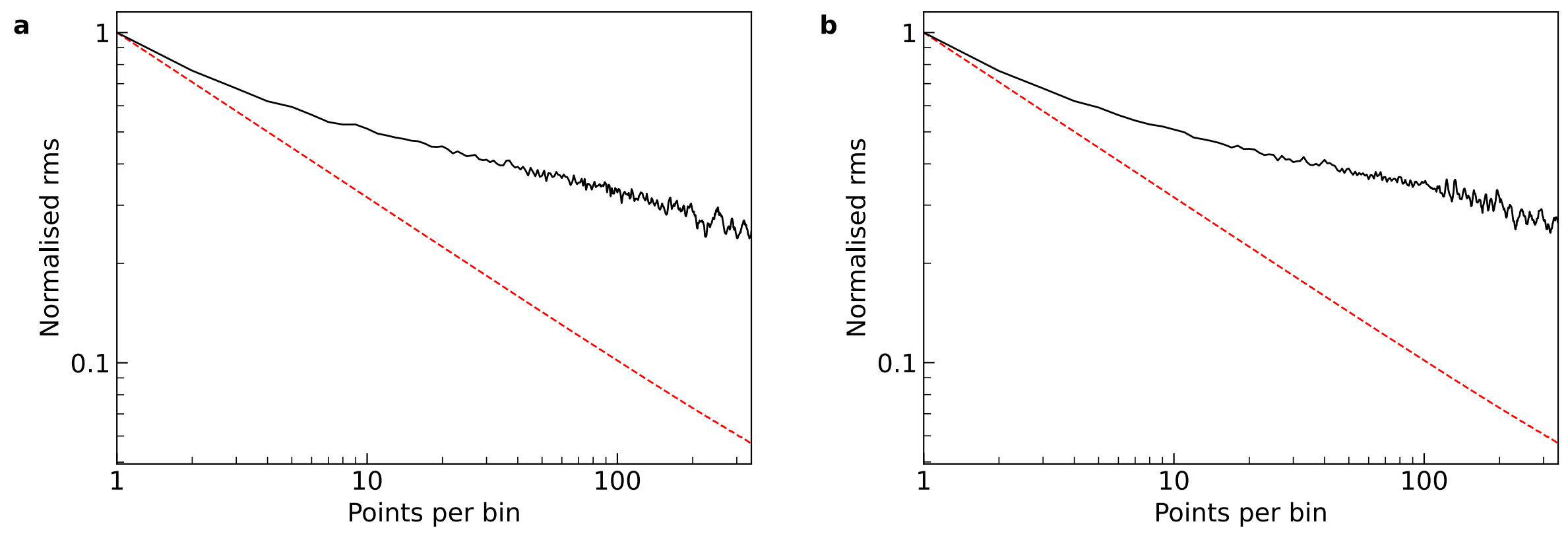

**Supplementary Fig. 2 | Root mean square (rms) of white phase-curve model residuals. a,** The black line shows how the normalized rms of the best-fit model residuals bins down as a function of the number of points per bin for the NRS1 white phase curve. The dashed red line indicates the expected rms scaling for uncorrelated ('white') noise. **b,** The same as **a**, but for the NRS2 white phase-curve fit.

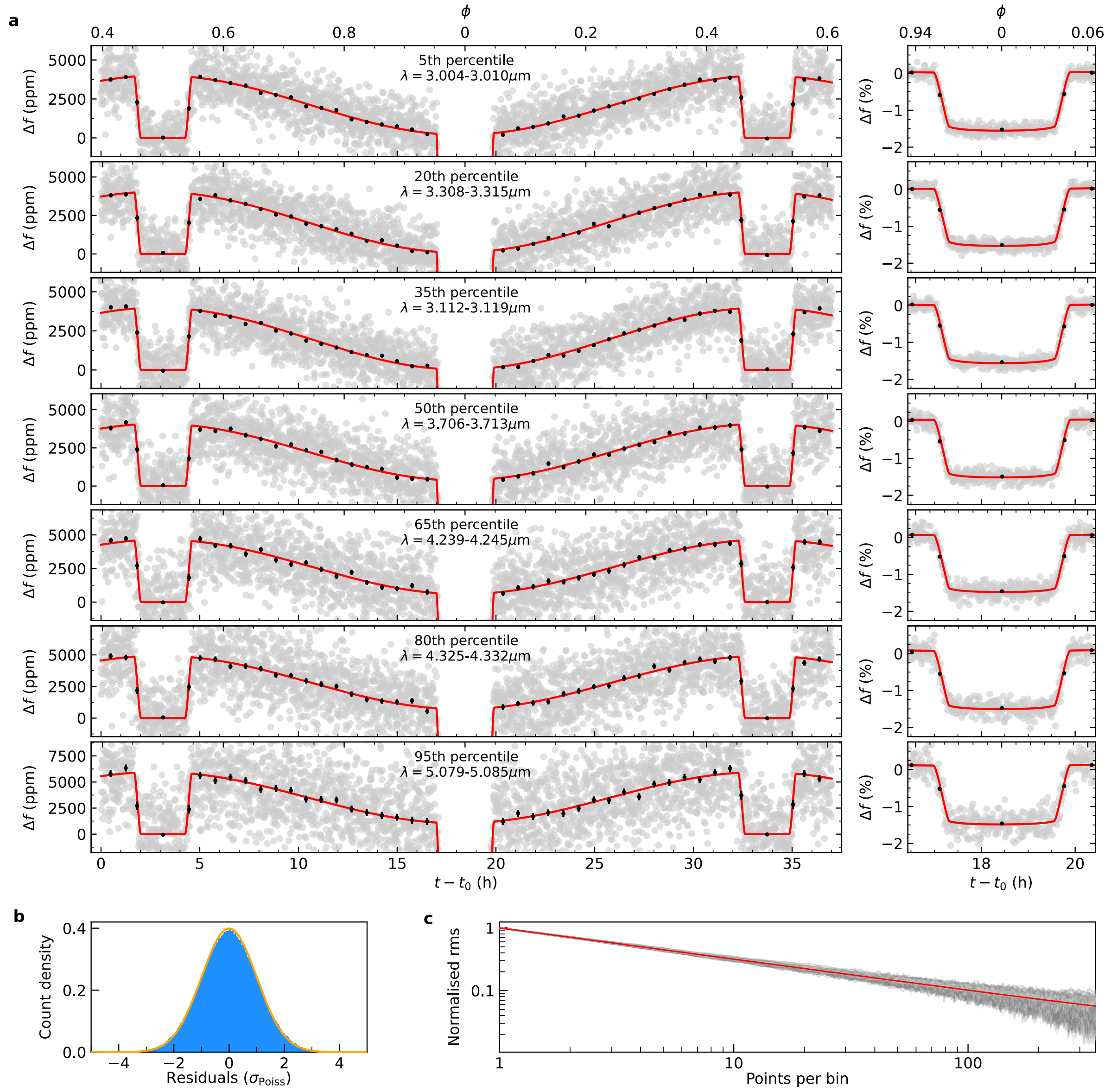

**Supplementary Fig. 3 | Spectroscopic phase-curve fits and residuals. a,** A selection of spectroscopic phase curves illustrating the range of fit qualities across all wavelength channels, ranging from the 5th to 95th percentiles as measured by the $\chi^2$ statistic. The left column highlights the planetary emission component of the signal and the right column gives a close-up of the primary transit. Grey circles are the measured emission relative to the stellar baseline level ($\Delta f$), after correcting for the instrument baseline trend and subtracting the stellar baseline level. Black circles show median values and error bars indicate standard deviations ($1\sigma$ uncertainties) based on $n = 5{,}000$ posterior samples, for the phase bins used to generate the phase-resolved emission spectra, as well as additional bins for the ingress, bottom and egress of the primary transit and each secondary eclipse. Red lines show the best-fit phase-curve models. **b,** Blue bars show a histogram of the binned model residuals for all spectroscopic phase curves, expressed in units of the pipeline Poisson noise ($\sigma_{\rm Poiss}$). The orange line shows a standard normal distribution. **c,** Grey lines show the rms of model residuals versus bin size for all spectroscopic phase curves. The red line shows how the root-mean-square (rms) of the model residuals would bin down for uncorrelated ('white') noise.

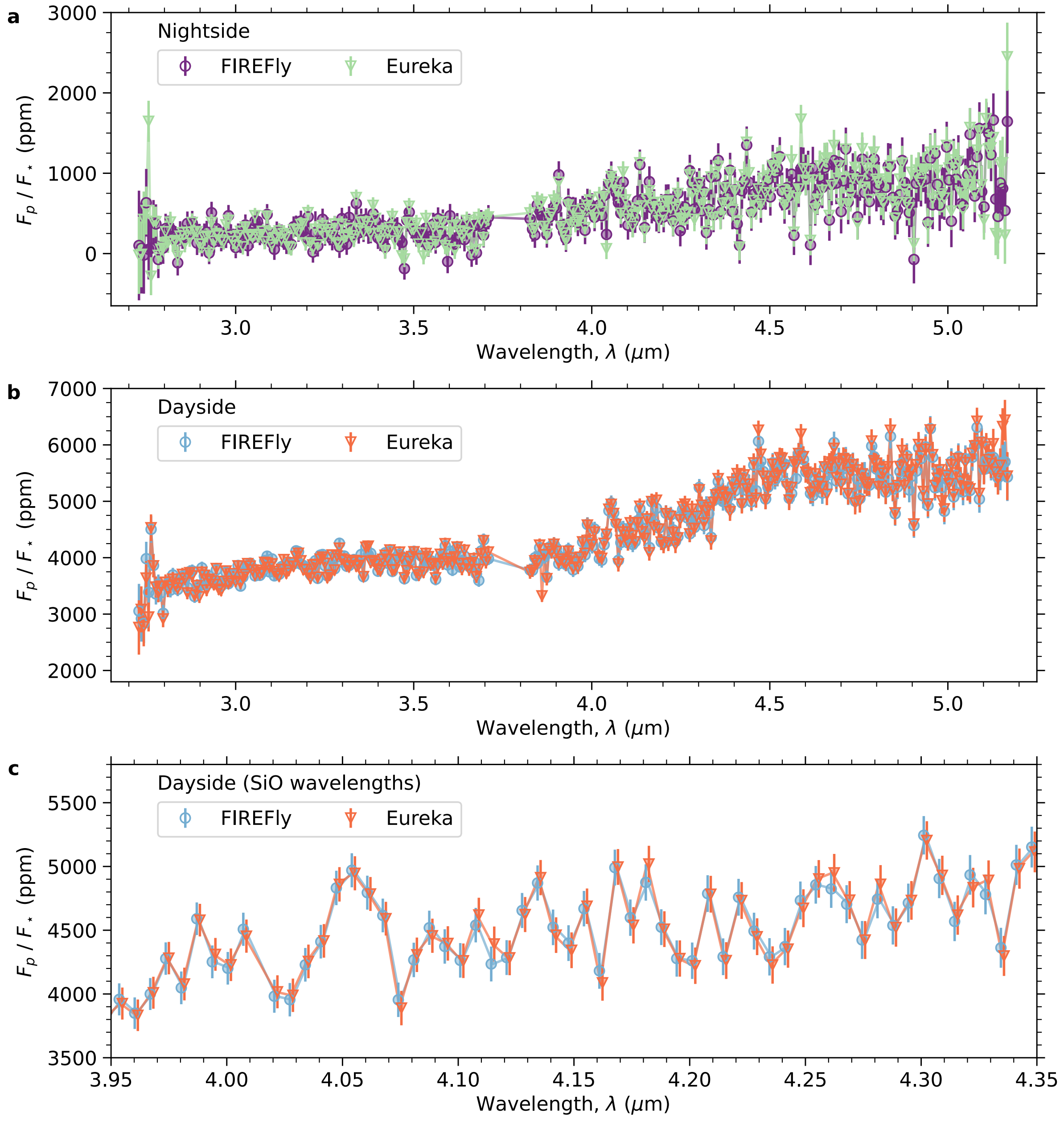

**Supplementary Fig. 4 | Spectra obtained with independent data reductions.** In our primary analysis, the raw data were reduced using the FIREFly pipeline. We also performed a second analysis, in which the independent Eureka! pipeline was used for the data reduction. **a,** A comparison of the resulting nightside ($n_\phi = 18$) emission spectra from FIREFly (purple circles) and Eureka! (green triangles). **b,** A comparison of the resulting dayside ($n_\phi = 34$) emission spectra from FIREFly (blue circles) and Eureka! (red triangles). **c,** The same as **b**, but displaying only the 3.95–4.35 µm wavelength range that encompasses the SiO signal. In **a–c**, data points show median values and error bars indicate standard deviations ($1\sigma$ uncertainties) based on $n = 5{,}000$ posterior samples.

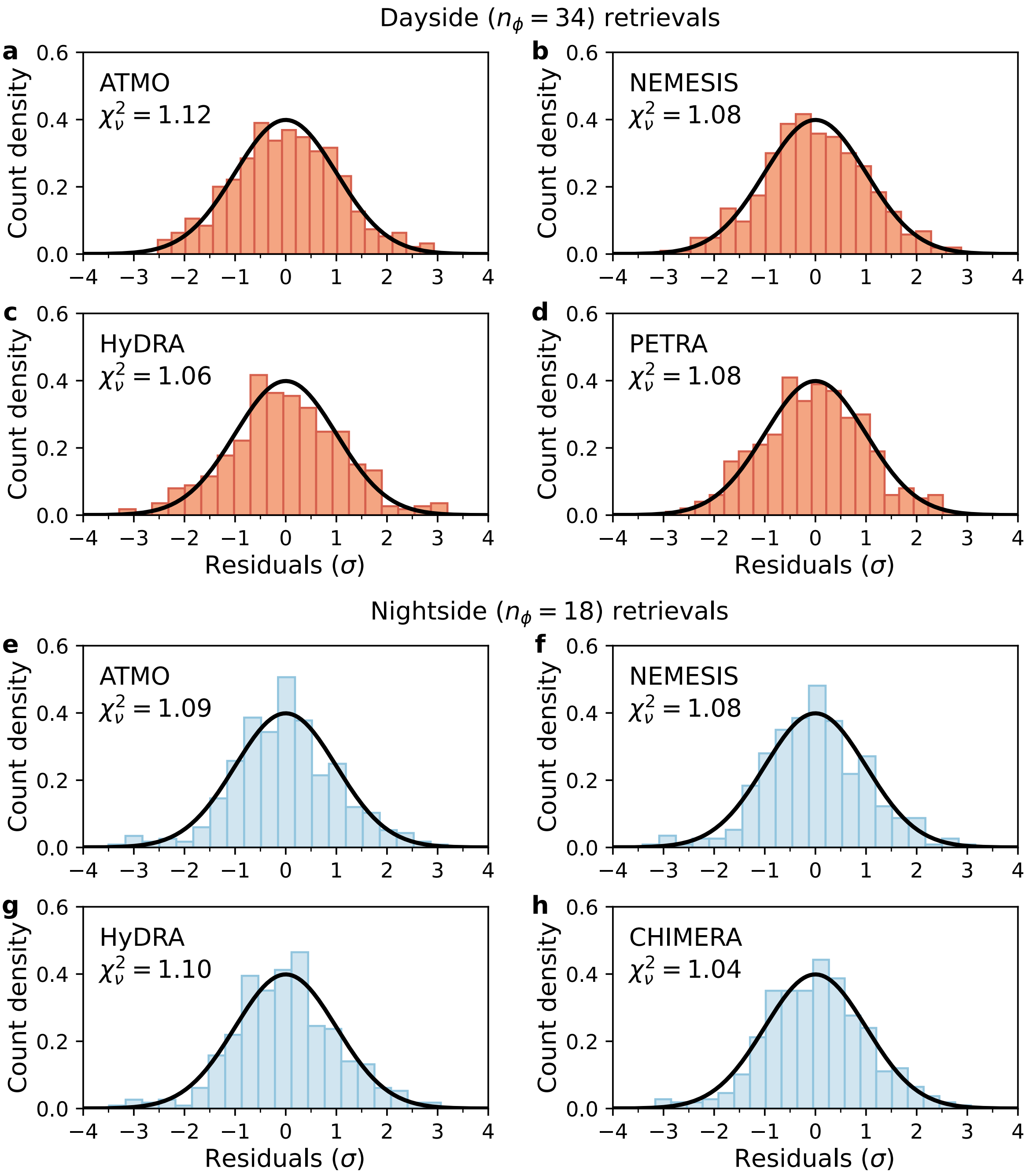

**Supplementary Fig. 5 | Normalized residuals for the retrieval best-fit models.** Histograms of residuals between the measured WASP-121b spectra and the best-fit retrieval models. Prior to binning, each residual was divided by the associated measurement uncertainty $\sigma$. Black lines show the standard normal distribution with zero mean and unity variance. Corresponding reduced $\chi^2$ values ($\chi^2_\upsilon$) are printed in each panel (that is, the $\chi^2$ fit metric divided by the number of degrees of freedom $\upsilon$). **a–d**, Residuals obtained by ATMO, NEMESIS, HyDRA and PETRA for the dayside ($n_\phi = 34$) spectrum. **e–h**, Residuals obtained by ATMO, NEMESIS, HyDRA and CHIMERA for the nightside ($n_\phi = 18$) spectrum.

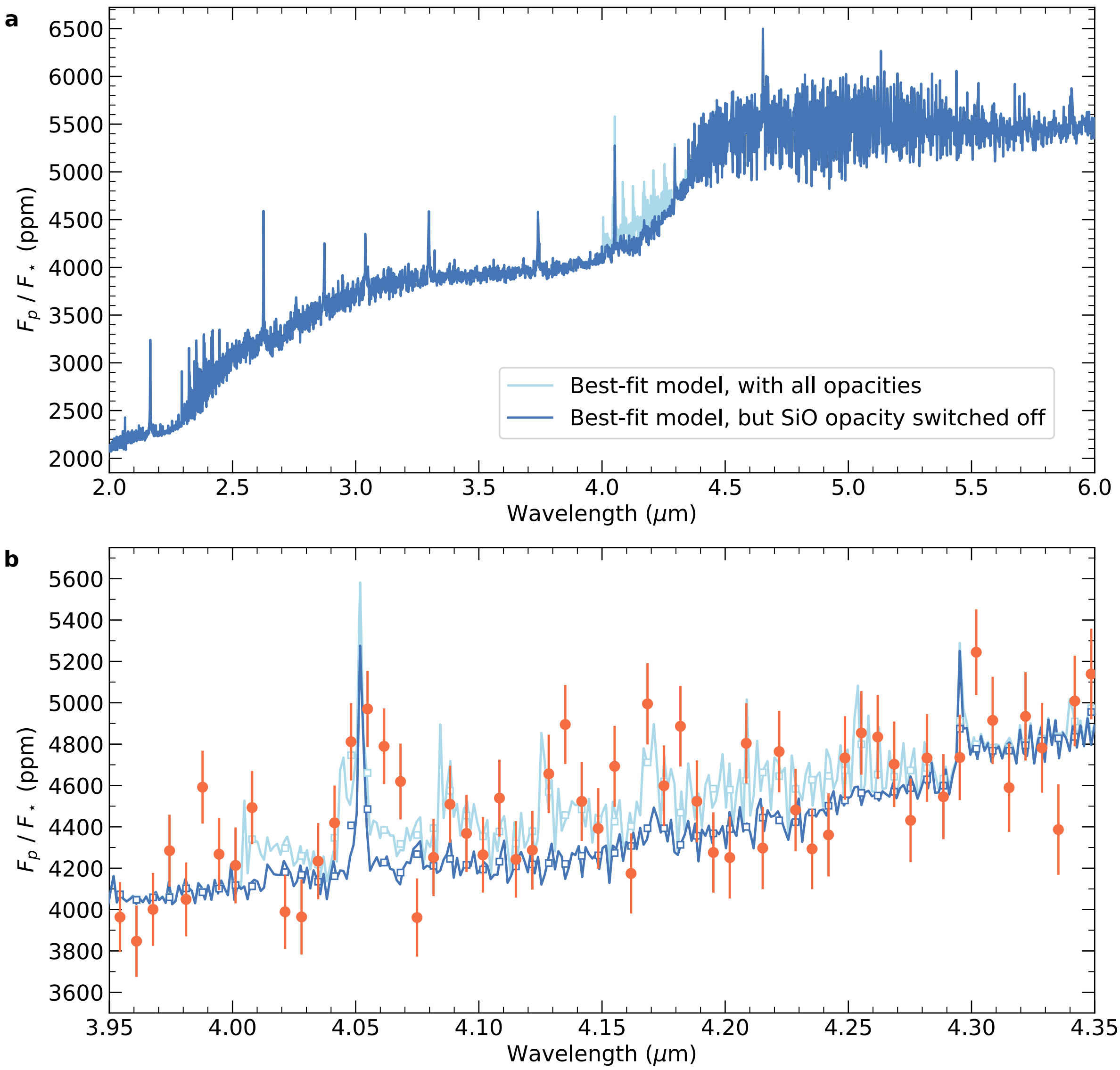

**Supplementary Fig. 6 | SiO emission signal in the dayside spectrum of WASP-121b. a,** The dark blue line shows the best-fit model obtained from the ATMO retrieval analysis, but with the SiO opacity switched off. The full best-fit model, including SiO opacity, is plotted underneath as a light blue line to highlight the SiO signal protruding across the 4.0–4.3 µm wavelength range. **b,** A zoom-in on the 4.0–4.3 µm wavelength range showing the same ATMO models as **a**, as well as the unbinned dayside emission data from Fig. 1 (red circles with error bars) and each model binned to the data wavelength channels (small squares).

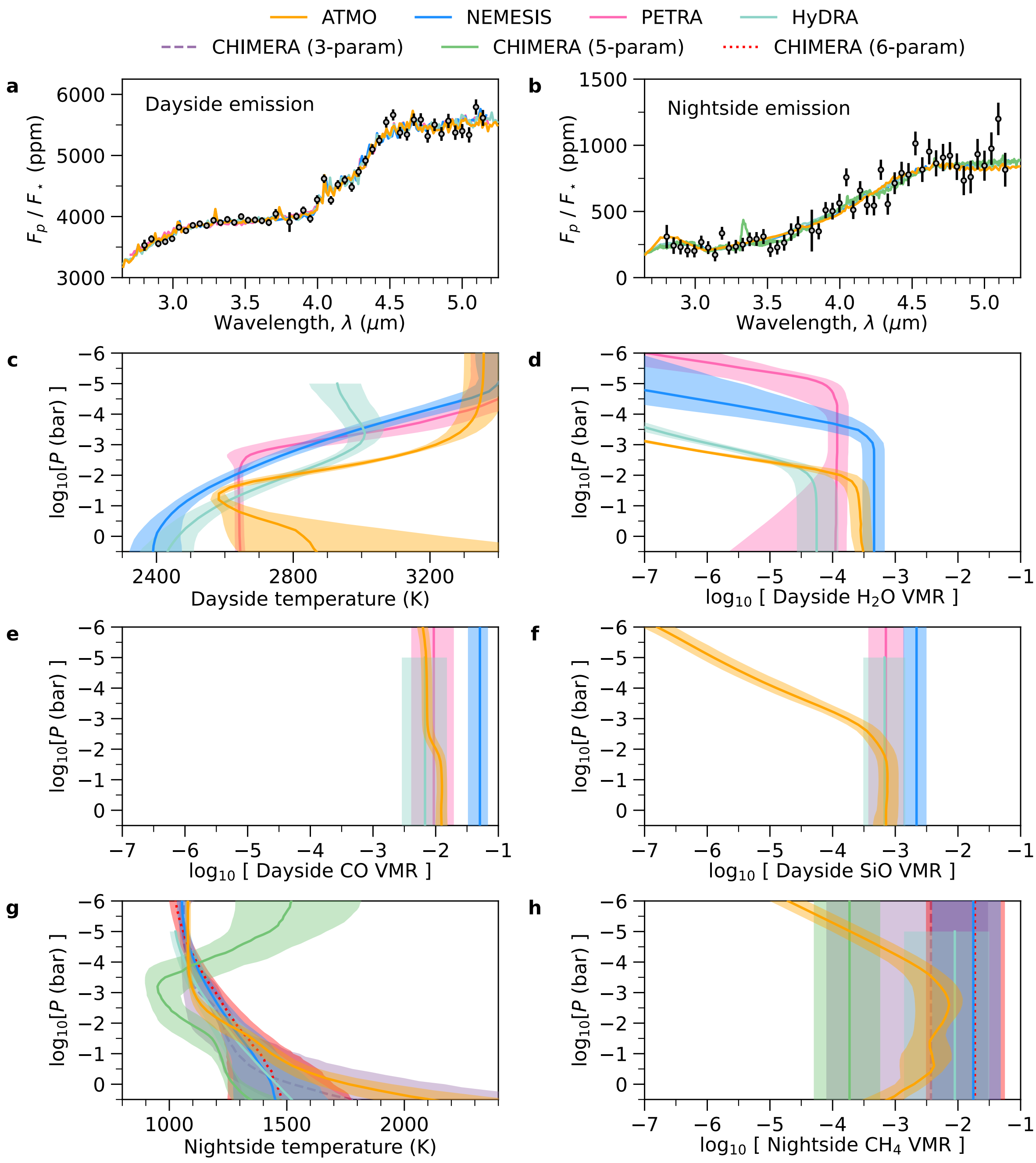

**Supplementary Fig. 7 | Comparison of retrieval results. a–b,** Black circles with grey filling show the binned emission measurements from Fig. 1 for the dayside ($n_\phi = 34$) and nightside ($n_\phi = 18$) hemispheres. Solid lines show the best-fit spectra from each retrieval code, with legend provided at the top of the figure. **c–h,** Inferred dayside PT profiles, dayside $H_2O$ VMR profiles, dayside CO VMR profiles, dayside SiO VMR profiles, nightside PT profiles and nightside $CH_4$ VMR profiles. Lines show median values and shading indicates $1\sigma$ uncertainties defined by the 16th and 84th percentiles at each pressure level, based on at least $n = 1{,}000$ posterior samples in all cases. Note that the HyDRA models were evaluated down to a pressure of 10 μbar, while the other models were evaluated down to 1 μbar or lower.

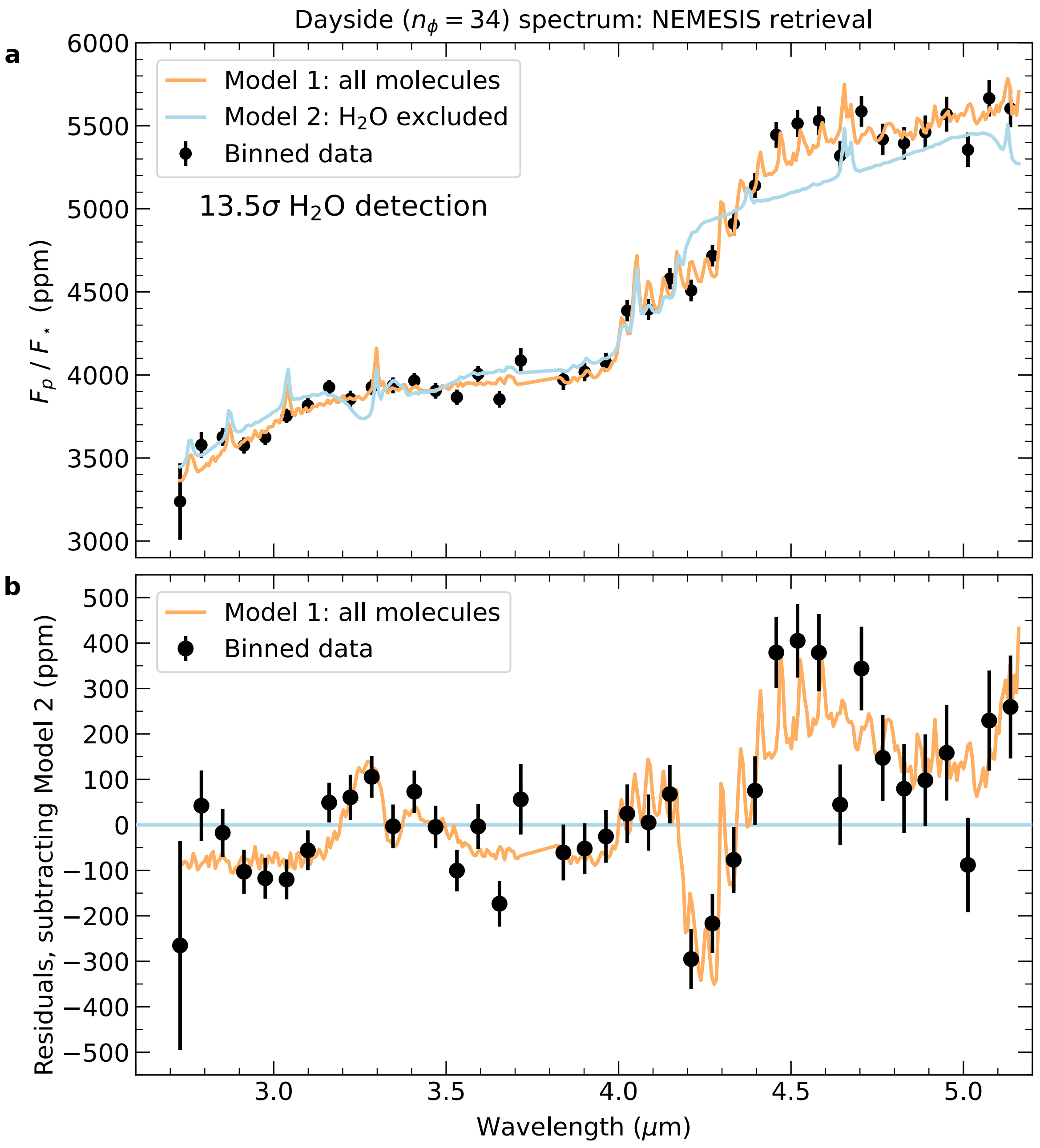

**Supplementary Fig. 8 | NEMESIS dayside $H_2O$ detection. a,** Black circles show the measured WASP-121b dayside ($n_\phi = 34$) planet-to-star emission spectrum, binned in wavelength for visual clarity. The plotted values correspond to the median of the dayside emission measurements shown in Fig. 1, within each of 40 equally spaced wavelength bins. Error bars indicate $\tilde{\sigma}_j/\sqrt{N_j}$, where $\tilde{\sigma}_j$ is the median $1\sigma$ uncertainty and $N_j$ is the number of points in the $j$th bin. The orange line shows the best-fit model obtained by NEMESIS with all chemical species considered, including $H_2O$ (Model 1). The blue line shows the best-fit model obtained by NEMESIS for a separate retrieval using the same model setup, but with $H_2O$ excluded as an opacity source (Model 2). As described in Methods, by comparing the Bayesian evidence of Models 1 and 2, the detection significance is determined to be $13.5\sigma$. **b,** The same as **a** after subtracting Model 2 from both Model 1 and the data, to emphasize the wavelengths at which the inclusion of $H_2O$ improves the fit to the data.

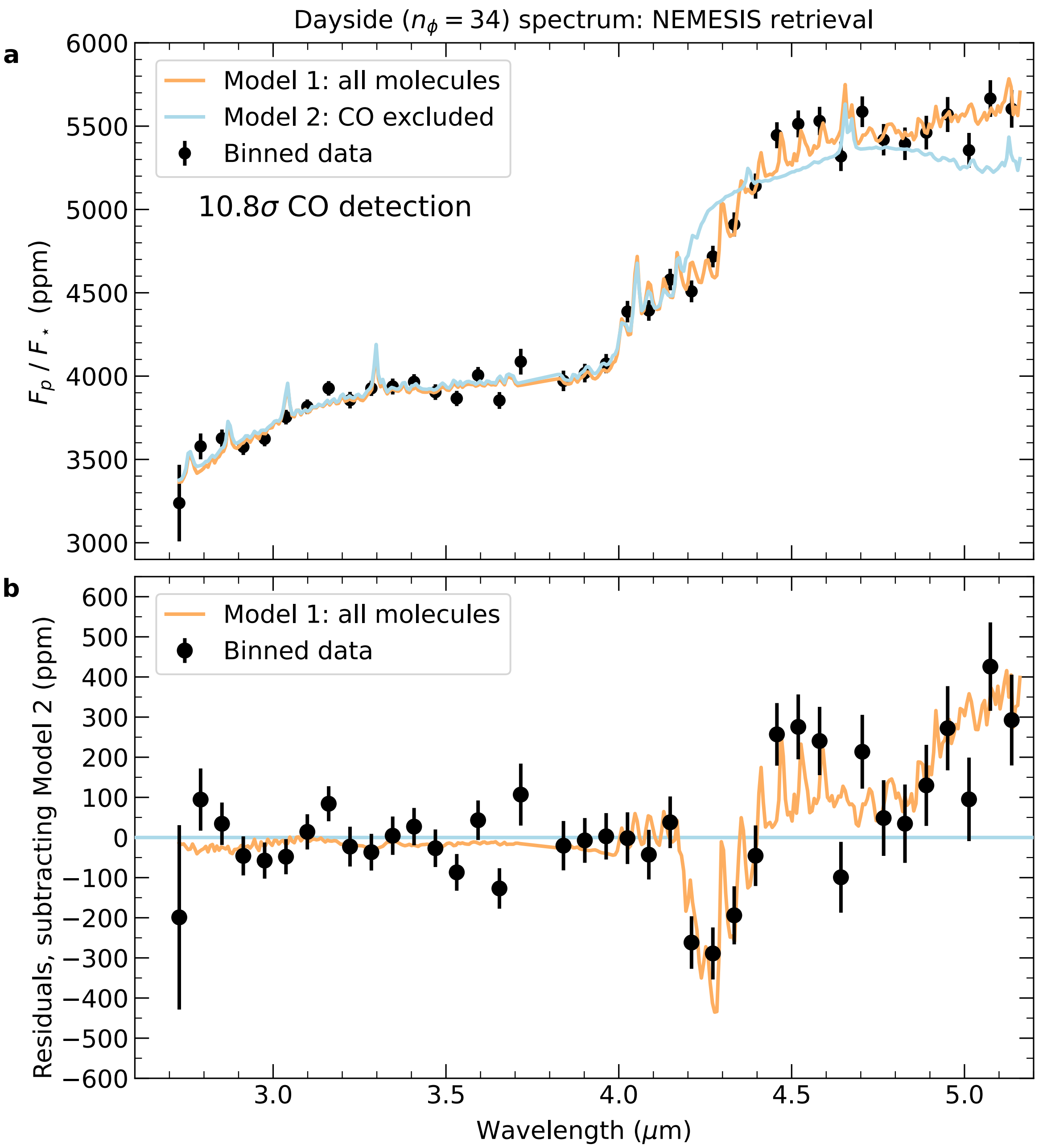

**Supplementary Fig. 9 | NEMESIS dayside CO detection.** The same as Supplementary Fig. 8 but showing the best-fit models for the WASP-121b dayside ($n_\phi = 34$) spectrum obtained by NEMESIS with and without CO.

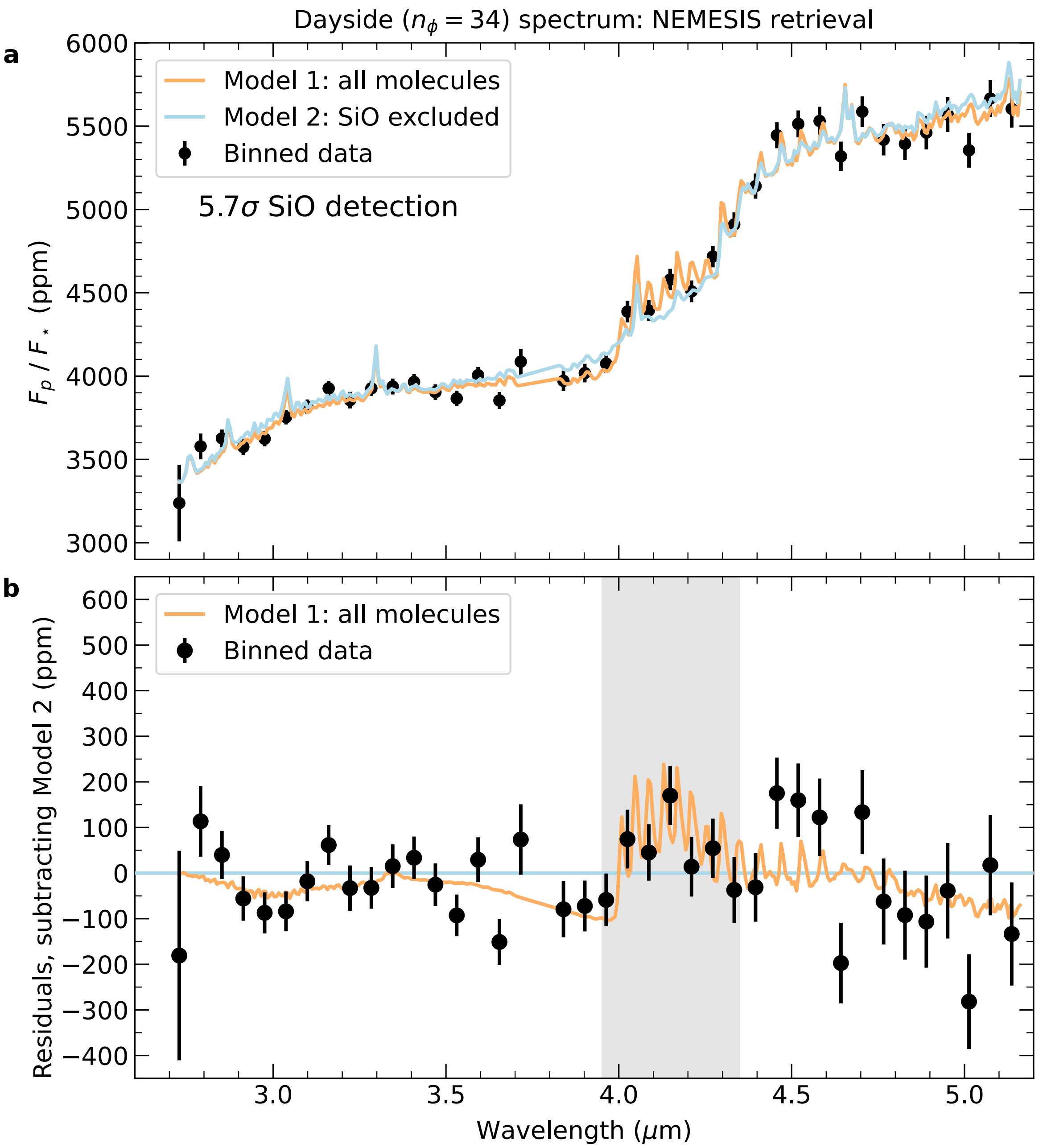

**Supplementary Fig. 10 | NEMESIS dayside SiO detection.** The same as Supplementary Fig. 8 but showing the best-fit models for the WASP-121b dayside ($n_\phi = 34$) spectrum obtained by NEMESIS with and without SiO.

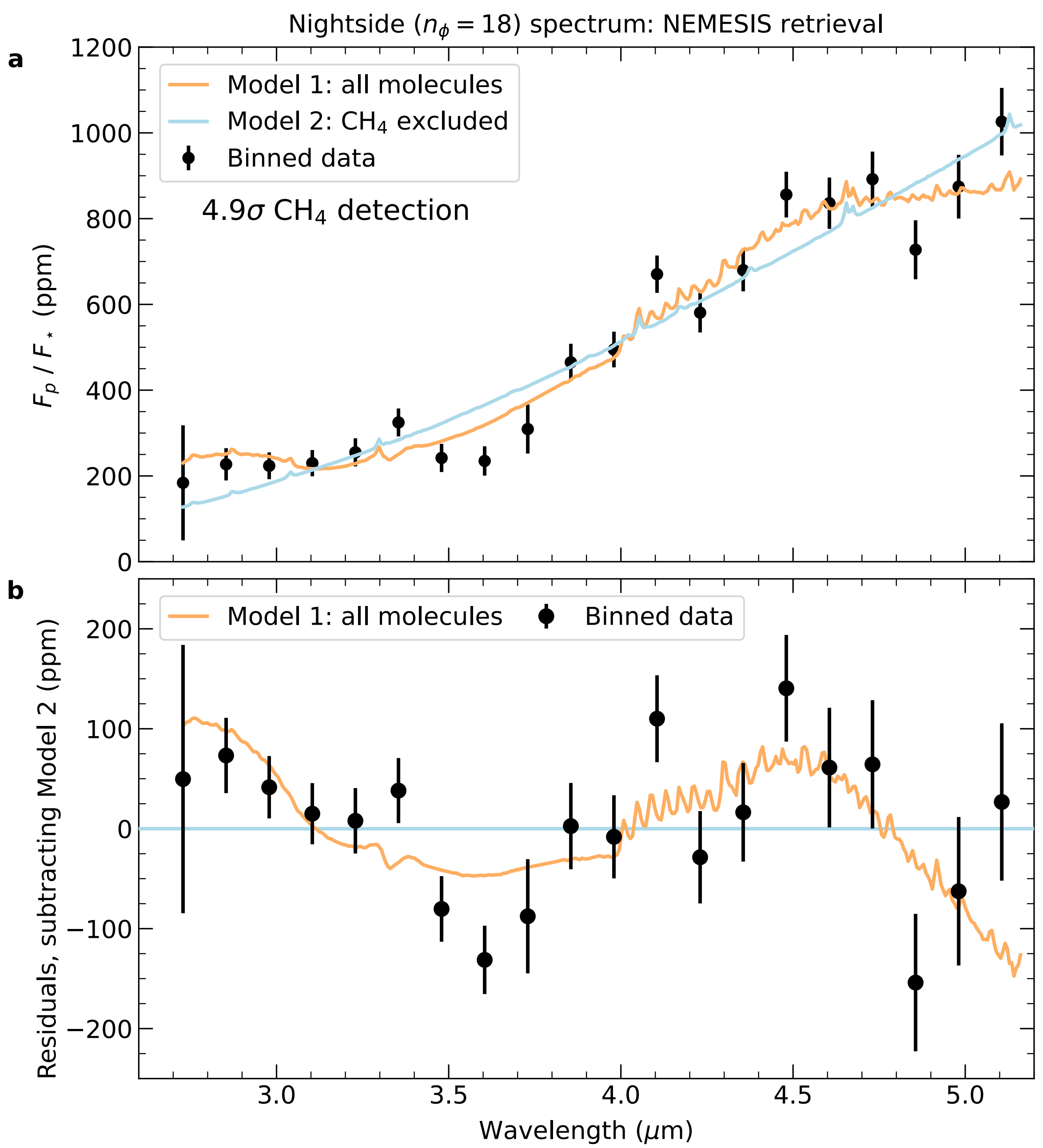

**Supplementary Fig. 11 | NEMESIS nightside $CH_4$ detection.** The same as Supplementary Fig. 8 but showing the best-fit models for the WASP-121b nightside ($n_\phi = 18$) spectrum obtained by NEMESIS with and without $CH_4$. Rather than 40 bins, only 20 bins have been used due to the relatively low signal-to-noise of the nightside spectrum compared to the dayside spectrum.

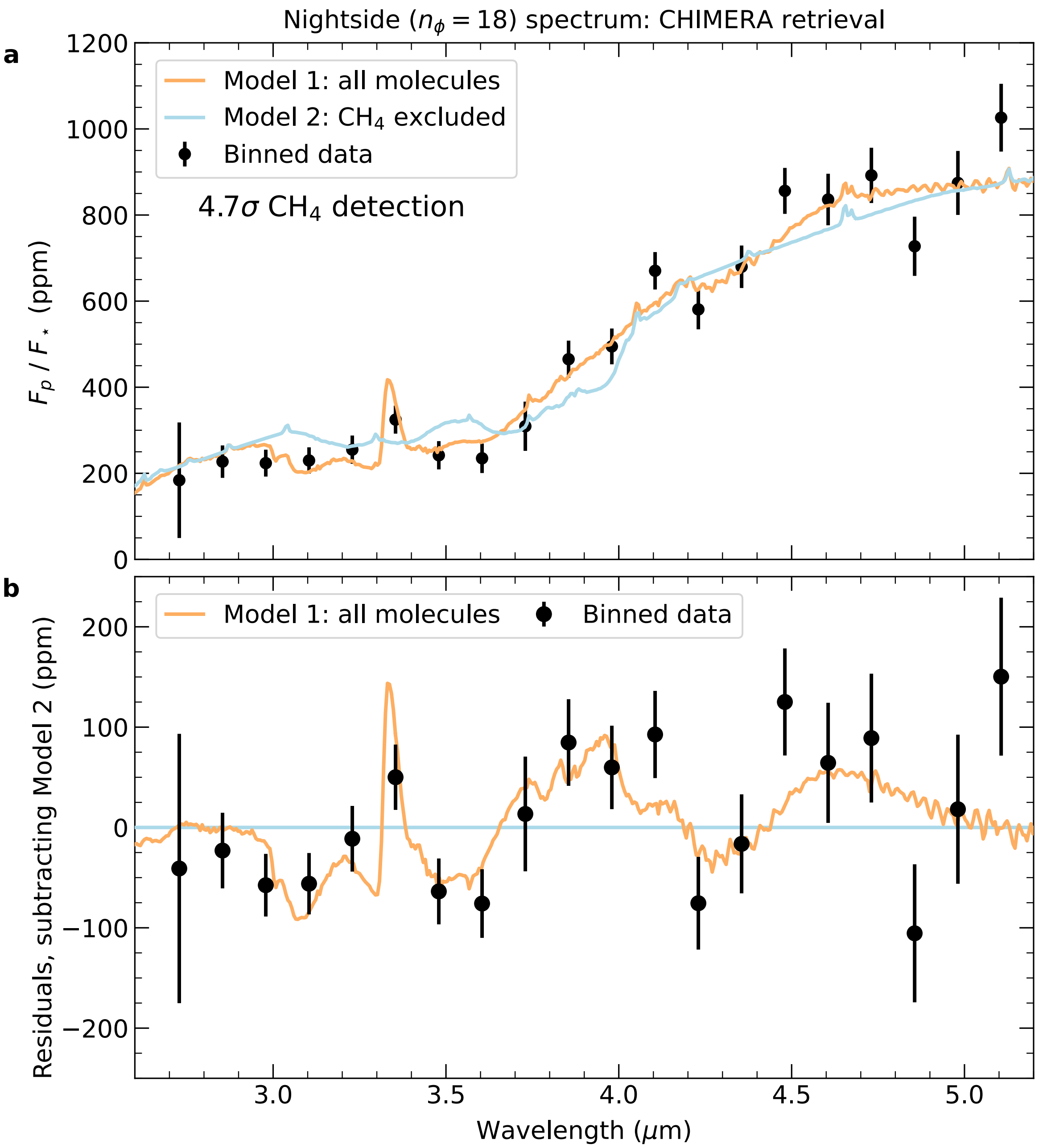

**Supplementary Fig. 12 | CHIMERA nightside $CH_4$ detection.** The same as Supplementary Fig. 11 but showing the best-fit models for the WASP-121b nightside ($n_\phi$ = 18) spectrum obtained by CHIMERA with and without $CH_4$.

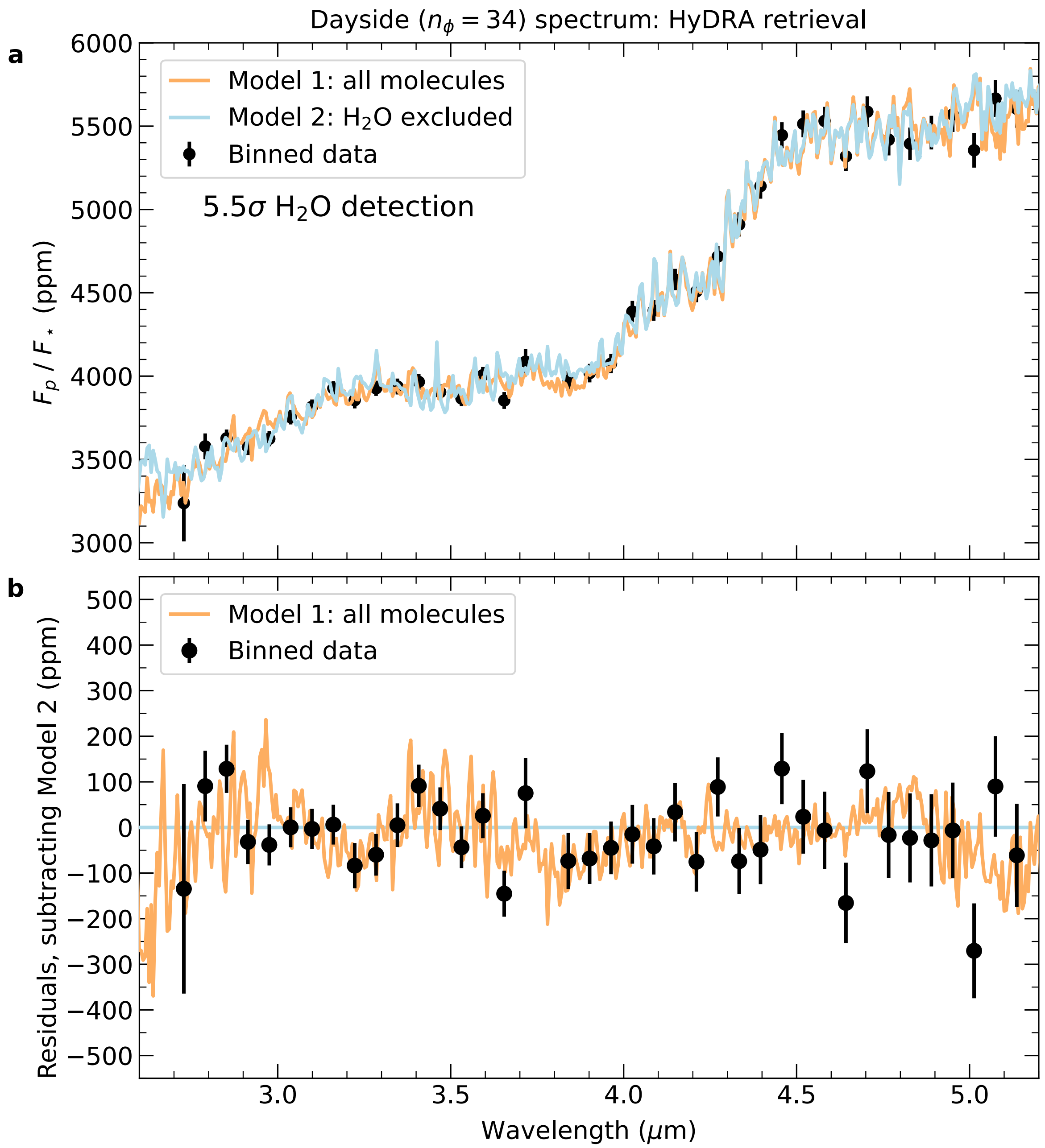

**Supplementary Fig. 13 | HyDRA dayside $H_2O$ detection.** The same as Supplementary Fig. 8 but showing the best-fit models for the WASP-121b dayside ($n_\phi = 34$) spectrum obtained by HyDRA with and without $H_2O$.

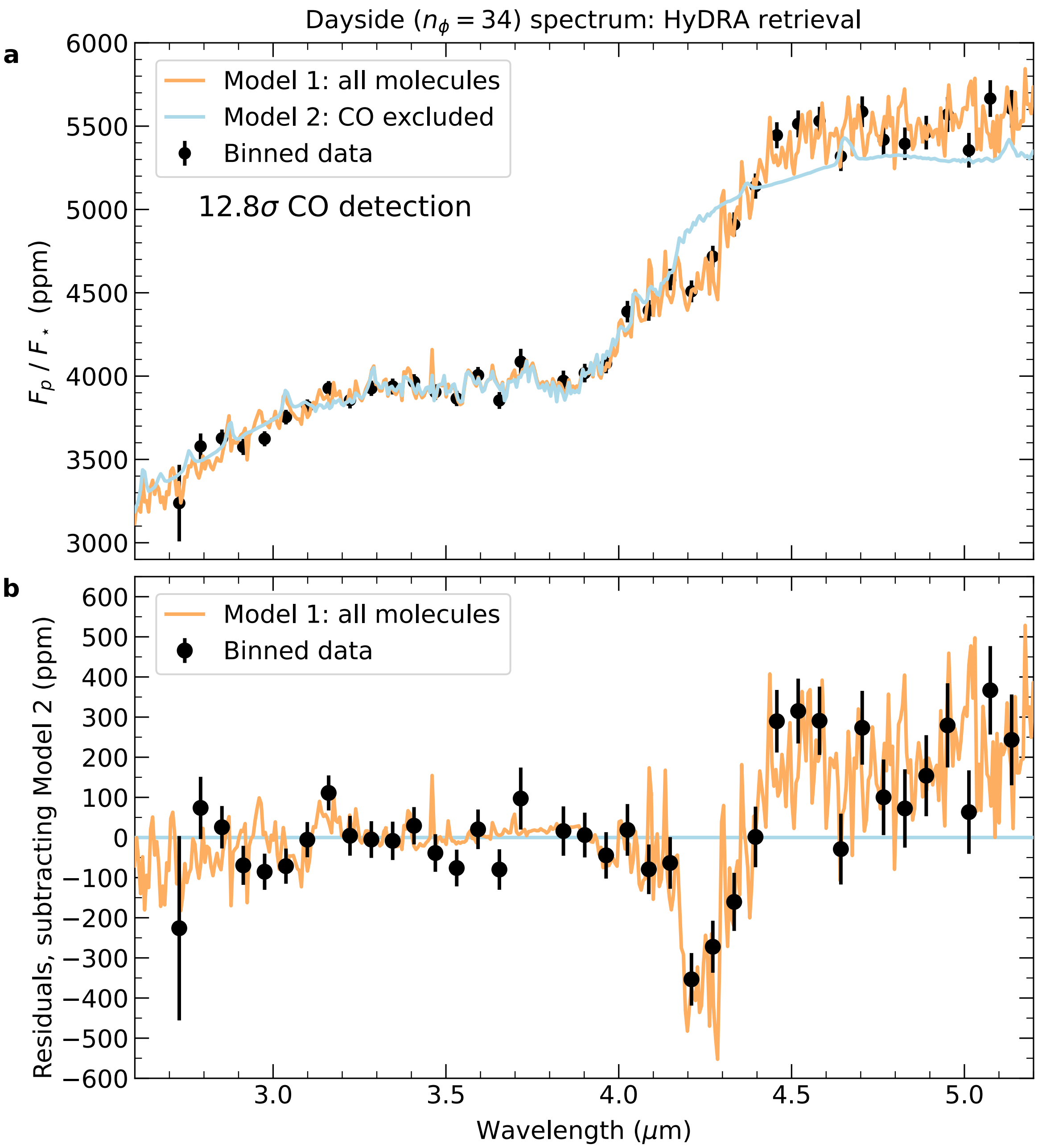

**Supplementary Fig. 14 | HyDRA dayside CO detection.** The same as Supplementary Fig. 8 but showing the best-fit models for the WASP-121b dayside ($n_\phi = 34$) spectrum obtained by HyDRA with and without CO.

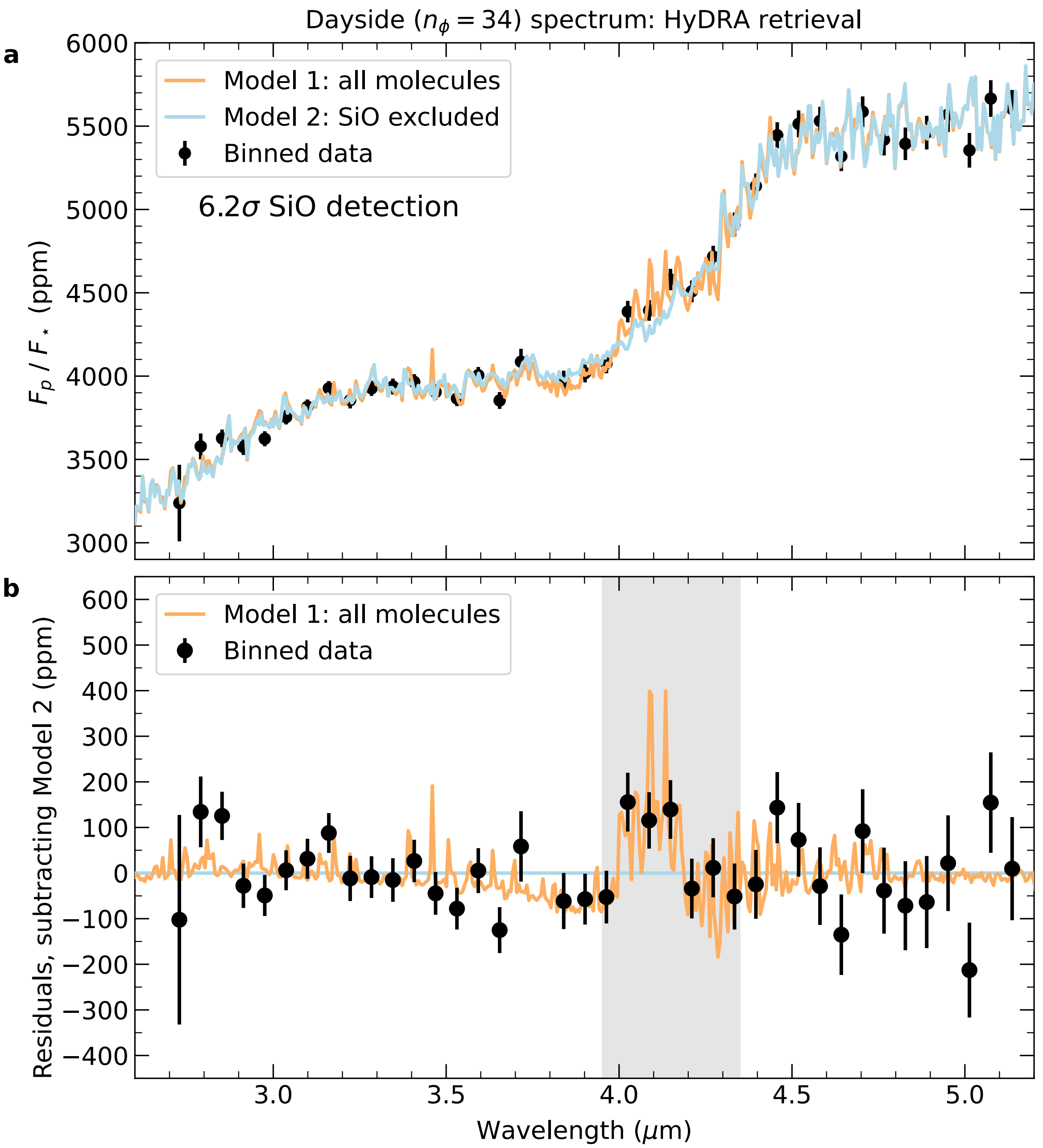

**Supplementary Fig. 15 | HyDRA dayside SiO detection.** The same as Supplementary Fig. 8 but showing the best-fit models for the WASP-121b dayside ($n_\phi = 34$) spectrum obtained by HyDRA with and without SiO.

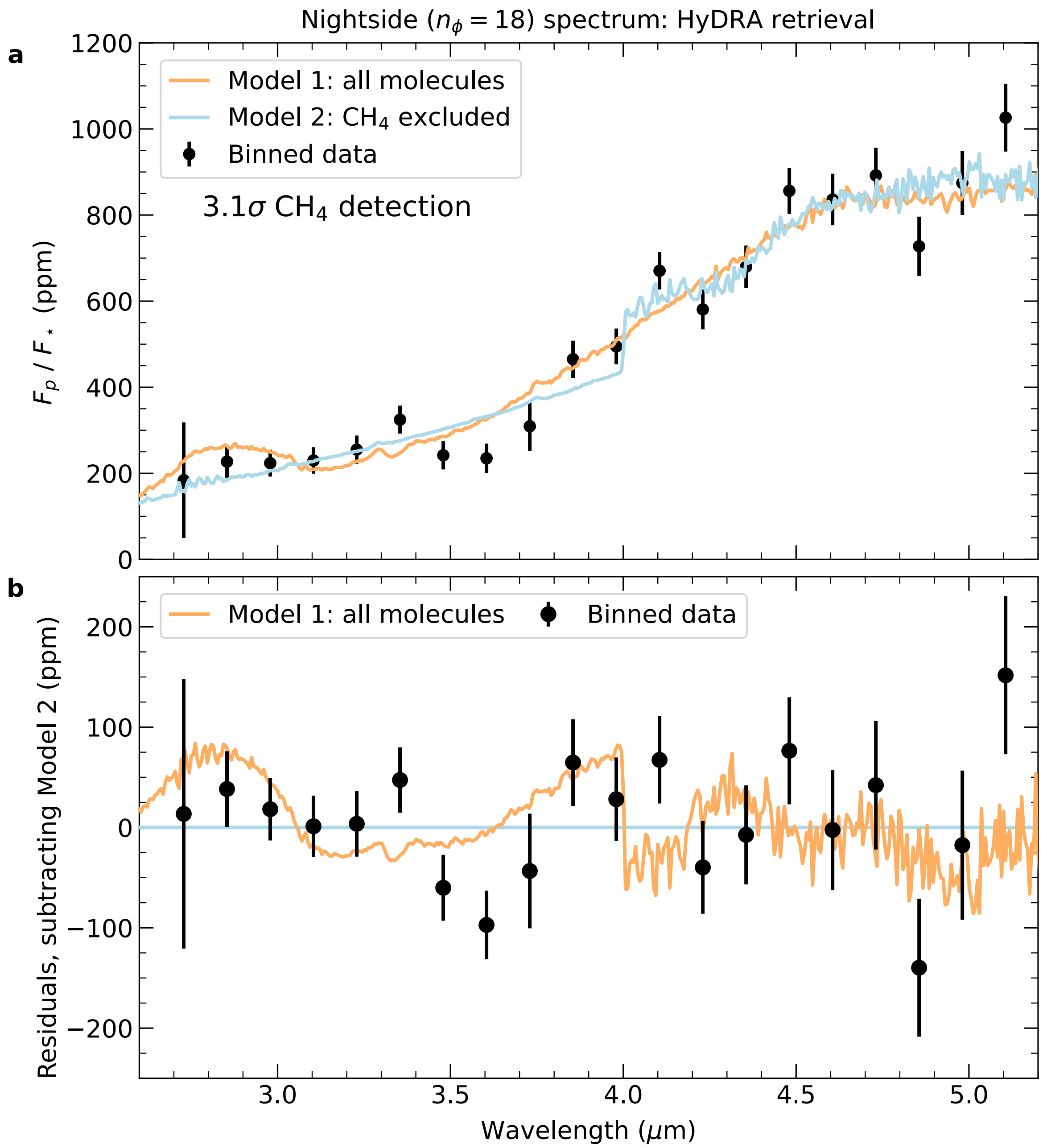

**Supplementary Fig. 16 | HyDRA nightside $CH_4$ detection.** The same as Supplementary Fig. 11 but showing the best-fit models for the WASP-121b nightside ($n_\phi = 18$) spectrum obtained by HyDRA with and without $CH_4$.

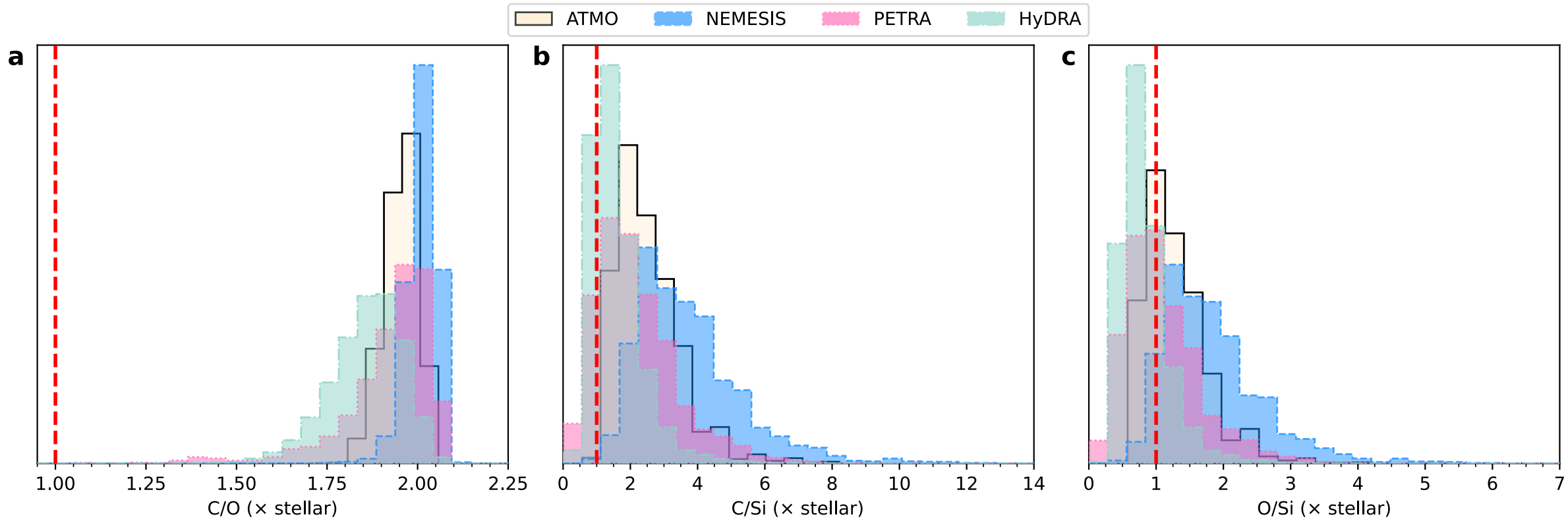

**Supplementary Fig. 17 | Element abundance ratios across retrievals.** Posterior distributions obtained for the **a**, C/O, **b**, C/Si, and **c,** O/Si ratios of the planetary atmosphere relative to the corresponding stellar ratio, as obtained from the dayside retrieval analyses. Vertical dashed lines in each panel coincide with a planetary value equal to the stellar value, where stellar abundance ratios were derived from high-resolution spectroscopy (see 'Stellar abundances and relative planetary abundances' in Methods).

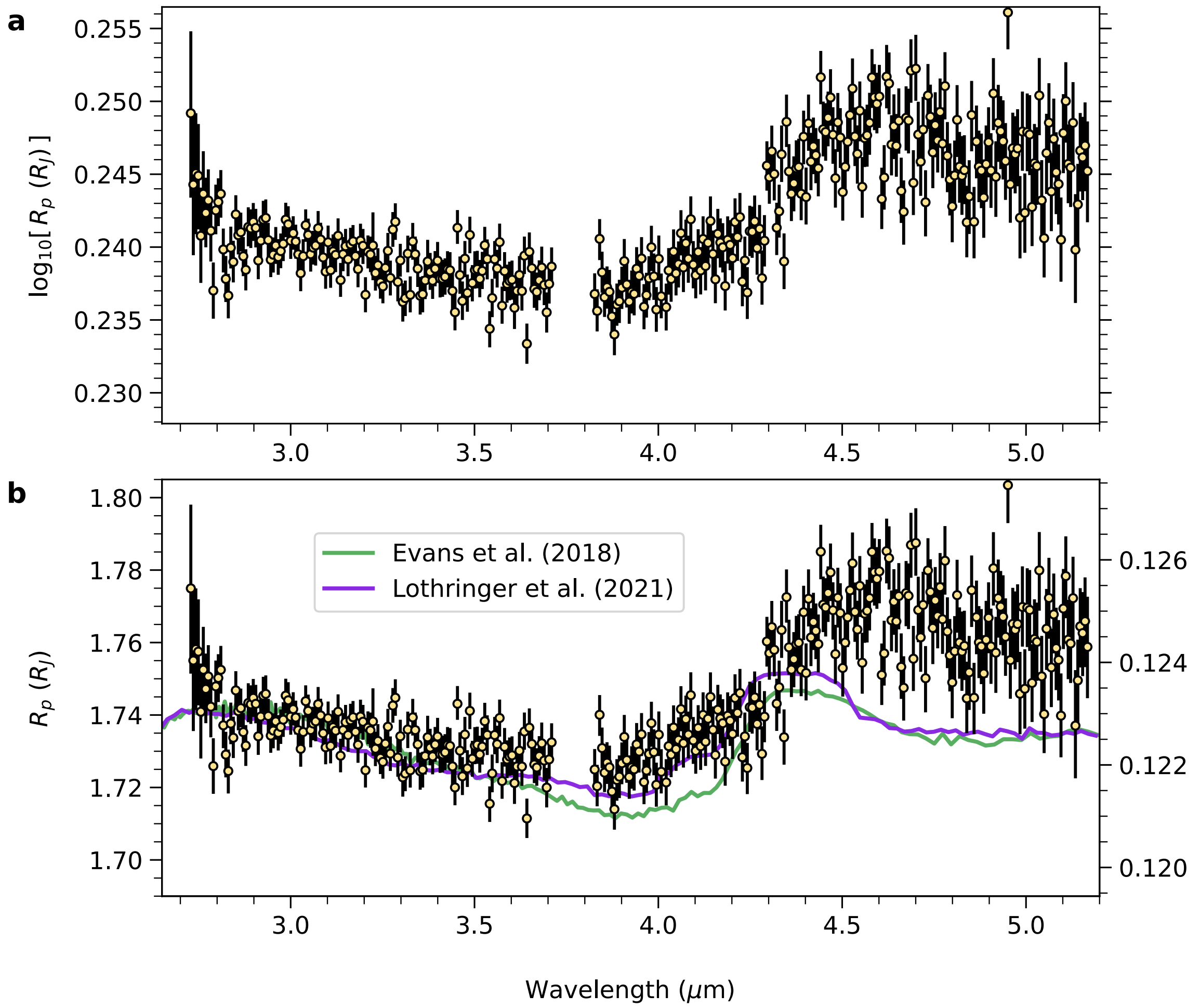

**Supplementary Fig. 18 | Planetary radius for each wavelength channel.** Circles show median values and error bars indicate $1\sigma$ uncertainties defined by the 16th and 84th percentiles of $n = 5{,}000$ posterior samples for each of the 349 spectroscopic phase-curve fits. **a,** Wavelength-dependent values for $\log_{10}[R_\mathrm{p}\ (R_\mathrm{J})]$ obtained directly from the spectroscopic phase-curve fits, where $R_\mathrm{J}$ denotes the radius of Jupiter. **b,** Wavelength-dependent values for $R_\mathrm{p}$ in units of Jupiter radii, derived from the $\log_{10}[R_\mathrm{p}\ (R_\mathrm{J})]$ values. The right vertical axis displays the corresponding planet-to-star radius ratio ($R_\mathrm{p}/R_\star$). Models from ref. [1] (purple line) and ref. [167] (green line) are also shown, both of which were obtained by fitting to HST data at shorter wavelengths and have not been fitted to the JWST data.

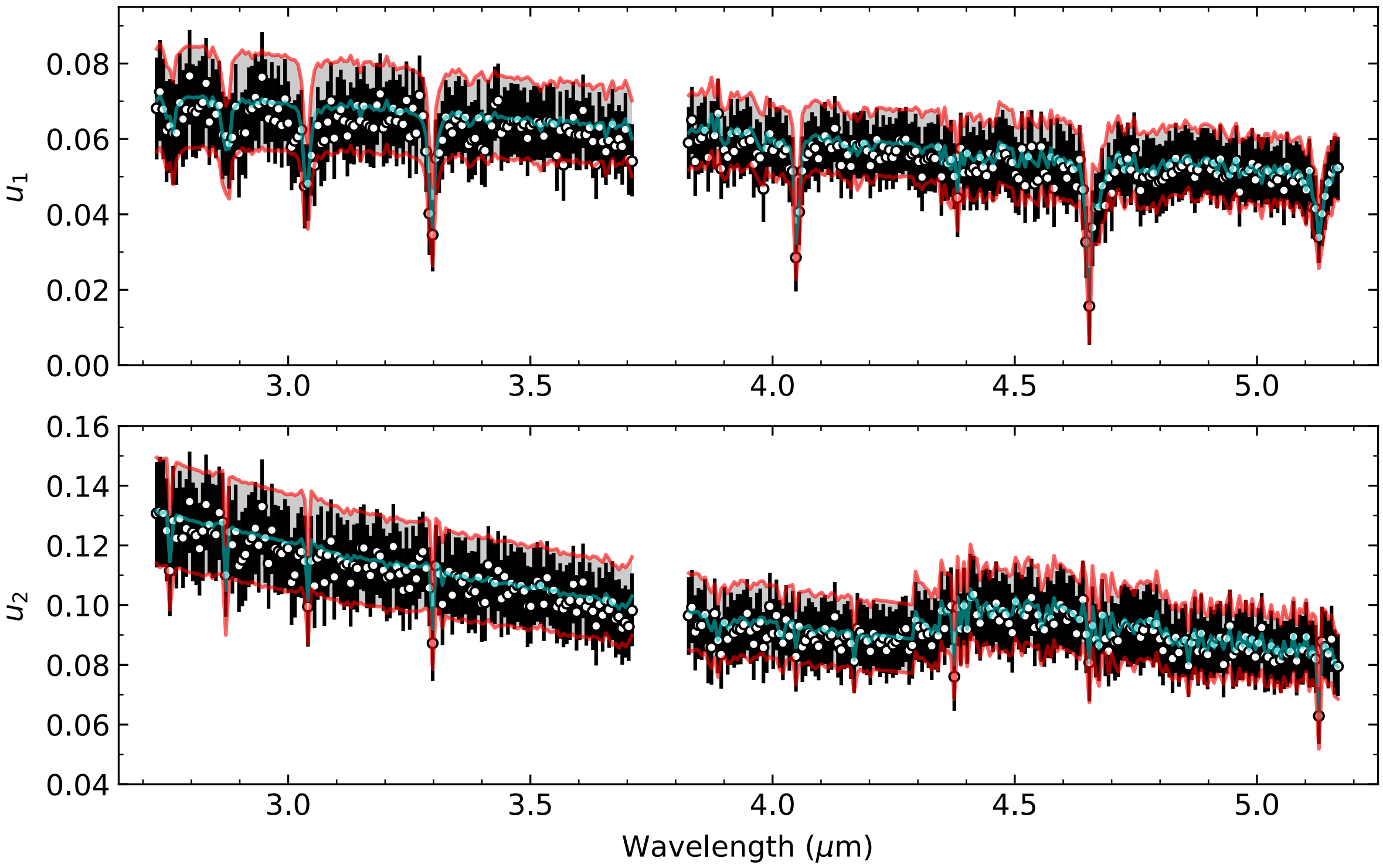

**Supplementary Fig. 19 | Stellar limb darkening.** Inferred coefficients $u_1$ (top panel) and $u_2$ (bottom panel) for the quadratic stellar limb-darkening law adopted in the phase-curve fits. In both panels, circles show median values and error bars indicate $1\sigma$ uncertainties defined by the 16th and 84th percentiles of $n = 5{,}000$ posterior samples for each of the 349 spectroscopic phase-curve fits. Light blue lines show the mean values and grey shaded regions with red boundaries indicate the standard deviations of the ExoTiC-LD normal priors adopted for each wavelength channel.

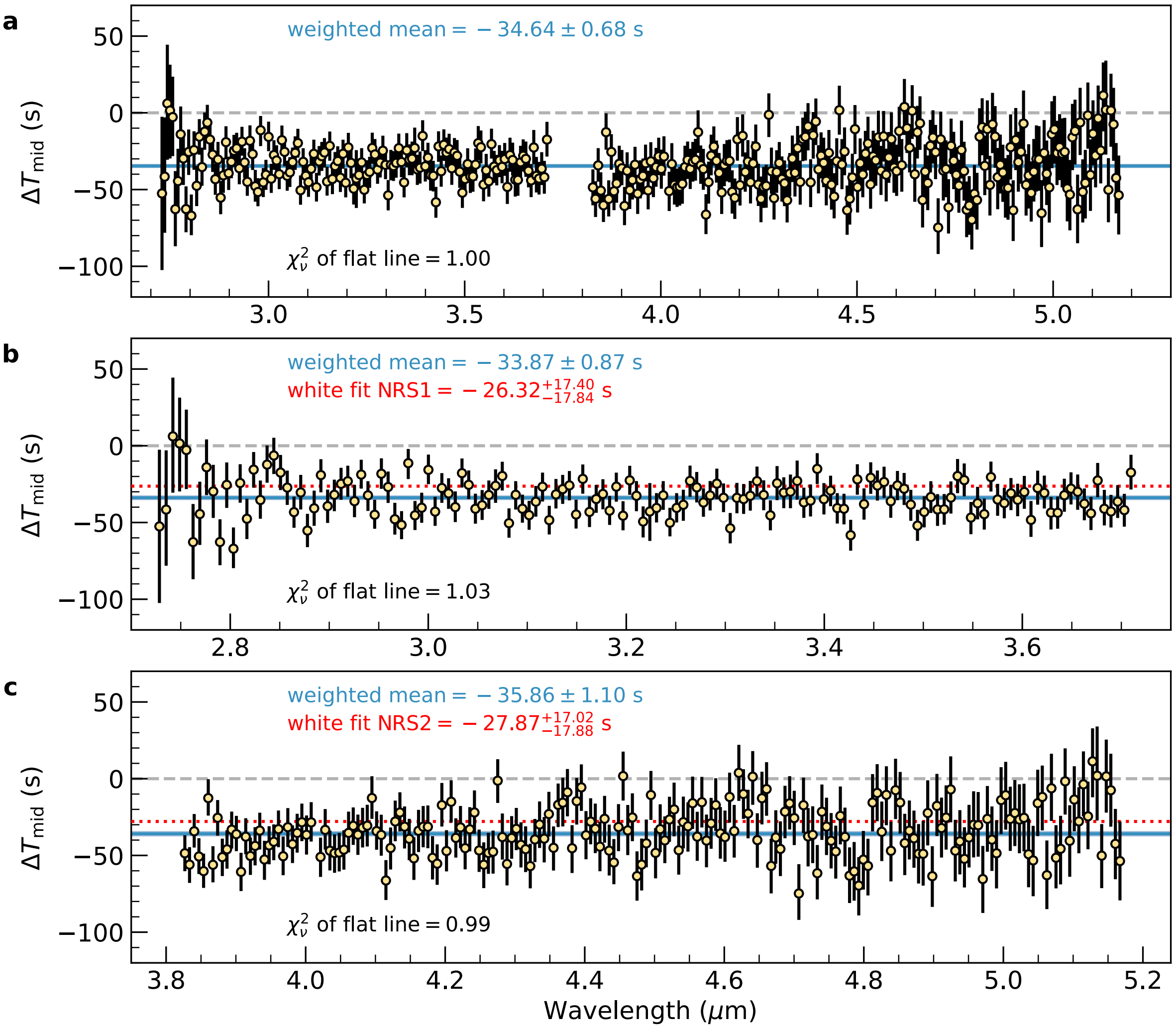

**Supplementary Fig. 20 | Transit mid-time offsets for each wavelength channel. a,** Measured transit mid-time offsets ($\Delta T_{\rm mid}$) relative to the transit mid-time reported in ref. [52], across the full G395H wavelength range. Circles show median values and error bars indicate $1\sigma$ uncertainties defined by the 16th and 84th percentiles of $n = 5{,}000$ posterior samples for each of the 349 spectroscopic phase-curve fits. **b–c,** The same as **a** but only showing the wavelength ranges of the NRS1 and NRS2 detectors. In each panel, the weighted-mean of the plotted data with uncertainty given by the associated standard deviation is printed in blue font at the top of the axis and plotted as a horizontal blue line. For the NRS1 and NRS2 panels, the values obtained from the white phase-curve fit (Supplementary Table 1) are displayed in red font for comparison and plotted as dotted red lines. Reduced $\chi^2$ values ($\chi^2_\nu$) calculated under the assumption of $\Delta T_{\rm mid}$ being constant (but not necessarily zero) at all wavelengths are printed at the bottom of each panel in black font. Dashed grey lines indicate $\Delta T_{\rm mid} = 0$ for reference.

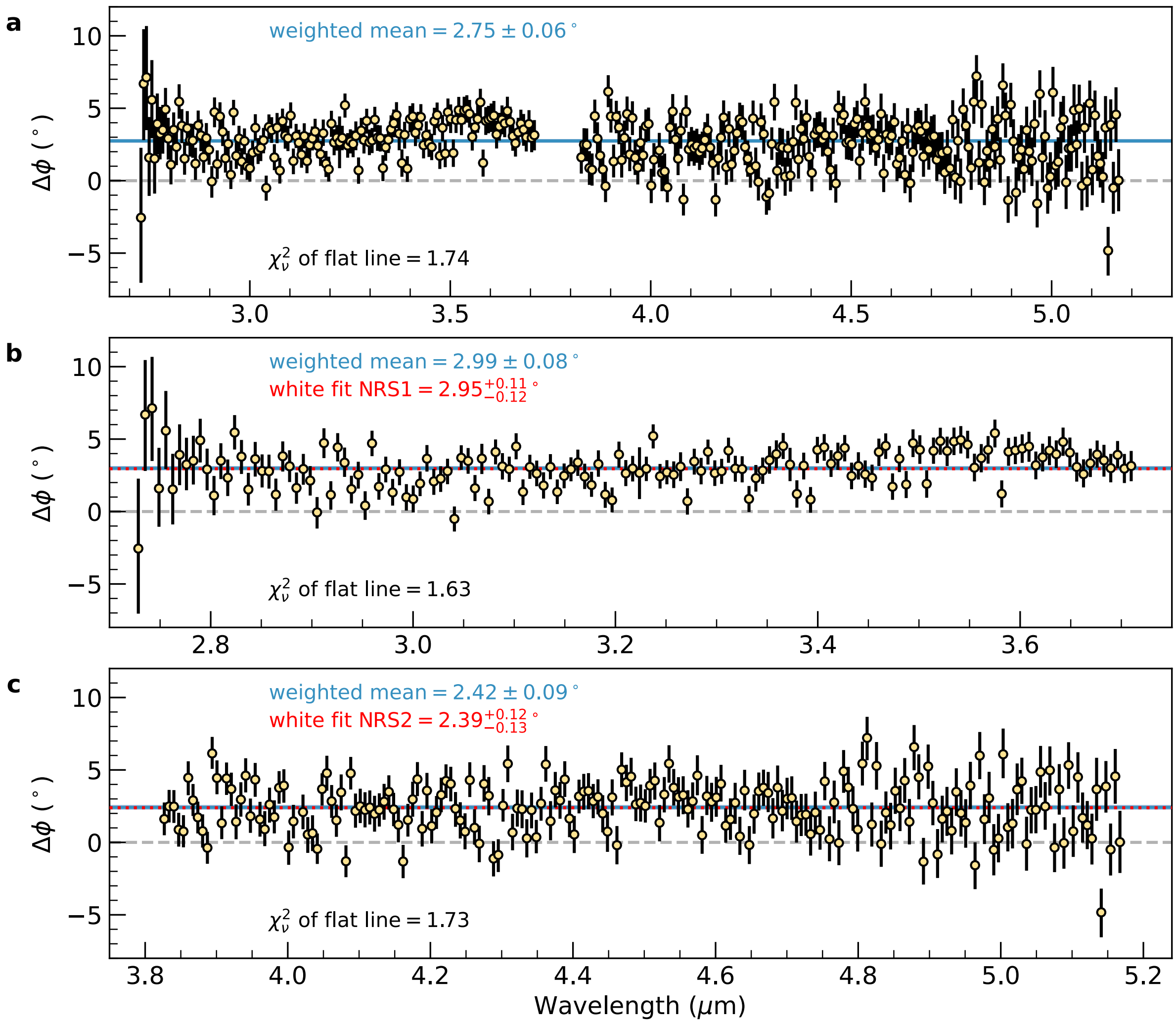

**Supplementary Fig. 21 | Planetary brightness map phase offsets for each wavelength channel.** Similar to Supplementary Fig. 20 but showing the measured planetary brightness map phase offsets ($\Delta\phi$).